\documentclass[11pt]{article}
\pdfoutput=1

\usepackage{jcappub}
\usepackage{caption}
\usepackage{subcaption}
\usepackage{hyperref}
\usepackage{graphicx}
\usepackage{amsmath}
\usepackage{mathtools}
\usepackage{physics}
\usepackage{siunitx}
\usepackage{tablefootnote}
\usepackage{soul}
\usepackage{color}
\usepackage{verbatim}
\usepackage{float}
\usepackage{bm}
\usepackage{ulem}
\usepackage[dvipsnames]{xcolor}
\usepackage{tikz}
\usetikzlibrary{shapes.geometric, backgrounds, arrows, calc, fit}
\tikzstyle{process} = [rectangle, rounded corners, minimum width=8cm, minimum height=1cm, text centered, draw=black, fill=blue!30]
\tikzstyle{arrow} = [thick,->,>=stealth]

\usepackage{color,appendix,latexsym,amsmath,amssymb,graphicx,booktabs,epsfig,hyperref,url}

\numberwithin{equation}{section}% numera le equazioni seconde le sezioni , e.g. 1.15 invece che consecutivamente; anche le appendici, eq.~(A.1) etc. Richiede amsmath
\usepackage[english]{babel}

\usepackage{letltxmacro}
\LetLtxMacro{\originaleqref}{\eqref}

\addto\extrasenglish{%
}

\definecolor{MyBlue}{rgb}{0.15,0.15,0.70}
\hypersetup{
colorlinks=true,
citecolor=MyBlue,
linkcolor=MyBlue,
urlcolor=MyBlue
}

\definecolor{orange}{rgb}{0.98, 0.6, 0.01}
\definecolor{darkolivegreen}{rgb}{0.33, 0.42, 0.18}
\definecolor{tealblue}{rgb}{0.21, 0.46, 0.53}

\usepackage{xspace}
\newcommand{\CLASS}{\texttt{CLASS}\xspace}
\newcommand{\CLASSGW}{\texttt{GW\textunderscore CLASS}\xspace}
\newcommand{\MontePy}{\texttt{MontePython}\xspace}
\newcommand{\OmGW}{\ensuremath{\bar{\Omega}_{\rm GW}}}
\newcommand{\ClGW}{\ensuremath{C_\ell^{\rm CGWB \times CGWB}}}
\newcommand{\fdec}{\ensuremath{f_{\rm dec}(\eta_\mathrm{in})}}

\usepackage{listings}
\definecolor{codegreen}{rgb}{0,0.6,0}
\definecolor{codegray}{rgb}{0.5,0.5,0.5}
\definecolor{codepurple}{rgb}{0.58,0,0.82}
\definecolor{backcolour}{rgb}{0.95,0.95,0.92}
\graphicspath{figures/} 

\title{\CLASSGW: Cosmological Gravitational Wave Background in the Cosmic Linear Anisotropy Solving System}

\author[1,2]{Florian Schulze,}
\author[3,4]{Lorenzo Valbusa Dall'Armi,}
\author[2]{Julien Lesgourgues,}
\author[5,6,3]{Angelo Ricciardone,}
\author[3,4,7]{Nicola Bartolo,}
\author[3,4,7]{Daniele Bertacca,}
\author[2]{Christian Fidler,}
\author[3,4,7,8]{Sabino Matarrese}

\affiliation[1]{Max-Planck-Institut f\"ur Kernphysik, Saupfercheckweg 1, 69117 Heidelberg, Germany}
\affiliation[2]{Institute for Theoretical Particle Physics and Cosmology (TTK), RWTH Aachen University, \\ D-52056 Aachen, Germany}
\affiliation[3]{Dipartimento di Fisica e Astronomia “Galileo Galilei”, \\Universit\`a degli Studi di Padova, via Marzolo 8, I-35131, Padova, Italy}
\affiliation[4]{INFN, Sezione di Padova, via Marzolo 8, I-35131, Padova, Italy}
\affiliation[5]{Dipartimento di Fisica “Enrico Fermi”, Universit\`a di Pisa, Pisa I-56127, Italy}
\affiliation[6]{INFN sezione di Pisa, Pisa I-56127, Italy}
\affiliation[7]{INAF- Osservatorio Astronomico di Padova, \\ Vicolo dell’Osservatorio 5, I-35122 Padova, Italy}
\affiliation[8]{Gran Sasso Science Institute, Viale F. Crispi 7, I-67100 L’Aquila, Italy}

\emailAdd{florian.schulze@mpi-hd.mpg.de}
\emailAdd{lorenzo.valbusadallarmi@phd.unipd.it}

\date{}

\abstract{The anisotropies of the Cosmological Gravitational Wave Background (CGWB) retain information about the primordial mechanisms that source the gravitational waves and about the geometry and the particle content of the universe at early times. In this work,
we discuss in detail the computation of the angular power spectra of CGWB anisotropies and of their cross correlation with Cosmic Microwave Background (CMB) anisotropies, assuming different processes for the generation of these primordial signals. We present an efficient implementation of our results in a modified version of \texttt{CLASS} which will be publicly available. By combining our new code \CLASSGW with \MontePy, we forecast the combined sensitivity of future gravitational wave interferometers and CMB experiments to the cosmological parameters that characterize the cosmological gravitational wave background. 
}

\begin{document}

%\hfill{\small TTK-23-30}
\begin{flushleft}
ET-0151A-23\\
TTK-23-30
\end{flushleft}

\maketitle
\flushbottom

\section{Introduction}
The Cosmological Background of Gravitational Waves (CGWB) is probably the most fascinating target of future Cosmic Microwave Background (CMB) and Gravitational Wave (GW) detectors. It can be produced by different mechanisms in the early universe such as inflation, phase transitions (PTs), cosmic strings (CSs), primordial black holes (PBHs), etc. (see \cite{Maggiore:2018sht,Guzzetti:2016mkm,Caprini:2018mtu, LISACosmologyWorkingGroup:2022jok} for reviews). Each of these sources leads to a peculiar frequency spectrum that can be used to discriminate among them \cite{Caprini:2019pxz, Flauger:2020qyi}. Thanks to their remarkable sensitivity, the next generation of detectors -- such as
the Einstein Telescope (ET)~\cite{Punturo:2010zz,Maggiore:2019uih,Branchesi:2023mws}, Cosmic Explorer (CE)~\cite{Reitze:2019iox,Evans:2021gyd} or  LISA~\cite{Audley:2017drz}, BBO~\cite{Corbin:2005ny} and DECIGO~\cite{Kawamura:2006up} -- will detect the sum of a large number of individual black hole and neutron star mergers plus various relevant sources of astrophysical and cosmological backgrounds. In such a situation, the detection of the stochastic background of gravitational waves from the data will be challenging. For this reason, it is important to introduce as many observables characterizing the Stochastic Gravitational Wave Background (SGWB) signal as possible. 

It has been shown in the literature that mapping anisotropies in the SGWB  \cite{Cornish:2001hg, Mingarelli:2013dsa, Contaldi:2020rht, KAGRA:2021mth} would provide a powerful way to distinguish between (different contributions to) the astrophysical background (generated by many overlapping or faint sources) \cite{Ferrari:1998jf,Ferrari:1998ut,Phinney:2001di,Farmer:2003pa,Regimbau:2009rk,Zhu:2011bd,Regimbau:2011rp} and the cosmological one~\cite{Guzzetti:2016mkm,Caprini:2018mtu,LISACosmologyWorkingGroup:2022jok} (see \cite{Mentasti:2023gmg} for a recent forecast about detection of anisotropies with future ground-based detectors.). 
Like for CMB photons, the energy density of GWs exhibits small spatial fluctuations, which carry information both on the GW production mechanism and on the properties of the universe along each line of sight. 

The astrophysical background has been thoroughly studied in the literature. A complete gauge invariant treatment of its anisotropies has been carried out in~\cite{Bertacca:2019fnt}. The authors of \cite{Bellomo:2021mer} have recently released a branch of the Einstein-Boltzmann solver (EBS) \texttt{CLASS}~\cite{Blas:2011rf} that allows to predict the statistics of anisotropies in the Astrophysical Gravitational Wave Background (AGWB), taking into account astrophysical dependencies and projection effects. Meanwhile, reference \cite{Alba:2015cms,Contaldi:2016koz,Bartolo:2019oiq,Bartolo:2019yeu} has provided a framework to describe the statistics of anisotropies in the CGWB, which shares many common features with CMB anisotropies. 
Reference \cite{ValbusaDallArmi:2020ifo} shows that the CGWB contains information about the abundance of ultra-relativistic relics in the very early universe and about pre-recombination physics that is not accessible with other means. It also shows that CGWB anisotropies are significantly correlated with CMB anisotropies~\cite{Ricciardone:2021kel,Braglia:2021fxn}. This can be used to test systematic errors and/or new physics in both CMB and SGWB measurements, or to assess the significance of CMB anomalies~\cite{Galloni:2022rgg}.

In this paper, we present and release\footnote{As soon as this paper is accepted, the code will be publicly accessible as a new branch \texttt{GW\textunderscore CLASS} on the \texttt{CLASS} repository \url{www.github.com/lesgourg/class_public}. 
The repository will include an explanatory jupyter notebook, allowing to reproduce the most important figures of this work.}\CLASSGW, a code able to compute the one-point and two-point statistics of the CGWB anisotropies generated by a list of plausible mechanisms. This code is an extension of the publicly available EBS \texttt{CLASS},\footnote{\url{http://www.class-code.net}} which has been developed to compute mainly CMB, cosmic shear and galaxy number count anisotropies~\cite{Lesgourgues:2011re,Blas:2011rf}.

The \CLASSGW code takes into account the production mechanism of the GW background in the very early universe (i.e. GW decoupling) and its propagation through large-scale cosmic inhomogeneities.  It incorporates the computation of the CGWB angular power spectrum $C_\ell^{\rm CGWB\times CGWB}$, of the CMB$\times$CGWB angular cross-correlation spectrum and of the spatially-averaged energy density spectrum $\OmGW(f)$.

The CGWB anisotropies do not carry directly an information on the GW monopole $\OmGW(f)$. However, since gravitons do not thermalise, the knowledge of $\OmGW(f)$ is necessary to compute the spectrum $C_\ell^{\rm CGWB\times CGWB}$ observable at a given GW detector frequency $f$~\cite{Alba:2015cms}. One can actually play with the $f$-dependency to optimize the detection of anisotropies~\cite{Allen:1996gp}.

Among many possible cosmological mechanisms of GW production, we focus here on inflationary models with a blue tilt ($n_{T} > 0$), first-order phase transitions and scalar-induced GWs that may lead to the formation of PBHs. According to current knowledge, these mechanisms are the most likely to be detectable by GW interferometers like ET~\cite{Punturo:2010zz,Maggiore:2019uih}, CE~\cite{Reitze:2019iox,Evans:2021gyd} and LISA~\cite{Audley:2017drz, Caprini:2015zlo, Bartolo:2016ami}. In a first step, the \CLASSGW code can be used to predict the monopole $\OmGW(f)$, to compare it with the sensitivity of the detector, and to check whether the chosen generation mechanism may lead to a detection. If this is the case, \CLASSGW is used again to compute the angular power spectrum $C_\ell^{\rm CGWB\times CGWB}$ and then, in combination with the parameter inference package \MontePy \cite{Audren:2012wb,Brinckmann:2018cvx},  to forecast the sensitivity of the detector to model parameters. 

An important feature of \CLASSGW is that it takes into account any possible ultra-relativistic relics that may have decoupled at very high energy scales -- much larger than those usually probed by CMB observations. This allows to use CGWB anisotropies as a new tool to probe physics beyond the Standard Model~\cite{ValbusaDallArmi:2020ifo}. Furthermore, \CLASSGW computes efficiently the angular power spectrum of the CGWB at different frequencies, including the auto- and cross-correlation of the spectra at a list of frequencies that can be specified by the user. This attribute would be extremely useful in any analysis that require a precise knowledge of the shape in frequency of the stochastic background, in order to do component separation with other signals~\cite{ValbusaDallArmi:2022htu} or to minimize the impact of instrumental noise on the estimate of the anisotropies.

In section \ref{sec:BoltzmannApproach}, we summarize the formalism describing the two-point statistics of CGWB anisotropies. In section \ref{sec:CGWB_sources}, we present our modelling of GW generation mechanisms. In section \ref{sec:Cross-correlation spectra}, we show how to compute the CMB$\times$CGWB angular cross-correlation spectrum. In section \ref{sec:forecast}, we show the results of a few sensitivity forecasts for the LIGO-Virgo-Kagra (LVK) network, for the ET+CE network and for even more futuristic experiments. Our conclusions are exposed in section \ref{sec:conclusions}. In appendix \ref{sec:notations} we relate our notation to previous work. Appendix \ref{sec:adiabaticIC} and \ref{sec:Doppler} detail our assumptions concerning the initial conditions for the GW anisotropies at the time of GW decoupling, while appendix \ref{sec:app_inflation} summarizes the formalism describing the generation of GWs during inflation. 
Appendix \ref{sec:Il} presents a calculation that helps to understand the shape of the CMB$\times$CGWB cross-correlation spectrum. Finally, appendix \ref{sec:CLASS_GW} contains technical details on the numerical implementation and accuracy of \CLASSGW.

\section{Boltzmann approach to the SGWB anisotropies}
\label{sec:BoltzmannApproach}

The summary presented in this section is based on previous results from references \cite{Bartolo:2019oiq,Bartolo:2019yeu,ValbusaDallArmi:2020ifo,Ricciardone:2021kel}. However, we switch here to notations closer to those of the \CLASS code and papers \cite{Blas:2011rf,Lesgourgues:2013bra}. Appendix~\ref{sec:notations} summarizes the correspondence between different sets of notations.

\subsection{Basic formalism}
\label{sec:basic}

In the limit of geometrical optics,
the propagation of GWs (or gravitons) can be described using the Boltzmann equation~\cite{Dodelson:2003ft}. The limit of geometrical optics is well justified if the shortwave approximation is valid, i.e., when the wavelength of the GWs under consideration is much smaller than the typical scales over which the background metric varies. In this work we are computing the anisotropies of cosmological GW background that can be detected by current and future GW interferometers, thus the smallest frequencies considered are in the Pulsar Timing Array (PTA) band (around the $\rm nHz$) and the angular scales that can be probed (typically up to a multipole $\ell_{\rm max} \approx 10-15$~\cite{LISACosmologyWorkingGroup:2022kbp, Mentasti:2023gmg}) are much larger than the wavelength of the GWs. Along each geodesic, GWs can be described by a distribution function $f_{\rm GW}(x^{\mu}, p^{\mu})$, where  $x^{\mu}$ denotes position and $p^{\mu} = d x^{\mu}/d \lambda$ momentum (which relates to the GW frequency). $f_{\rm GW}$ evolves according to the Boltzmann equation
\begin{equation}
    \label{eq:Boltzgeneric}
    \mathcal{L}[f_{\rm GW}] = \mathcal{C}[f_{\rm GW}(\lambda)] + \mathcal{I}[f_{\rm GW}(\lambda)]\,,
\end{equation}  
where  $\mathcal{L}\equiv d/ d\lambda$ is the Liouville operator along the geodesic, while $\mathcal{C}$ is the collision operator and $\mathcal{I}$ accounts for emissivity from cosmological and astrophysical sources~\cite{Contaldi:2016koz}.
In the case of the CGWB, the emissivity term can be treated as an initial condition on the GW distribution, while, in the case of the AGWB, it would be related, e.g., to black hole merging along the geodesic~\cite{Contaldi:2016koz}.
Typically, the GW collision term can be neglected, since it only affects the distribution at higher orders in the gravitational strength $M_{\rm Pl}^{-1}$ (where $M_{\rm Pl}$ is the reduced Planck mass). See e.g., \cite{Pizzuti:2022nnj} for a study of the AGWB anisotropies in presence of collision, accounting for gravitational Compton scattering off of massive objects.

We assume that our universe is well described by a perturbed Friedmann-Lemaitre-Robertson-Walker (FLRW) metric, and we adopt the Newtonian gauge in which
\begin{equation}
    \label{eq:metric}
    ds^2 = a^2(\eta)
    \left[ -e^{2\psi} d\eta^2 + 
    (e^{-2\phi}\delta_{ij} + h_{ij}) dx^i dx^j\right]\, ,
\end{equation}
where $a(\eta)$ is the scale factor as a function of conformal time $\eta$, $\phi(\eta, x^i)$ and $\psi(\eta, x^i)$ account for scalar perturbations, and the transverse-traceless (TT) stress tensor $h_{ij}(\eta, x^i)$ for large-scale tensor perturbations. Then, we can solve the Boltzmann equation \eqref{eq:Boltzgeneric} at the background and linear levels. 

Once the background distribution $\bar{f}_{\rm GW}$ is expressed as a function of $\eta$ and of the comoving momentum $q = a p$ (left invariant by the expansion, while the physical momentum scales like $p\propto a^{-1}$), the background Boltzmann equation simply reads $\partial \bar{f}_{\rm GW}/ \partial \eta=0$. It is solved by any distribution that does not explicitly depend on time, namely $f_{\rm GW} = {\bar f}_{\rm GW} \left( q \right)$.
Including the linear level, the equation becomes \cite{Contaldi:2016koz, Bartolo:2019oiq, Bartolo:2019yeu}
\begin{equation}
    \label{eq:Dfc2}
    \frac{\partial f_{\rm GW}}{\partial \eta} +
    n^i \, \frac{\partial f_{\rm GW}}{\partial x^i} +
    \left[ \frac{\partial \phi}{\partial \eta} - n^i \, \frac{\partial \psi}{\partial x^i} - \frac{1}{2} n^i n^j \frac{\partial h_{ij}}{\partial \eta} \right] q \, \frac{\partial f_{\rm GW}}{\partial q} = 0 \,,
\end{equation}
where the unit vector $\hat{n}$ of components $n^i = p^i/p$ denotes the direction of propagation of GWs.

The perturbed GW energy at a given time and location is given by integrating over momentum (or frequency) and direction,
\begin{eqnarray}
    \label{eq:rho-GW}
    \rho_{\rm GW} \left( \eta ,\, \vec{x} \right) &=& \frac{1}{a^4(\eta)}  \int \!d^2\hat{n} \int \!dq \,\, q^3 \, f_{\rm GW} \left( \eta ,\, \vec{x} ,\, q ,\, {\hat n} \right)~.
\end{eqnarray}
It is useful to introduce a dimensionless quantity $\omega_{\rm GW}$ that accounts for the contribution of GWs propagating at a given time, location, frequency and direction to the critical density:
\begin{equation}
\label{eq:omega-GW}
\omega_{\rm GW} \left( \eta ,\, \vec{x} ,\, q ,\, {\hat n} \right) =
\frac{4\pi}{\rho_{\rm crit}(\eta)} \frac{q^4}{a^4(\eta)} \, f_{\rm GW} \left( \eta ,\, \vec{x} ,\, q ,\, {\hat n} \right)~,
\end{equation}
where $\rho_{\rm crit}= 3 H^2 M_{\rm Pl}^2$ and $H$ the Hubble rate. The monopole of $\omega_{\rm GW}$ is denoted as
\begin{equation}
\label{eq:Omega-GW}
\Omega_{\rm GW} \left( \eta ,\, \vec{x} ,\, q  \right)=
\int \! \frac{d^2\hat{n}}{4\pi} \,\, \omega_{\rm GW} \left( \eta ,\, \vec{x} ,\, q ,\, {\hat n} \right)~,
\end{equation}
and the average value of this monopole as
\begin{equation}
\label{eq:Omega-bar-GW}
\bar{\Omega}_{\rm GW} \left( \eta ,\, q  \right)= \langle \Omega_{\rm GW} \left( \eta ,\, \vec{x} ,\, q  \right) \rangle_{\vec{x}}~.
\end{equation}
With such definitions, the anisotropy of the CGWB density at the detector's time $\eta_0$,  location $\vec{x}_0$ and momentum/frequency $q$ can be simply defined as:
\begin{equation}
    \label{eq:gwdensitycontrast1}
    \delta_{\rm GW}(\eta_0, \vec{x}_0, q, \hat{n}) \equiv
    \frac{\omega_{\rm GW} ( \eta_0 ,\, \vec{x}_0 ,\, q ,\, {\hat n} ) -\Omega_{\rm GW}(\eta_0 ,\, \vec{x}_0 ,\, q)}{\Omega_{\rm GW}(\eta_0,\, \vec{x}_0 ,\, q)} \,.
\end{equation}
In very good approximation, we can assume that the detector occupies a typical position in the universe where $\Omega_{\rm GW}(\eta_0,\, \vec{x}_0 ,\, q)= \bar{\Omega}_{\rm GW}(\eta_0,\, q)$ and redefine the anisotropy as
\begin{equation}
    \label{eq:gwdensitycontrast}
    \delta_{\rm GW}(\eta_0, \vec{x}_0, q, \hat{n}) \equiv
    \frac{\omega_{\rm GW} ( \eta_0 ,\, \vec{x}_0 ,\, q ,\, {\hat n} ) -\bar{\Omega}_{\rm GW}(\eta_0,\, q)}{\bar{\Omega}_{\rm GW}(\eta_0,\, q)} \,.
\end{equation}
Given that $[\Omega_{\rm GW}(\vec{x})-\bar{\Omega}_{\rm GW}]$ is a linear perturbation -- with random values at each $\vec{x}$ much smaller than $\bar{\Omega}_{\rm GW}$, typically by five orders of magnitude -- the definitions given in (\ref{eq:gwdensitycontrast1}) and (\ref{eq:gwdensitycontrast}) only differ by a stochastic multiplicative factor very peaked near one and a stochastic monopole term very peaked near zero, none of which are detectable. Thus, we can rely on the second definition, for which making theoretical predictions is more straightforward.\footnote{The same approximation is always implicitly performed in the case of CMB anisotropies. For instance, the temperature anisotropy map $\delta T/\bar{T}$ should in principle be defined with respect the temperature monopole $T_{\rm CMB} = \langle T(\vec{x}_0, \hat{n}) \rangle_{\hat{n}}$ at the detector's location. However, for the purpose of making theoretical predictions, the map is always implicitly assumed to be defined with respect to the spatially averaged temperature monopole $\bar{T}=\langle T(\vec {x}, \hat{n}) \rangle_{\vec{x},\hat{n}}$.}

At this point, as shown in \cite{Contaldi:2016koz, Bartolo:2019oiq, Bartolo:2019yeu}, it is useful to re-define the perturbed graviton distribution function in terms of the phase-space relative perturbation $\Gamma$, 
\begin{equation}
  \delta f_{\rm GW} = f_{\rm GW} - \bar{f}_{\rm GW} \equiv  - q \, \frac{\partial {\bar f}_{\rm GW}}{\partial q} \, \Gamma \left( \eta ,\, \vec{x} ,\, q ,\, {\hat n} \right),
\end{equation}
such that the first order Boltzmann equation reads in Fourier space
\begin{equation}
    \label{eq:Boltfirstgamma1}
    \Gamma'+ i \, k \, \mu\, \Gamma =  \phi' - i k \, \mu \, \psi -  \frac{1}{2}n^i n^j  \, h_{ij}' \, ,
\end{equation}
where the terms on the rhs define the so-called source function $S (\eta ,\, \vec{k} ,\, {\hat n})$.
A prime denotes a derivative with respect to conformal time, and $\mu$ is the cosine of the angle between $\vec{k}$ and ${\hat n}$. $\vec{k}$ and $k$ are the wave vector and wave number of the (large-scale) cosmological perturbations and should not be confused with the graviton comoving momentum $\vec{q}=a\,\vec{p}$ which is orders of magnitudes larger.
The solution of Eq.~\eqref{eq:Boltfirstgamma1} can be decomposed as 
\begin{equation} 
    \label{eq:dec}
    \Gamma ( \eta ,\, \vec{k} ,\, q ,\, {\hat n} ) =  \Gamma_I ( \eta ,\, \vec{k} ,\, q ,\, {\hat n} ) +  \Gamma_S ( \eta ,\, \vec{k} ,\,  {\hat n} ) +  \Gamma_T ( \eta ,\, \vec{k} ,\,  {\hat n} ) \;, 
\end{equation}
where $I$, $S$, and $T$ stand for {\it Initial}, {\it Scalar} and {\it Tensor}. These indices refer to the mechanism {\it sourcing} the graviton perturbation. The {\it Initial} term is related to the anisotropies generated by the GW production mechanism before the free propagation of the waves. The scalar and tensor terms correspond to additional anisotropies induced by the propagation of GWs in a background with large-scale perturbations of scalar type (sourced by $\phi' - i k \, \mu \, \psi$) and/or tensor type (sourced by $-  \frac{1}{2}n^i n^j  \, h_{ij}'$). The scalar and tensor source terms are independent of the GW momentum/frequency $q$, and so are the contributions $\Gamma_S$, $\Gamma_T$. On the other hand, as stressed in \cite{Bartolo:2019oiq,Bartolo:2019yeu} the initial anisotropies can have a large (order unity) dependence on the frequency (contrary to what happens for CMB photons at linear order)\footnote{The initial anisotropies could be sourced by both scalar and tensor perturbations, depending on the mechanism considered.}. 

The GW energy density contrast $\delta_{\rm GW}$ of Eq.~\eqref{eq:gwdensitycontrast} is related to the phase-space relative perturbation $\Gamma$ and to the fractional background energy density contribution $\bar{\Omega}_{\rm GW}$~\cite{Bartolo:2019oiq, Bartolo:2019yeu} through
\begin{equation}
    \delta_{\rm GW} \left( \eta_0 ,\, \vec{x}_0 ,\, q ,\, {\hat n} \right) = \left( 4 -  \frac{\partial \ln \, {\bar \Omega}_{\rm GW}\left( \eta_0 ,\, q \right)}{ \partial \ln \, q }\right) \, \Gamma \left( \eta_0 ,\, \vec{x}_0 ,\, q ,\, {\hat n} \right) \equiv \left( 4 - n_{\rm gwb} \right) \Gamma \,,
    \label{eq:prefactor_delta}
\end{equation}
where we define the Gravitational Wave Background (GWB) spectral index 
\begin{equation}
  n_{\rm gwb}(q) \equiv \frac{\partial \ln \, {\bar \Omega}_{\rm GW}\left( \eta_0 ,\, q \right)}{ \partial \ln \, q } . 
\end{equation}
In this relation, $n_{\rm gwb}(q)$ should be evaluated at the detector momentum/frequency $q$ at which $\delta_{\rm GW}$ is measured.
Many cosmological GW production mechanisms (see Sec. \ref{sec:source_power_law}) have a GW frequency spectrum well described by a simple power law (i.e., $\bar{\Omega}_{\rm GW}\propto q^{n_{\rm GWB}}$) such that $n_{\rm gwb}$ is independent of the considered frequency.

Like in the case of the CMB temperature anisotropies,  we can expand the GW density contrast in spherical harmonics $Y_{\ell m} ( {\hat n} )$,
\begin{equation}
    \delta_{\rm GW} (\eta_0, \vec{x}_0, q, {\hat n} ) = \sum_\ell \sum_{m=-\ell}^\ell \delta_{{\rm GW}\!,\, \ell m}(\eta_0 ,\, \vec{x}_0 ,\, q) \, Y_{\ell m} ( {\hat n} ) \,.
\end{equation}
As we will see in the next sections, the scalar and tensor contributions are ubiquitous for all cosmological production mechanisms, since they are generated by the propagation itself, while the initial contribution should depend on each specific scenario.
Following~\cite{Bartolo:2019oiq, Bartolo:2019yeu}, in harmonic space, the three contributions to the solution of the linear Boltzmann equation read 
\begin{eqnarray}
    \delta_{{\rm GW}\!,\,\ell m,\,I} \left( q \right)  &=& 4 \pi \left( - i \right)^\ell 
    %\left( 4 -  \frac{\partial \ln \, {\bar \Omega}_{\rm GW}}{ \partial \ln \, q }\right)\, 
    \left( 4 - n_{\rm gwb} \right)\,
    \int \frac{d^3 k}{\left( 2 \pi \right)^3} \,  {\rm e}^{i \vec{k} \cdot \vec{x}_0} \, 
    Y_{\ell m}^* ( \hat{k} ) \,
    \Gamma ( \eta_\mathrm{in} ,\, \vec{k} ,\, q  ) \,
    j_\ell \left[ k \left( \eta_0 - \eta_\mathrm{in} \right] \right) \,,\nonumber
    \\
    \delta_{{\rm GW}\!,\,\ell m,\,S} &=&  4 \pi \left( - i \right)^\ell \, 
    %\left( 4 -  \frac{\partial \ln \, {\bar \Omega}_{\rm GW}}{ \partial \ln \, q }\right) 
    \left( 4 - n_{\rm gwb} \right)\,
    \int \frac{d^3 k}{\left( 2 \pi \right)^3} \,  {\rm e}^{i \vec{k} \cdot \vec{x}_0} \, 
    Y_{\ell m}^* ( \hat{k} ) \, \mathcal{R} ( \vec{k} )\; \Delta_\ell^S \left(  k ,\, \eta_0 ,\, \eta_\mathrm{in} \right)  \,,\nonumber
    \\
    \delta_{{\rm GW}\!,\,\ell m,\,T}  &=&  4 \pi \left( - i \right)^\ell \,  
    %\left( 4 -  \frac{\partial \ln \, {\bar \Omega}_{\rm GW}}{ \partial \ln \, q }\right)
    \left( 4 - n_{\rm gwb} \right)\,
    \int \frac{d^3 k}{\left( 2 \pi \right)^3} \,  {\rm e}^{i \vec{k} \cdot \vec{x}_0} 
    \sum_{\lambda=\pm2}  \,  _{-\lambda}Y_{\ell m}^* ( \hat{k} ) \, h_\lambda ( \vec{k} ) \;
    \Delta_\ell^T \left(  k ,\, \eta_0 ,\, \eta_\mathrm{in} \right)\,, \nonumber\\
    \label{eq:solutionlm}
\end{eqnarray}
where $\eta_\mathrm{in}$ is an initial time after which the GWs propagate freely (defined in the next subsection) and $j_\ell(x)$ are the spherical Bessel functions. For concision, on the left-hand side, we omitted the detector's time and location $(\eta_0, \vec{x}_0)$ in the argument of the multipoles.

In order to derive the first integral, one needs to assume that the initial phase-space perturbation -- that would read in general $\Gamma ( \eta_\mathrm{in},\, \vec{k} ,\, q,\, \hat{n} )$ -- does not depend on the direction $\hat{n}$. Such a dependence might arise in the case of a breaking of statistical isotropy, as mentioned in Ref.~\cite{Bartolo:2019yeu}, even though this is not the only possibility. In any case, we do assume here that $\Gamma ( \eta_\mathrm{in},\, \vec{k} ,\, q ,\, \hat{n}  )$ does not depend on $\hat{n}$, giving a plausible physical example on how this condition can be satisfied in Section \ref{sec:ContributionsPowerSpectrum} and Appendix \ref{sec:Doppler}.

In the next two terms accounting for scalar and tensor contributions, we have factorized out the primordial scalar (comoving curvature) perturbation $\mathcal{R}(\vec{k})$ and the primordial tensor perturbation $h_\lambda( \vec{k})$ (where $\lambda=\pm2$ accounts for the two polarisation degrees of freedom) at a given wavevector $\vec{k}$, defined at very early times on super-Hubble scales. The scalar and tensor transfer functions $\Delta_\ell^{X}$, with $X= S, T$, account for the sourcing of graviton fluctuations by metric perturbations at a given wavenumber $k$ between $\eta_\mathrm{in}$ and $\eta_0$. They depend on the metric transfer functions $T_\phi$, $T_\psi$, $T_h$ normalised to
\begin{eqnarray}
\phi(\eta,\vec{k}) &=& T_\phi(\eta,k) \mathcal{R}(\vec{k})\,,\\
\psi(\eta,\vec{k}) &=& T_\psi(\eta,k) \mathcal{R}(\vec{k})\,,\\
h_{ij}(\eta,\vec{k}) &=& \sum_{\lambda = \pm 2} e_{ij, \lambda}(\hat{k}) \, T_h(\eta ,\, k) \, h_\lambda(\vec{k}) \,,
\end{eqnarray}
where $e_{ij, \lambda}(\hat{k})$ is the polarisation tensor normalised to one, $e_{ij,\lambda}(\hat{k})e_{ij,\lambda^\prime}(\hat{k})=\delta_{\lambda\lambda^\prime}$. The scalar and tensor transfer functions can be expressed as line-of-sight integrals,
\begin{eqnarray}
    \Delta_\ell^S \left( k ,\, \eta_0 ,\, \eta_\mathrm{in} \right) &\equiv& T_\psi \left( \eta_\mathrm{in} ,\, k \right) \, j_\ell \left( k \left( \eta_0 - \eta_\mathrm{in} \right) \right)
    + \int_{\eta_\mathrm{in}}^{\eta_0} d \eta \, 
    \frac{\partial \left[ T_\phi \left( \eta ,\, k \right) +  T_\psi \left( \eta ,\, k \right) \right] }{\partial \eta} \, 
    j_\ell \left( k \left( \eta_0 - \eta \right) \right) \,,\nonumber\\ 
    \Delta_\ell^T \left( k ,\, \eta_0 ,\, \eta_\mathrm{in} \right) &\equiv&  \sqrt{ \frac{\left(\ell +2 \right)!}{\left( \ell - 2 \right)!}} \, \frac{1}{4} \int_{\eta_\mathrm{in}}^{\eta_0} d \eta \, \frac{ \partial T_h \left( \eta ,\, k \right)}{\partial \eta} \,  \frac{j_\ell \left( k \left( \eta_0 - \eta \right) \right)}{k^2 \left( \eta_0 - \eta \right)^2}\,. \,
    \label{eq:transf}
\end{eqnarray}
Under the assumption of statistical isotropy and in absence of statistical correlations between the three GW density multipoles $(\delta_{{\rm GW}\!,\,\ell m,\,I} \,\, ,\,\, \delta_{{\rm GW}\!,\,\ell m,\,S} \,\, ,\,\, \delta_{{\rm GW}\!,\,\ell m,\,T})$, we can decompose the harmonic power spectrum of GW density as %\nicola{I think Kronecker deltas are missing on the left hand side of the following equation}
\begin{eqnarray} 
    \delta_{\ell \ell'} \,\, \delta_{mm'}C_\ell^{\rm CGWB \times \rm CGWB} \equiv\left\langle \delta_{{\rm GW},\ell m}   \delta_{{\rm GW},\ell' m'}^*  \right\rangle \equiv \delta_{\ell \ell'} \,\, \delta_{mm'} \, \left[C_{\ell,I} \left( q \right) + C_{\ell,S} + C_{\ell,T}\right]\;, 
    \label{eq:Cell-Bell-def}
\end{eqnarray} 
where the three contributions are given by
\begin{eqnarray}
    C_{\ell,I} \left( q \right)   &=&  4 \pi 
    %\left( 4 -  \frac{\partial \ln \, {\bar \Omega}_{\rm GW}}{ \partial \ln \, q }\right)^2 
    \left( 4 - n_{\rm gwb} \right)^2
    \int \frac{d k}{k} \,  \left[  j_\ell \left(k \left( \eta_0 - \eta_\mathrm{in} \right) \right) \right]^2 
    \,  P_{I} \left( q ,\, k \right) \,, \label{eq:ClIq} \\ 
    C_{\ell,S}  &=&  4 \pi  
    %\left( 4 -  \frac{\partial \ln \, {\bar \Omega}_{\rm GW}}{ \partial \ln \, q }\right)^2  
    \left( 4 - n_{\rm gwb} \right)^2
    \int \frac{dk}{k} \,  \left[ \Delta_\ell^{S} \left( k ,\, \eta_0 ,\, \eta_\mathrm{in} \right) \right]^2
    \, P_{\mathcal{R}} \left( k \right)  \;, \\ 
    C_{\ell,T}  &=&  4 \pi 
    %\left( 4 -  \frac{\partial \ln \, {\bar \Omega}_{\rm GW}}{ \partial \ln \, q }\right)^2  
    \left( 4 - n_{\rm gwb} \right)^2
    \int \frac{d k}{k} \, \left[ \Delta_\ell^{T}   \left( k ,\, \eta_0 ,\, \eta_\mathrm{in} \right) \right]^2
    \sum_{\lambda=\pm2}  P_{h_\lambda} \left( k \right)   \;. 
    \label{eq:Cell-res}
\end{eqnarray}
Here, we have introduced the primordial power spectrum $ P_{I} \left( q ,\, k \right)$ of the initial phase-space perturbation $\Gamma ( \eta_\mathrm{in},\, \vec{k} ,\, q  )$, the primordial scalar spectrum $P_{\mathcal{R}} \left( k \right)$,  and the primordial tensor spectra $P_{h_\lambda} \left( k \right)$ for $\lambda=\pm2$. In a FLRW universe, for standard models, the latter are equal to each other and usually parametrized as $P_{h_1}(k) = P_{h_2}(k) = \frac{1}{4} P_T(k)$, where $P_T(k)$ is the primordial tensor spectrum (defined in more details in Appendix~\ref{sec:app_inflation}). 
Since the two-point function $\langle \mathcal{R}(\vec{k}) h_\lambda(\vec{k}^\prime)\rangle$ vanishes as a consequence of statistical isotropy conservation, the scalar and tensor contributions are uncorrelated and their spectra can be summed in quadrature.
On the other hand, the initial anisotropy of the stochastic background could depend on the scalar perturbation $\mathcal{R}$, leading to a cross-correlation between $\delta_{{\rm GW}\!,\,\ell m,\,I}$ and $ \delta_{{\rm GW}\!,\,\ell m,\,S}$ that was neglected above for simplicity. A more general analysis of this correlation is performed in Section \ref{sec:ContributionsPowerSpectrum}. We will see that in the case of pure adiabatic initial conditions, the initial and scalar-induced contributions are actually maximally correlated, which implies that the angular power spectrum must be written differently than in Eq.~\eqref{eq:Cell-Bell-def}. 

The expression of the GW scalar transfer function in Eq.~\eqref{eq:transf} allows to draw an analogy with the CMB: GWs are affected by a Sachs-Wolfe (SW) contribution, which represents the energy lost by a graviton escaping from a potential well $\psi(\eta_\mathrm{in},\vec{x})$, and by an Integrated Sachs-Wolfe (ISW) contribution that depends on the variation of the potentials $(\psi+\phi)$ along the line of sight.
However, for all GW production mechanisms, the decoupling time is considerably smaller than the photon decoupling time, leading to very different values of the SW and ISW terms in the GW and CMB cases.

The main goal of our new code {\CLASSGW} is to evaluate the angular power spectra of Eqs.~(\ref{eq:Cell-Bell-def} - \ref{eq:Cell-res}) for different CGWB production mechanisms.

\subsection{Initial time}
\label{sec:initial_time}

In principle, the choice of initial time $\eta_{\rm in}$ should be very different for the purpose of CMB and CGWB calculations.
For the CMB, it is sufficient to pick up an initial time corresponding to a few e-folds of expansion before the smallest observable wavelength crosses the Hubble radius. This condition is met with an initial conformal time $\eta_\mathrm{min} \sim \mathcal{O}(10^{-1})$~Mpc, that corresponds roughly to a redshift $z_\mathrm{max} \sim \mathcal{O}(10^6)$.\footnote{\texttt{CLASS} makes a conservative choice of integrating background and thermodynamical equations starting from a much earlier time, but what really matters is the time at which the perturbation equations start to be followed. For simple cosmologies, default precision and CMB calculations, this time is indeed close to $\eta_\mathrm{min} \simeq 10^{-1}$~Mpc.} 

Instead, the CGWB allows us to  probe much earlier times in the cosmic history.
In this work, we consider a CGWB of frequency $f$ produced at some time $\eta_{\rm prod}$ and decoupled afterwards (e.g. $\eta_{\rm prod}$ could be the end of inflation or of a phase transition).
If at this time the CGWB of frequency $f$ is already inside the horizon, $\eta_{\rm in}$ should in principle be set to $\eta_{\rm prod}$. Otherwise, since GWs start to propagate and transport energy when they enter the Hubble radius, $\eta_{\rm in}$ should in principle be set to the Hubble crossing time $\eta_{\rm h.c.}$. For a GW of momentum $q$ -- related to the observed frequency today $f$ through $q = 2\pi \, a_0 \, f/c$ -- this time reads
\begin{equation}
    \frac{q}{\mathcal{H}(\eta_{\rm h.c.})}=1 \qquad \Rightarrow \qquad \eta_{\rm h.c.} = 
    \begin{cases}
    %    \frac{1}{q} 
        1/q \hspace{3em} \rm Radiation\, \, \, Domination \\
    %   \frac{2}{q}
        2/q \hspace{3em} \rm Matter\, \, \, Domination 
    \end{cases}
\end{equation}
where $\mathcal{H}$ is the conformal Hubble rate.
Thus, in principle, the initial time should be adjusted to 
\begin{equation}
    \eta_\mathrm{in}(f) = \max\left[\eta_{\rm prod}, \eta_{\rm h.c.}(f)\right]\,.
    \label{eq:def_eta_i}
\end{equation}
For instance, the network ET$+$CE is expected to probe a frequency range $f \in [1,3000]\, \rm Hz$, leading approximately to
\begin{equation}
    \eta_{\rm h.c.} \in \left[10^{-18}, 10^{-15}\right]\, \rm Mpc \,, \hspace{3em}
\end{equation}
which is considerably smaller than the time of equality $\eta_{\rm eq} \approx 112\, \rm Mpc$, or even than the initial time of CMB calculations $\eta_\mathrm{min} \simeq 10^{-1}$~Mpc. This is also true for other GW detectors such as LISA, BBO, DECIGO and Taiji, which are expected to work in the milli-Hertz frequency range.

In practice, at the code level, the situation is different for two reasons:
\begin{itemize}
\item On super-Hubble scales $k\ll {\cal H}$, the metric fluctuations $\phi(\eta,\vec{k})$, $\psi(\eta,\vec{k})$, $h_{ij}(\eta, \vec{k})$ have a trivial evolution that can be computed analytically. The same holds for the GW perturbations $\Gamma(\eta, \vec{k}, q)$ and their power spectrum $P_I(q,k)$. Thus, it would be a waste of time to integrate numerically some perturbation equations between $\eta_{\rm in}$ and the usual EBS initial time $\eta_\mathrm{min}$. %\nicola{Please define what EBS means} 
\item Besides the evolution of perturbations, $\eta_{\rm in}$ appears either in the integral boundary or in the argument of the spherical Bessel function in Eqs.~\eqref{eq:transf}, \eqref{eq:ClIq}, and thus, plays a role in projection effects from Fourier to multipole space. However, from this point of view, as long as $\eta_{\rm in}$ is very small compared to $\eta_0$, its precise value does not affect the final angular power spectra. In particular, we checked that for all the production mechanisms considered in the next section, $\eta_{\rm in}$ can be substituted in the Bessel function by any value smaller or equal to $\eta_\mathrm{min}$ without changing the spectra.
\end{itemize}
The conclusion is that, in the code, it is possible to integrate all perturbation equations starting from the usual time $\eta_\mathrm{min} \simeq 10^{-1}$~Mpc and to set $\eta_{\rm in}$ in spherical Bessel functions to any arbitrary value smaller or equal to $\eta_\mathrm{min}$ -- as long as the super-Hubble evolution of perturbations between the time $\eta_{\rm in}$ and $\eta_\mathrm{min}$ is correctly modelled analytically. The next sections will show how to account for this evolution in different cases. As a result of this discussion, we see that the smallness of $\eta_{\rm in}$ has no impact on the structure nor computational cost of EBSs like \texttt{CLASS}\footnote{Still notice that, as already mentioned at the end of the previous section, and as we will detail later, the fact that $\eta_{\rm prod}$ and $\eta_{\rm in}$ for the CGWB are much smaller that the decoupling time of CMB photons, {\it does} indeed lead to significant differences in the corresponding final angular power spectra.}.

\subsection{Adiabatic initial conditions for GWs}
\label{sec:ContributionsPowerSpectrum}

We now come back to the expression of the first (initial) contribution to the angular power spectrum in Eq. \eqref{eq:ClIq}.

Let us assume for a while that the initial perturbation $\Gamma$ of the graviton phase-space distribution $f_{\rm GW}$ has non-negligible anisotropies that depend on the angle between the wavevector $\vec{k}$ and the direction of propagation $\hat{n}$, such that $\vec{k}\cdot \hat{n} = k \, \mu$. It can then be expanded in multipoles, 
\begin{equation}
    \Gamma(\eta_\mathrm{in},\vec{k},q,\hat{n}) = \sum_\ell (-i)^\ell (2\ell+1)\,  \Gamma_l(\eta_\mathrm{in},\vec{k},q)\, \mathcal{P}_\ell(\mu)\,,
\end{equation}
where $\mathcal{P}_\ell(\mu)$ stands for Legendre polynomials.
The evolution of these multipoles obeys to a Boltzmann hierarchy. We can make progress under the assumption that, at some very early times, only the first multipoles survive, for instance because of some tight coupling regime between GW modes. (One can draw an analogy with the CMB for which, as long as photons are tightly coupled, higher multipoles are suppressed by powers of $k\eta/\kappa$, where $\kappa$ is the photon optical depth, see e.g. \cite{Dodelson:2003ft}). Later on, during the free-streaming regime, higher multipoles grow, but remain very small on super-Hubble scales, since the structure of the Boltzmann hierarchy requires  
\begin{equation}
    \Gamma_{\ell}(\eta,\vec{k},q) \approx k\eta \, \Gamma_{\ell-1}(\eta,\vec{k},q) \, .
\end{equation}
At the initial time $\eta_\mathrm{in}$ defined in section \ref{sec:initial_time}, Fourier modes of interest are particularly far outside the Hubble scale, $k\eta_\mathrm{in} \ll 1$, and thus the multipoles $\ell \geq 2$ can be safely neglected. This argument does not hold for the dipole $\Gamma_1$, which is a gauge-dependent quantity, and which also depends on momentum exchanges during a possible early tight coupling regime. It does hold for the quadrupole $\Gamma_2$, which could be sourced by tensor perturbations on cosmological scales, but remains very small as long as these scales are super-Hubble.

In a next step, we can decompose as usual the possible solutions of the system of perturbation equations (including the Boltzmann hierarchy for $\Gamma_\ell$) in one growing adiabatic mode and several non-adiabatic and/or decaying modes. If we consider only the growing adiabatic mode (and neglect any decaying mode) we are compatible with the ``separate universe assumption'' \cite{Wands:2000dp}. As a matter of fact, in this case, each Hubble patch evolves like a separate universe, where it is possible to foliate the space-time in spatial hypersurfaces of uniform density in which the curvature perturbation is conserved. Then, on super-Hubble scales,  all quantities, such as e.g. the density of GWs $\rho_\mathrm{GW}(\eta, \vec{x})$, have spatial fluctuations (whatever gauge one chooses) related to a unique time-shifting function $\delta \eta(\vec{x})$ via the time-derivative of the background solution, like in
\begin{equation}
\delta \rho_\mathrm{GW}(\eta, \vec{x})
= \dot{\bar{\rho}}_\mathrm{GW}(\eta) \,\, \delta \eta(\vec{x})\,.
\end{equation}

If a single-clock mechanism generates primordial curvature perturbations $\mathcal{R}(\vec{k})$ in the universe, the presence of the adiabatic mode is unavoidable. Non-adiabatic modes may appear in cases where the generation of GWs leads additionally to intrinsic primordial fluctuations in $\Gamma$ that are not captured by the ``separate universe assumption''. Such a GW generation mechanism should involve a local time-shifting function (that is, a local random process) on top of the time-shifting function that describes primordial curvature perturbations. Such a mechanism is difficult to realize on super-Hubble scales. However, we will see later an explicit example based on GWs generated by the formation of primordial black holes triggered by non-Gaussian perturbations \cite{Bartolo:2019zvb}.

We already argued that multipoles $\ell \geq 2$ should be vanishingly small on super-Hubble scales obeying $k \eta_\mathrm{in} \ll 1$. In the case of the adiabatic mode, it is also possible to relate the monopole $\Gamma_0$ to metric perturbations (in the Newtonian gauge, to $\psi$, see appendix \ref{sec:adiabaticIC}) and to prove that $\Gamma_1 \ll \Gamma_0$ (see Appendix \ref{sec:Doppler} for a proof in the same gauge). For non-adiabatic modes, it is not obvious that the dipole can be neglected, but for simplicity, we assume that this is true in this work and in our \texttt{CLASS} implementation. Thus, we always assume that $\Gamma$ reduces to its monopole component $\Gamma_0$, and we can expand the initial perturbation $\Gamma$ of the graviton phase-space distribution into an adiabatic and non-adiabatic contribution,
\begin{equation}
    \label{eq:Gamma_ini_split}
    \Gamma(\eta_\mathrm{in}, \vec{k}, q, \hat{n}) =
    \Gamma_0(\eta_\mathrm{in}, \vec{k}, q) =
    T_\Gamma^\mathrm{AD}(\eta_\mathrm{in}, k, q) \,\mathcal{R}(\vec{k}) 
    + \Gamma_0^\mathrm{NAD}(\eta_\mathrm{in}, \vec{k}, q)  \, .
\end{equation}
In the Newtonian gauge and for the adiabatic contribution, we can use the relation 
\begin{equation}
\Gamma_0(\eta_\mathrm{in}, \vec{k}, q) = -\frac{2}{4 - n_{\rm gwb}(q)} \psi(\eta_\mathrm{in},\vec{k})
\label{eq:gamma_ad}
\end{equation}
shown in appendix \ref{sec:adiabaticIC}, and express the adiabatic transfer function of $\Gamma$ at a given momentum/frequency $q$ as
\begin{equation}
T_\Gamma^\mathrm{AD}(\eta_\mathrm{in},k, q)
= -\frac{2}{4 - n_{\rm gwb}(q)}  T_\psi(\eta_\mathrm{in},k)~.
\end{equation}
Finally, the total angular power spectrum of the CGWB can be expressed in a handy form,
\begin{align}
    \frac{C_\ell^{\rm CGWB \times \rm CGWB}}{\left( 4 -  n_\mathrm{gwb}\right)^2} =  4 \pi \int \frac{dk}{k} & \Big\{  \left[
    \Delta_\ell^\mathrm{AD}(k, \eta_0, \eta_\mathrm{in}, q) +\Delta_\ell^\mathrm{SW}(k, \eta_0, \eta_\mathrm{in}) +\Delta_\ell^\mathrm{ISW}(k, \eta_0,\eta_\mathrm{in})\right]^2
    P_\mathcal{R}(k)  \nonumber \\
    & + 
    \left[ j_\ell \left( k \left( \eta_0 - \eta_\mathrm{in} \right) \right) \right]^2 P_\Gamma^\mathrm{NAD}(k,q)
    \nonumber \\
    & +
    j_\ell \left( k \left( \eta_0 - \eta_\mathrm{in} \right) \right) 
    \left[
    \Delta_\ell^\mathrm{AD} +\Delta_\ell^\mathrm{SW} +\Delta_\ell^\mathrm{ISW}\right]
    P^{\times}(k,q)
    \nonumber \\
    & +  \left[\Delta_\ell^{T}(k, \eta_0,\eta_\mathrm{in})\right]^2 \sum_{\lambda = \pm 2} P_{h_\lambda}(k) \Big\} \, ,
    \label{eq:tot_cl_bis}
\end{align}
where, in addition to Eq.~\eqref{eq:transf}, we defined the GW anisotropy transfer functions
\begin{eqnarray}
    \Delta_\ell^\mathrm{AD}(k, \eta_0, \eta_\mathrm{in},q)
    &\equiv& 
    -\frac{2}{4 - n_{\rm gwb}(q)} \,\,
    T_\psi \left( \eta_\mathrm{in} ,\, k \right) \, 
    j_\ell \left( k \left( \eta_0 - \eta_\mathrm{in} \right) \right) \,,\nonumber\\ 
    \Delta_\ell^\mathrm{SW}(k, \eta_0, \eta_\mathrm{in})
    &\equiv& 
    T_\psi \left( \eta_\mathrm{in} ,\, k \right) \, 
    j_\ell \left( k \left( \eta_0 - \eta_\mathrm{in} \right) \right) \,,\nonumber\\ 
    \Delta_\ell^\mathrm{ISW}(k, \eta_0, \eta_\mathrm{in})
    &\equiv& \int_{\eta_\mathrm{in}}^{\eta_0} d \eta \, 
    \frac{\partial \left[ T_\phi \left( \eta ,\, k \right) +  T_\psi \left( \eta ,\, k \right) \right] }{\partial \eta} \, 
    j_\ell \left( k \left( \eta_0 - \eta \right) \right) \, ,
    \label{eq:transf2}
\end{eqnarray} 
and the non-adiabatic and cross-correlation primordial spectra
\begin{eqnarray}
P_\Gamma^\mathrm{NAD}(k,q)
&=& \left\langle \, \left| \Gamma_0^\mathrm{NAD}(\eta_\mathrm{in}, \vec{k}, q) \right|^2 \, \right\rangle~,
\nonumber \\
P^\times(k,q)
&=& \left\langle \, {\cal R} (\vec{k}) \,
\Gamma_0^{\mathrm{NAD}*} (\eta_\mathrm{in}, \vec{k}, q) \, 
+
{\cal R}^* (\vec{k}) \,
\Gamma_0^{\mathrm{NAD}} (\eta_\mathrm{in}, \vec{k}, q)
\right\rangle~.
\end{eqnarray}
The re-writing of Eq.~(\ref{eq:Cell-Bell-def}) in the form of Eq.~(\ref{eq:tot_cl_bis}) can be seen as a change of basis from the modes ``initial, scalar'' ($I$,$S$) to the modes ``adiabatic, non-adiabatic'' ($\rm AD$, $\rm NAD$), leaving the tensor contribution unchanged. The latter decomposition better captures the statistical correlation between the adiabatic component of the initial contribution $\delta_{\mathrm{GW}\!,\, \ell m, I}$ and the scalar-induced contribution $\delta_{\mathrm{GW}\!,\, \ell m,\, S}$.

\begin{figure}[t!]
    \centering
    \begin{subfigure}{0.49\textwidth}
        \includegraphics[scale=0.79]{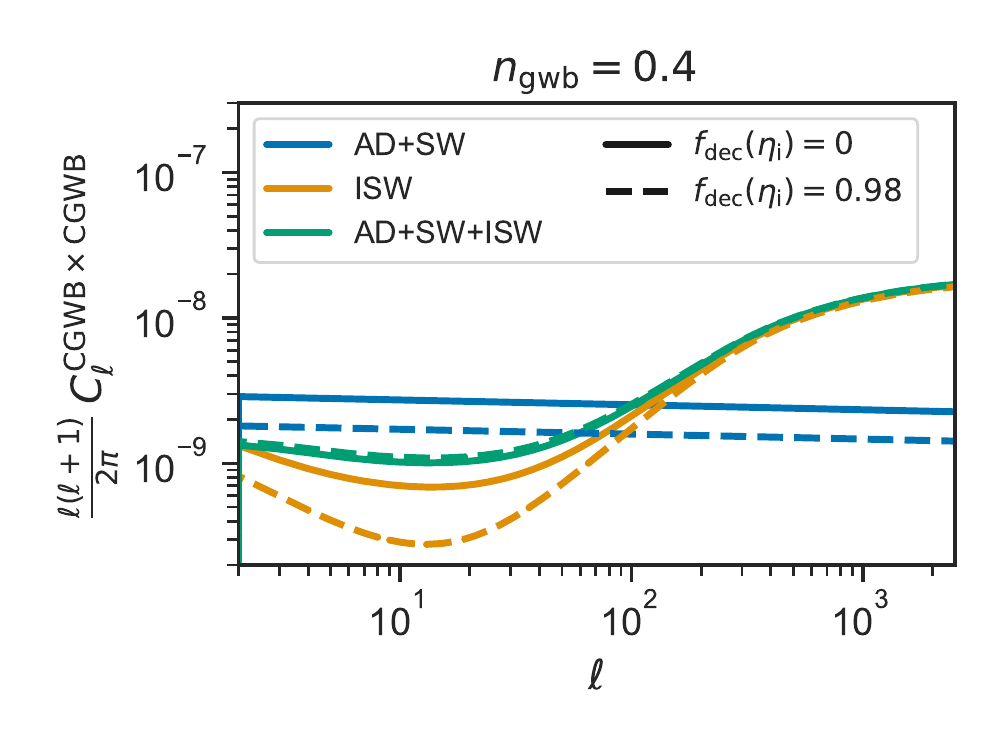}
    \end{subfigure}
    \begin{subfigure}{0.49\textwidth}
        \includegraphics[scale=0.79]{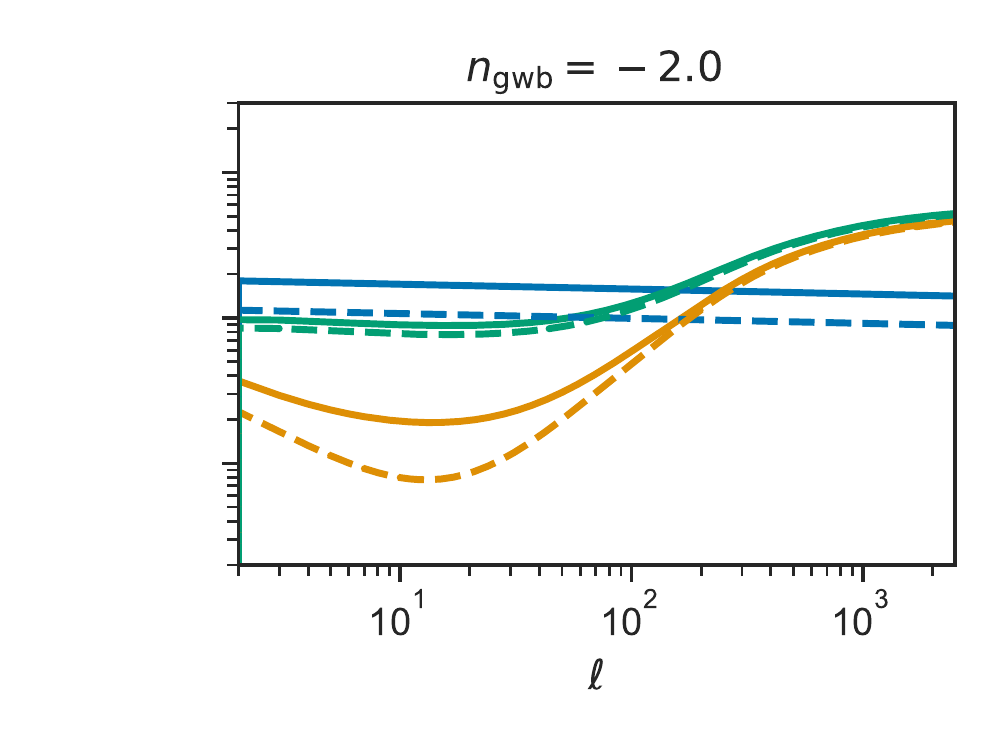}
    \end{subfigure}
    \caption{Plot of the contributions to the dimensionless angular power spectrum of CGWB anisotropies seeded by curvature perturbations: AD+SW contribution (blue), ISW contribution (orange) and total AD+SW+ISW contribution (green). Solid lines assume a fraction of relativistic decoupled species -- defined in Eq.~(\ref{eq:fdec_def}) -- $\fdec=0$, and dashed lines $\fdec=0.98$. We show two examples of GWB with different spectral indices,  $n_\mathrm{gwb}=0.4$ (left panel) and $n_\mathrm{gwb}=-2.0$ (right panel), and adiabatic initial conditions.}
    \label{fig:cgwb_spectrum}
\end{figure}

In Fig. \ref{fig:cgwb_spectrum}, we show two examples of CGWB angular power spectra generated by scalar perturbations with adiabatic initial conditions -- that is, considering only the first line in \eqref{eq:tot_cl_bis} -- with two different tilts $n_\mathrm{gwb}$.
As already explained in~\cite{ValbusaDallArmi:2020ifo}, these spectra do not feature acoustic peaks like the CMB ones. As a matter of fact, GW anisotropies arise from metric perturbations, rather than fluctuations in a fluid featuring pressure and acoustic waves.
The ISW contribution and the total angular power spectrum are enhanced at small angular scales ($\ell>100$), because Fourier modes crossing the Hubble radius during radiation domination experience a variation of metric fluctuations that boosts the ISW term. At larger angular scales, the ISW term also picks up a contribution from the variation of metric perturbations around the time of equality between radiation and matter. Indeed, during the transition from a radiation-dominated universe to a matter-dominated one, the large-scale scalar metric perturbations get damped by a factor close to $9/10$~\cite{Ma:1995ey,Dodelson:2003ft,Bashinsky:2003tk}. In the CMB case, this variation does not contribute to the ISW term since it occurs before recombination.

\begin{figure}
    \centering
    \includegraphics[scale=1.1]{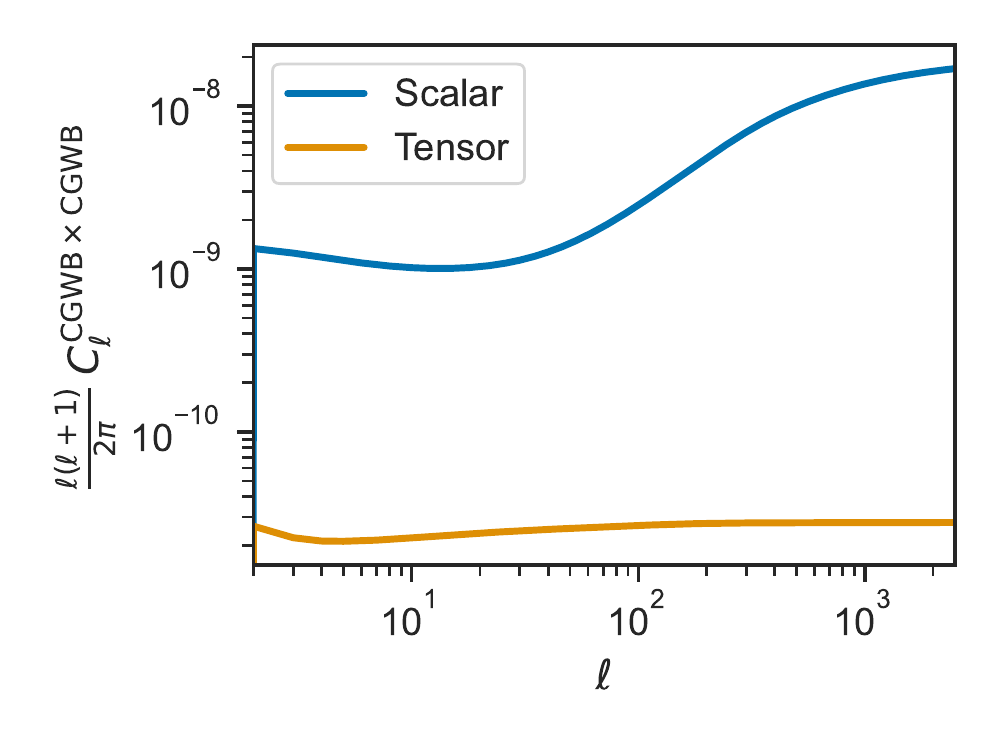}
    \caption{Plot of the contributions to the angular power spectrum of the CGWB for adiabatic initial conditions and $n_\mathrm{gwb}(q)=0.4$: scalar (blue), tensor (orange). The primordial tensor spectrum has been computed for the tensor-to-scalar ratio (defined in \eqref{eq:PTpl}) $r=0.03$  and $n_\mathrm{t}(k)=-0.003$. All the spectra assume $f_\mathrm{dec}(\eta_\mathrm{in})=0$.}
    \label{fig:cgwb_tensor_spectrum}
\end{figure}

In Fig. \ref{fig:cgwb_tensor_spectrum}, we show the angular power spectrum of the CGWB generated by scalar and tensor perturbations with adiabatic initial conditions and $n_\mathrm{gwb}(q)=0.4$. The tensor power spectrum $P_T(k)$ has been computed by assuming the Planck constraints on the tensor-to-scalar ratio, $r=0.03$, and $n_\mathrm{t}(k)=-r/8(2-r/8-n_S)\approx -0.003$.\footnote{Note that in general $n_\mathrm{gwb}\neq n_t$, see the discussion after Eq. \eqref{eq:PTpl}.}
In analogy with the ISW term of the scalar part, $C_{\ell,T}$ is not suppressed at small angular scales, because the CGWB is sensitive to the variations of $T_h$ also during the radiation-dominated era.  As expected~\cite{ValbusaDallArmi:2020ifo}, the anisotropies induced by large-scale tensor perturbations are subdominant w.r.t. the scalar ones.

\subsection{Relativistic decoupled species at early times}
\label{sec:rel_dec_spe}

The transfer functions of Eq.~\eqref{eq:transf2} show that the angular power spectrum of the CGWB depends on the value of the scalar metric fluctuations $(\phi, \psi)$ -- or, equivalently, of the transfer functions $(T_\phi, T_\psi)$ -- at the very early time $\eta_\mathrm{in}$ and at all subsequent times. In section \ref{sec:initial_time}, we argued that the evolution of such perturbations on super-Hubble scales can be followed analytically between $\eta_\mathrm{in}$ and $\eta_\mathrm{min} \sim 0.1$~Mpc.

This evolution is discussed in reference~\cite{ValbusaDallArmi:2020ifo}. It depends mainly on variations in the fractional energy density of relativistic and decoupled particles species $f_{\rm dec}(\eta)$, defined as
\begin{equation}
    f_{\rm dec}(\eta) \equiv \frac{\bar{\rho}_\mathrm{r}^{\rm dec}(\eta)}{\bar{\rho}_{\rm tot}(\eta)}\, .\label{eq:fdec_def}
\end{equation}
As a matter of fact, these species have a non-vanishing anisotropic stress $\sigma_{\rm dec}$ that determines the difference between the scalar metric perturbations, as shown by the transverse-traceless part of the Einstein equation in the Newtonian gauge, 
\begin{equation}
    k^2(\phi-\psi)=16\pi G \, a^2 \bar{\rho}_\mathrm{r}^{\rm dec} \, \sigma_{\rm dec} \, .
\end{equation}

The solution of the perturbation equations on super-Hubble scales and for the adiabatic mode gives~\cite{Ma:1995ey,ValbusaDallArmi:2020ifo}
\begin{eqnarray}
    \psi(\eta,\vec{k})&=&-\frac{2}{3}\Bigl[1+\frac{4}{15}f_{\rm dec}(\eta)\Bigl]^{-1}\mathcal{R}(\vec{k})\, ,\nonumber\\
    \phi(\eta,\vec{k})&=&-\frac{2}{3}\Bigl[1+\frac{4}{15}f_{\rm dec}(\eta)\Bigl]^{-1}\Bigl[1+\frac{2}{5}f_{\rm dec}(\eta)\Bigl]\mathcal{R}(\vec{k})\, .
    %\nonumber
    \label{eq:psi_phi_f_dec}
\end{eqnarray}
The knowledge of $f_\mathrm{dec}(\eta_\mathrm{in})$ is important to set properly the value of the transfer function $T_\psi(\eta_\mathrm{in},k)$ in $\Delta_\ell^\mathrm{AD}$ and $\Delta_\ell^\mathrm{SW}$. In addition, any variation of $f_\mathrm{dec}(\eta)$ over time leads to a non-zero derivative $(\phi'+\psi')$, and thus to a contribution to the ISW transfer function $\Delta_\ell^\mathrm{ISW}$ of Eq.~\eqref{eq:transf2}. Note that the ISW transfer function of GW anisotropies features an integral from the very early time $\eta_\mathrm{in}$ defined in section~\ref{sec:initial_time} until today. This is very different from the ISW transfer function of CMB anisotropies, in which the integral runs only from the time of photon decoupling $\eta_\mathrm{dec}$ until today.\footnote{To be precise, the ISW integral of CMB anisotropies is performed over $e^{-\kappa}(\phi'+\psi')$, where $\kappa$ is the photon optical depth. In very good approximation, $e^{-\kappa}$ vanishes for $\eta < \eta_\mathrm{dec}$, which means that the lower boundary of the integral can be set effectively to $\eta_\mathrm{dec}$ -- although EBSs do not perform such an approximation.}  

In its standard version, the \texttt{CLASS} code infers from user input the value $f_{\rm dec}(\eta_\mathrm{min})$ at the initial time at which perturbations are integrated, $\eta_\mathrm{min} \sim 0.1$~Mpc. At this time, which corresponds to a temperature much smaller than that of neutrino decoupling, $T_\mathrm{min} \ll T_\nu^\mathrm{dec}\sim 1$~MeV, neutrinos are expected to free-stream. For a standard cosmology with three neutrinos, a simple calculation involving the neutrino-to-photon temperature ratio gives approximately $f_{\rm dec}(\eta_\mathrm{min}) = \bar{\rho}_\nu(\eta_\mathrm{min}) /  \bar{\rho}_\mathrm{r}(\eta_\mathrm{min}) \simeq 0.4$. For models with a non-standard neutrino density or with additional free-streaming relics (e.g. dark radiation particles originating from a dark sector), \texttt{CLASS} adapts this value consistently as a function of input parameters (like the neutrino temperature, the effective number of free-streaming ultra-relativistic relics, etc.\footnote{In the code, $f_{\rm dec}(\eta_\mathrm{min})$ is dubbed \texttt{fracnu}, but it does take into account any free-streaming relativistic relics beyond standard neutrinos.}) The value of $f_{\rm dec}(\eta_\mathrm{min})$ is used to set initial conditions for metric perturbations, and then, these fluctuations are evolved according to the Einstein equations. This approach is clearly sufficient for the calculation of CMB anisotropies, but not for that of GW anisotropies, since it neglects any variation of $f_{\rm dec}(\eta)$ between $\eta_\mathrm{in}$ and $\eta_\mathrm{min}$.

In \CLASSGW, the user passes the value of  $f_{\rm dec}(\eta_\mathrm{in})$ in a given model as an input parameter \texttt{f\textunderscore dec\textunderscore ini}. This value is taken into account for calculating $T_\psi(\eta_\mathrm{in},k)$ 
in $\Delta_\ell^\mathrm{AD}$ and $\Delta_\ell^\mathrm{SW}$.
The code also computes $f_{\rm dec}(\eta_\mathrm{min})$ like in the standard version: this number is still used to initialise all perturbations at $\eta_\mathrm{min}$. Finally, the integral in $\Delta_\ell^\mathrm{ISW}$ is decomposed in two parts. The first part, referred to as the primordial (prim) ISW -- with an integral from $\eta_\mathrm{in}$ to $\eta_\mathrm{min}$ -- is performed analytically \cite{ValbusaDallArmi:2020ifo}, giving
\begin{align}
    \Delta_\ell^{\rm ISW-prim} &\equiv 
    \int_{\eta_\mathrm{in}}^{\eta_\mathrm{min}}d\eta \,j_\ell[k(\eta_0-\eta)] \, \left[T_\psi^\prime(\eta,k)+T^\prime_\phi(\eta,k)\right] \nonumber \\
    & \simeq j_\ell[k(\eta_0-\eta_\mathrm{in})] \, \bigg[ T_\psi(\eta,k)+T_\phi(\eta,k) \bigg]_{\eta_\mathrm{in}}^{\eta_\mathrm{min}}\, \nonumber \\
    & = j_\ell[k(\eta_0-\eta_\mathrm{in})] \frac{2}{15}\frac{f_{\rm dec}(\eta_\mathrm{in})-f_{\rm dec}(\eta_\mathrm{min})}{1+\frac{4}{15}f_{\rm dec}(\eta_\mathrm{min})}T_{\rm \psi}(\eta_\mathrm{in},k) \, ,
    \label{eq:prim_isw}
\end{align}
where in the second line we used the fact that $k\eta_\mathrm{in}$ and $k\eta_\mathrm{min}$ are both much smaller than one (see section~\ref{sec:initial_time}) to approximate $j_\ell[k(\eta_0-\eta)]$ as $j_\ell[k (\eta_0-\eta_\mathrm{in})]$, in order to pull this factor out of the integral. The second part -- with an integration from $\eta_\mathrm{min}$ to $\eta_0$ -- is performed numerically like in the standard version of the code, and both terms are added up to form $\Delta_\ell^{\rm ISW}$.

From a theoretical prospective, the most natural value for $f_{\rm dec}(\eta_{\rm in})$ should be zero, because standard model particles are expected to be strongly interacting in the early universe, at the time when the GW background forms. This is the default setting in \CLASSGW.

The CGWB is sensitive not just to the relativistic and decoupled degrees of freedom at high temperatures, $f_{\rm dec}(\eta_{\rm in})$, but in principle also to any ingredient changing the evolution of metric perturbations in the very early universe. Another example is provided by the equation of state of the cosmic fluid driving the expansion of the Universe at $\eta_{\rm prod}$~\cite{Malhotra:2022ply}. More specifically, deviations from the standard equation of state of a relativistic fluid at early times, $w=1/3$, would affect both the SW and the primordial ISW with expressions analogous to Eqs. \eqref{eq:psi_phi_f_dec} and \eqref{eq:prim_isw}.

We should stress that the sensitivity of the anisotropies of the CGWB to parameters that influence the evolution of the metric perturbations -- like $f_{\rm dec}(\eta_\mathrm{in})$ -- strongly depends on initial conditions and on the frequency dependence of the monopole. In the case of adiabatic initial conditions, the dependence of the CGWB on the $f_{\rm dec}(\eta_\mathrm{in})$ can be summarized as
\begin{equation}
     T_\Gamma^{\rm AD}(\eta_\mathrm{in},k,q)+T_\Phi(\eta_\mathrm{in},k)+\frac{\Delta_\ell^{\rm ISW-prim}(\eta_\mathrm{in},k)}{j_\ell(k(\eta_0-\eta_\mathrm{in}))} \propto \frac{\frac{6-n_\mathrm{gwb}(q)}{4-n_\mathrm{gwb}(q)}+\frac{2}{5}f_\mathrm{dec}(\eta_\mathrm{in})}{1+\frac{4}{15}f_\mathrm{dec}(\eta_\mathrm{in})}\, .
\end{equation}
In the limit $n_\mathrm{gwb}\rightarrow 0$, the factor on the right-hand side is independent of the number of relativistic and decoupled degrees of freedom (or of the equation of state of the universe at early times, as pointed out by~\cite{Malhotra:2022ply}). 
However, we will explicitly show in further sections that, in the majority of the frequency bins accessible by future interferometers, the tensor tilt $n_\mathrm{gwb}$ is different from zero (in particular it must be larger than zero if we focus on inflationary mechanisms which respect CMB bounds) for signals that are detectable by current and future GW interferometers. 

Figure~\ref{fig:cgwb_spectrum} shows the impact of varying $\fdec$ on two examples of CGWB spectra. In the left panel, assuming purely adiabatic initial conditions with $n_\mathrm{gwb}=0.4$, we see that the effect of increasing $\fdec$ from 0 to 0.98 is large on the individual AD+SW and ISW contributions, but nearly cancels out in the total AD+SW+ISW  spectrum. In the right panel, assuming purely adiabatic initial conditions, but with a different tilt $n_\mathrm{gwb}=-2.0$, the effect of $\fdec$ on the total spectrum is enhanced.

The enhancement factor can be easily inferred from our code, but it is difficult to predict analytically. Indeed, on the one hand, the scaling of the AD+SW contribution with $f_{\rm dec}(\eta_i)$ is exactly given by $[1+4/15\, f_{\rm dec}(\eta_\mathrm{in})]^{-2}$, because this term is proportional to the square of $T_\psi$ at the time $\eta_{\mathrm{in}}$. On the other hand, for the ISW term, the factor found in Eq. \eqref{eq:prim_isw}, which described the primordial ISW contribution, needs to be summed up at large scales with the early ISW contribution. The latter is caused by the variation of scalar metric perturbations around equality, as discussed at the end of Section \ref{sec:ContributionsPowerSpectrum}, and needs to be computed numerically. In any case, the most important message that can be inferred from Figure \ref{fig:cgwb_spectrum} is that, even in the case of adiabatic initial conditions, the total (AD+SW+ISW) angular power spectrum is sensitive to $f_{\rm dec}(\eta_i)$, with a dependence that can be enhanced for non-zero values of the monopole tilt $n_\mathrm{gwb}$ of the CGWB. In~\cite{ValbusaDallArmi:2020ifo}, the adiabatic initial condition of the CGWB has not been considered, in order to do a model-independent discussion of the imprint of the relativistic and decoupled species on the CGWB anisotropies. This is the reason why the damping of the angular power spectrum for larger values of $f_\mathrm{dec}(\eta_\mathrm{in})$ is more enhanced in~\cite{ValbusaDallArmi:2020ifo} than in Figure~\ref{fig:cgwb_spectrum}.

\section{Cosmological Gravitational Wave Background Sources}
\label{sec:CGWB_sources}

As stated above, the physical observable that  we can measure with interferometers is the SGWB density contrast~\cite{Bartolo:2019oiq, Bartolo:2019yeu,LISACosmologyWorkingGroup:2022kbp} defined in Eqs.~(\ref{eq:gwdensitycontrast}, \ref{eq:prefactor_delta}),
\begin{equation}
    \delta_{\rm GW} \left( \eta_0 ,\, \vec{x}_0 ,\, f ,\, {\hat n} \right) = \left(4 -  \frac{\partial  \ln \bar{\Omega}_{\rm GW} \left( \eta_0 ,\, f \right)}{\partial \ln  f}  \right) \, \Gamma \left( \eta_0 ,\, \vec{x}_0 ,\, f ,\, {\hat n} \right),
    \label{delta-Gamma}
\end{equation} 
where we used the observed GW frequency $f = \frac{c}{2\pi \, a_0} \, q$ instead of the comoving graviton momentum $q$.
The value of the pre-factor in Eq.~(\ref{delta-Gamma}) is model dependent, so we need to know the underlying source of GWs.
Our goal is to estimate the angular power spectrum of $\delta_{\rm GW}$ in models that have a monopole amplitude detectable by the future ground-based interferometer network ET+CE.

Among several possible cosmological sources of GWs, three mechanisms are very promising because they relate to different aspects of early universe models: inflation with a blue tilt, first-order phase transitions, and second-order-induced GWs.
The detection of inflationary GWs with interferometers would bring information on the inflationary potential at completely
different scales than the measurement of tensor modes in the CMB; GWs from a first-order phase transition would probe physics beyond the Standard Model, at energies not accessible with colliders;
Finally, second-order-induced GWs could be related to Primordial Black Holes, which may explain a fraction or (for some mass range) the totality of the Dark Matter~\cite{Sasaki:2018dmp, Bartolo:2018evs}.

Below we briefly describe the GW sourcing mechanisms implemented in \CLASSGW. In principle, it is straightforward to implement in \CLASSGW other exotic mechanism that could generate a CGWB, such as cosmic strings~\cite{Caprini:2018mtu}.

\subsection{CGWB from inflation with adiabatic initial conditions}
\label{sec:sources_inflation}

For GWs produced by quantum fluctuations during inflation, the current value of the average GW energy density $\bar{\Omega}_{\rm GW}$ is related to the primordial tensor spectrum $P_T$ through (see Appendix~\ref{sec:app_inflation} for details)
\begin{equation}
    \bar{\Omega}_{\rm GW}(q) = \frac{1}{12 H_0^2 a_0^2} \frac{\eta_{\rm eq}^2}{2 \eta_0^4} P_T(q) \, .
    \label{eq:OGW_inf}
\end{equation}
Here we expressed the average GW energy density as function of comoving momentum $q$ instead of frequency $f=\frac{c}{2 \pi q}$.
This relation takes into account the evolution of the tensor modes that re-entered the Hubble scale during radiation domination~\cite{Caprini:2006jb}. It depends on the value of conformal time at equality between matter and radiation, $\eta_{\rm eq}$, and today, $\eta_0$. Note that Eq.~(\ref{eq:OGW_inf}) takes into account the evolution of GWs during radiation and matter domination, but not during dark energy domination. However, as explained~\cite{Watanabe:2006qe}, the impact of the latter stage is negligible at high frequencies (as long as $f \gg 10^{-18}\, \rm Hz$). Besides, Eq.~(\ref{eq:OGW_inf}) neglects the damping of tensor modes propagating in a universe containing free-streaming particles with non-zero anisotropic stress \cite{Weinberg:2003ur,Dicus:2005rh,Stefanek:2012hj}. This additional effect should lead to a suppression factor (close to $0.8^2$ in the minimal cosmological model, when the damping is only due to active neutrinos).

The primordial tensor spectrum $P_T(k) = 4\,  P_{h_\lambda}(k)$ is a familiar object in CMB physics, usually expressed as a function of a Fourier wavenumber $k$, since it is related to the Fourier transform of the tensor mode of metric fluctuations, $h_{ij}(\eta_\mathrm{in},\vec{x})$. The quantity $P_T$ in Eq.~(\ref{eq:OGW_inf}) is the same quantity, evaluated however at a much smaller wavenumber,  matching the wavelength of GWs probed by GW detectors. Like in the rest of this paper, we use $k$ to denote comoving wavenumbers associated to inhomogeneities on cosmological scales, and $q$ to denote comoving wavenumbers describing GW wavelengths, that is, comoving momenta of gravitons. However, the fluctuations of tensor perturbations on cosmological scales, whose variance is encoded on $P_T(k)$, comes from the existence of very large wavelengths in the graviton phase-space distribution $\Gamma$. Thus, in this case, $k$ and $q$ have the same physical interpretation. This means that $P_T(k)$ in Eq.~\eqref{eq:Cell-res} and $P_T(q)$ in Eq.~\eqref{eq:OGW_inf} represent fundamentally the same function, just evaluated on different scales (cosmological scales in the $P_T(k)$ case, and wavelengths to which GW detectors are sensitive in the $P_T(q)$ case).

The primordial tensor spectrum is commonly parametrized in terms of a tensor-to-scalar ratio $r$ and tensor tilt $n_\mathrm{t}$,
\begin{equation}
    P_T(k) = r A_s \left(\frac{k}{k_*}\right)^{n_\mathrm{t}} , 
    \label{eq:PTpl}
\end{equation}
where $A_s$ is the amplitude of scalar perturbations at the CMB pivot scale $k_* = 0.01\, \rm Mpc$. The most recent bounds on these parameters have been evaluated by combining several CMB and GW experiments~\cite{Galloni:2022mok}, finding $r<0.028$ and $-1.37<n_\mathrm{t}< 0.42$ at $95\%$ CL. 

Since $P_T(k)$ and $P_T(q)$ are fundamentally the same function, we could assume the same value for the $P_T(k)$ spectral index $n_\mathrm{t}$ and for the $P_T(q)$ spectral index $n_\mathrm{gwb}$. We note however that the power-law ansatz of Eq.~\eqref{eq:PTpl} is not necessarily valid across the huge interval ranging from CMB scales to detectable GW wavelengths. To deal with situations in which the tensor tilt is scale-dependent, we defined $n_\mathrm{gwb}$ as a free input parameter independent of $n_\mathrm{t}$ in \CLASSGW. Depending on physical assumptions, the user can either set $n_\mathrm{gwb}=n_\mathrm{t}$ (subject to the Planck bounds) or $n_\mathrm{gwb} \neq n_\mathrm{t}$ (accounting for the variation of the tensor tilt between cosmological scales and detector scales).\footnote{In the parametrization described in Appendix~\ref{sec:appendix_CLASSGW_sources}, the option $n_\mathrm{gwb}=n_\mathrm{t}$ corresponds to the case \texttt{inflationary\textunderscore gwb} and the option $n_\mathrm{gwb} \neq n_\mathrm{t}$ to the case \texttt{analytic\textunderscore gwb}.} 
We recall that the value of $n_\mathrm{gwb}$ matters because it enters the overall pre-factor in the expression of $C_\ell^{\rm CGWB \times \rm CGWB}$, see Eq.~\eqref{eq:tot_cl_bis}.

In a given cosmological model with known cosmological parameters, including $f_\mathrm{dec}(\eta_\mathrm{in})$, no further assumptions are needed to compute the SGWB power spectrum induced by single-field inflation: the code can readily evaluate Eq.~\eqref{eq:tot_cl_bis} (with the non-adiabatic power spectrum $P_\Gamma^\mathrm{NL}$ set to zero).

\subsection{Generic non-adiabatic CGWB}
\label{sec:source_power_law}

Among others, \CLASSGW offers a generic parametrization of a possible non-adiabatic contribution to the CGWB. This parametrization does not necessarily relate to known physical mechanisms, but is useful for tests and order-of-magnitude estimates.

With non-adiabatic perturbations, the initial GW spectrum $P_
\Gamma^\mathrm{NAD}(k,q)$ may depend on two independent arguments $k$ and $q$. Indeed, in the general case, $k$ refers to spatial modulations of the GW phase-space density on cosmological scales, and $q$ to the frequency spectrum of GWs. If we do not assume that tensor fluctuations are entirely generated by inflation, there is no general reason to assume that the dependence on $k$ and $q$ are the same.

In this case, we will assume that at the detector frequency $q=2\pi f/c$, the initial GW spectrum $P_
\Gamma^\mathrm{NAD}(k,q)$ depends on cosmological wavenumbers through
\begin{equation}
    P_\Gamma^\mathrm{NAD}(k,q) = A_\mathrm{gwi}(q) \, \exp\left[n_\mathrm{gwi}(q)\,  \log\frac{k}{k_*} + \frac{1}{2} \alpha_\mathrm{gwi}(q) \left(\log\frac{k}{k_*}\right)^2\right] \, ,
    \label{eq:ic_nad}
\end{equation}
where $A_\mathrm{gwi}(q)$ is the spectrum amplitude, $n_\mathrm{gwi}(q)$ the spectral index and $\alpha_\mathrm{gwi}(q)$  the running (gwi stands for Gravitational Wave Initial), all evaluated at the pivot scale $k_*$. A more general parametrization of the initial GW spectrum could be easily implemented in \CLASSGW. 
The spectral index $n_\mathrm{gwi}$, referring to $k$-dependence of the $\Gamma$ power spectrum, should not be confused with $n_\mathrm{gwb}$, which refers to frequency dependence of the background GW density (or monopole) $\bar{\Omega}_\mathrm{GW}(q)$. 

If we assume that this non-adiabatic CGWB contribution is not correlated with the adiabatic contribution, the non-adiabatic spectrum $C_\ell^{\rm CGWB \times \rm CGWB}$ featured in the second line of Eq.~(\ref{eq:tot_cl_bis}) (defined at the detector frequency $q$) can be computed for a given set of parameters 
$\{A_\mathrm{gwi}, n_\mathrm{gwi}, \alpha_\mathrm{gwi}, k_*,  n_\mathrm{gwb}, f_\mathrm{dec}(\eta_\mathrm{in}) \}$.

\subsection{Primordial Black Holes}
\label{sec:PrimordialBlackHoles}

Given the nonlinear nature of gravity, secondary GWs are produced by quadratic contributions in the scalar perturbations that act as a source in the transverse and traceless part of the Einstein equations~\cite{Tomita,Matarrese:1992rp, Matarrese:1993zf, Matarrese:1997ay,Mollerach:2003nq,Saito:2009jt,Ananda:2006af} (see \cite{Domenech:2020kqm} for a recent review). A significant amount of GWs can be generated only when the amplitude of the scalar perturbation (spectrum) at small scales is much larger than at CMB scales. This can happen, e.g., if the inflation evolution shows some deviation from scale invariance \cite{Zeldovich:1967lct}, for instance in an ultra-slow-roll phase (see e.g., \cite{Kristiano:2022maq, Riotto:2023gpm, Firouzjahi:2023ahg} for recent discussions about the possibility to generate PBH in single-field models.), or in multi-field models~\cite{Garcia-Bellido:2016dkw,Domcke:2017fix,Fumagalli:2020adf}. When such large density perturbations collapse they may lead to the formation of Primordial Black Holes (PBHs) with masses $\sim (0.001-1000) M_{\odot}$, which encompass the actual mass range probed by present GW detectors.
The SGWB energy density has been computed for different kinds of the primordial scalar spectra~\cite{Kohri:2018awv}. In the (idealized) monochromatic case, with a primordial scalar spectrum featuring a Dirac delta-function, ${\cal P}_{\mathcal{R}_s} \left( q \right) = A_* \, q_* \, \delta \left( q - q_* \right)$, it has been shown that the SGWB energy density can be computed analytically and results \cite{Espinosa:2018eve,Bartolo:2019zvb}\footnote{This expression is valid during radiation domination.}
\begin{equation} 
    \label{eq:PBH_monopole}
    \bar{\Omega}_{\rm GW} (\eta_0, f) = \frac{1}{ a_0^2 H_0^2 \eta_0^2 } \frac{A_*^2}{15552} \, 
    \frac{f^2}{f_*^2} \left[   \frac{4 f_*^2}{f^2}-1  \right]^2
    \theta \left( 2 f_* - f \right) \; \mathcal{I}^2 \left( \frac{f_*}{f} , \frac{f_*}{f}\right) \; 
 \end{equation} 
where $\theta$ is the Heaviside step function, and 
\begin{align}
    &\mathcal{I}^2\left( \frac{f_*}{f} , \frac{f_*}{f}\right) 
    % \equiv \mathcal{I}_c^2\left( \frac{f_*}{f} , \frac{f_*}{f}\right) + % \mathcal{I}_s^2\left( \frac{f_*}{f} , \frac{f_*}{f}\right)
    \nonumber \\
    & =
    \frac{729}{16} \left(\frac{ f}{ f_*} \right)^{12}
    \left(
    3-\frac{2 f_*^2}{f^2}\right)^4
    \left\{
    \left[ 4 \left(2 -3 \frac{f^2}{f_*^2}\right)^{-1} \!\!\!\!\! -\log \left( \left|   1-\frac{4 f_*^2}{3 f^2}\right| \right)\right]^2 \!\!\!
    +\pi ^2 \theta \left(\frac{2 f_*}{\sqrt{3} f}-1\right)
    \right\}. 
\end{align}
The peak GW frequency $f_*$ is related to the spike scale $q_*$ by
\begin{equation}
    f_* = \frac{c}{2 \pi a_0} \, q_* \, .
\end{equation}
In absence of primordial non-gaussianity,
 this model would generate negligible GW anisotropies beyond the unavoidable adiabatic initial conditions discussed previously. However, the authors of \cite{Bartolo:2019zvb} have shown that if some (local) underlying non-Gaussianity is present in the primordial curvature perturbation, then this generates intrinsic primordial GW anisotropies -- corresponding to non-adiabatic modes in the parametrization of Eq.~\eqref{eq:tot_cl_bis}. Reference \cite{Bartolo:2019zvb} assumes curvature perturbations with a local non-Gaussianity parametrized during matter domination by,
\begin{equation}
    \mathcal{R}( \vec{k}) = \mathcal{R}_g( \vec{k}) + \frac{3}{5} \, f_{\rm NL} \, \int \frac{d^3 p}{\left( 2 \pi \right)^3} \, 
    \mathcal{R}_g \left( \vec{p} \right) \,  \mathcal{R}_{g} ( \vec{k} - \vec{p}) \, ,
    \label{eq:fNL-def}
\end{equation}
where the subscript $g$ refers to the Gaussian part of the perturbations, and $f_{\rm NL}$ is assumed to be scale-independent. In this case,
the CGWB energy density acquires large-scale (beyond those following from the ``separate universe'' picture, which accounts for the adiabatic mode), captured by~\cite{Bartolo:2019zvb}
\begin{equation}
	\Omega_{\rm GW} ( \eta, \vec{x}, q) = \bar{\Omega}_{\rm GW} \left( \eta ,\, q \right)  
	\left[ 1 + \frac{24}{5} \, f_{\rm NL} \int \frac{d^3 k}{\left( 2 \pi \right)^3} \, {\rm e}^{i \vec{k} \cdot \vec{x}} \, \mathcal{R}_{g} \left( \vec{k} \right)  \right],
\end{equation}
where the term $\bar{\Omega}_{\rm GW} \left( \eta ,\, q \right)$ is given by \eqref{eq:PBH_monopole}.
Then the fluctuations of the graviton phase-space distribution have a non-adiabatic monopole term
\begin{equation} 
    \Gamma_0^\mathrm{NAD} ( \eta_{\rm in}, \vec{x},q) 
	=  \frac{3}{5} {\tilde f}_{\rm NL} \left( q \right)  \, \int \frac{d^3 k}{\left( 2 \pi \right)^3} \, {\rm e}^{i \vec{k} \cdot \vec{x}} 
	\, \mathcal{R}_g \left( \vec{k} \right)  \, ,
\end{equation}
where
\begin{equation}
    {\tilde f}_{\rm NL} \left( q \right) \equiv  \frac{8 \, f_{\rm NL}}{4-\frac{\partial \ln {\bar \Omega}_{\rm GW} }{\partial \ln q}} =
    \frac{8 \, f_{\rm NL}}{4-n_{\rm gwb} (q )}
    \, .
	\label{eq:GammaI-time}
\end{equation} 
In Fourier space, this initial condition reads 
\begin{equation}
    \Gamma_0^\mathrm{NAD}(\eta_{\rm in},\vec{k},q) = \frac{3}{5}\tilde{f}_{\rm NL}(q) \mathcal{R}_g(\vec{k}) \, ,
\end{equation}
where $\mathcal{R}_g(\vec{k})$ represents the Gaussian curvature perturbation on large (cosmological) scales, whose power spectrum is given by $P_{\cal R}(k)$ in the notations of previous sections.

Interestingly, since this non-adiabatic contribution depends on the curvature perturbation, it is fully correlated with the adiabatic mode. Then the adiabatic, non-adiabatic and cross-correlation spectra are related to each other through 
\begin{eqnarray}
P_\Gamma^\mathrm{NAD}(k,q) &=& \frac{9}{25}{\tilde{f}_{\rm NL}}^2(q) P_\mathcal{R}(k)~,
\label{eq:PBH_NAD} \\
P^\times(k,q) &=& \frac{6}{5}\tilde{f}_{\rm NL}(q) P_\mathcal{R}(k)~.
\label{eq:PBH_cross}
\end{eqnarray}
The non-adiabatic initial condition induced by $f_{\rm NL}$ may amplify GW anisotropies in a very significant way, since for the PBH scenario the ratio of non-adiabatic to standard AD+SW contributions to the anisotropy spectrum scales like 
\begin{equation}
    \frac{C_\ell^{\rm NAD}}{C_\ell^{\rm AD+SW}} \sim \left(\frac{\frac{3}{5} \frac{8 \, f_{\rm NL}}{4-n_{\rm gwb}(q ) }}{\frac{2}{3}\left[-\frac{2}{4-n_{\rm gwb}(q ) }+1\right]}\right)^2 = \left\{\frac{36 f_{\rm NL}}{5[2-n_{\rm gwb}(q)]}\right\}^2,
    \label{eq:C_ell_pbh_nad_enhancement}
\end{equation}
where we used $T_\psi=-\frac{2}{3}$ during radiation domination with $f_\mathrm{dec}=0$ and Eq.~(\ref{eq:transf2}). 
In the case in which $f_{\rm NL}=1$ and $n_{\rm gwb}(q)=0$, we find 
\begin{equation}
  \frac{C_\ell^{\rm NAD}}{C_\ell^{\rm AD+SW}} \approx 13\, .
  \label{eq:pbh_nad_over_ad}
\end{equation}
This qualitative relation shows that even with $f_{\rm NL}$ of order one, the non-adiabatic contribution dominates the spectrum $C_\ell^{\rm CGWB \times \rm CGWB}$.

\subsection{Phase Transition}
\label{sec:PhaseTransition}

When a PT takes place, the Universe goes from a metastable to a stable state, which represent the configurations of minimal potential energy at high and low temperatures respectively. If latent heat is involved, the PT is of the first order and the phases of the Universe are converted from the false to the true vacuum in a discontinuous way, through the nucleation of bubbles~\cite{Witten:1984rs}. Such first order PTs can happen in many extensions of the Standard Model (e.g., with additional scalar singlet or doublet, spontaneously broken conformal symmetry, or phase transitions in a hidden sector). In~\cite{Hogan:1986qda}, it has been realized for the first time that a large CGWB could be produced during a first-order PT and this is potentially detectable by present and future GW interferometers \cite{Caprini:2015zlo,Branchesi:2023mws}. In general, three main mechanisms contribute to the generation of GWs \cite{Caprini:2018mtu}, by acting as a source in the transverse-traceless part of the Einstein equations:
\begin{itemize}
    \item \textbf{Bubble wall collisions} creating distortions in the plasma. Their action is usually accounted with a method called \textit{envelope approximation}~\cite{Kosowsky:1992vn,Kamionkowski:1993fg,Caprini:2009fx,Huber:2008hg,Weir:2016tov}, consisting in approximating the bubble motion with an infinitesimally thin spherical layer. This is the backbone of the scalar field $\phi$ contribution to the SGWB signal.
    \item \textbf{Sound waves} generated by the coupling of the scalar field to the plasma during the expansion of the bubbles. These compressional modes constitute an important source of GWs also long after the collision of the bubbles~\cite{Hindmarsh:2013xza,Hindmarsh:2015qta,Hindmarsh:2016lnk}.
    \item \textbf{Turbulence phenomena} after the bubble collision, which generate vortices in the fluid with a non-vanishing quadrupole moment. The amount of GWs sourced by these eddies from Magneto-Hydro-Dynamics (MHD) turbulence has been computed for instance in~\cite{Caprini:2009yp,Binetruy:2012ze}.
\end{itemize}
As a consequence, the density of GWs generated by phase transitions can be split into three contributions,
\begin{equation}
   \bar{\Omega}_{\textrm{GW}}(f) = \bar{\Omega}_\phi(f) + \bar{\Omega}_{\textrm{sw}}(f) + \bar{\Omega}_{\textrm{turb}}(f) \, .
\end{equation}

\subsubsection*{Broken Power-law}
Each of these three contributions can be well described by a broken power law (BPL) spectrum.
In \CLASSGW we use the same parametrization as in the LIGO analysis of Ref.~\cite{Romero:2021kby},
\begin{equation}
    \bar{\Omega}^{\rm BPL}_{\textrm{GW}}(f) =
    \bar{\Omega}_* \left( \frac{f}{f_*} \right)^{n_1}
    \left[
    1 + \left( \frac{f}{f_*} \right)^{\Delta}
    \right]^{\frac{ n_2- n_1 }{\Delta}}
    \label{eq:DBPL}
\end{equation}
with $n_1 = 3$ to account causality, $n_2$ takes the value $-4$ (resp. $-1$) for sound waves (resp. bubble collisions),  and the contribution from turbulence (MHD) is neglected.

\subsection{Spectra for a few examples\label{sec:example}}

\begin{figure}[!ht]
    \centering
    \begin{subfigure}{0.50\textwidth}
        \includegraphics[width=\textwidth]{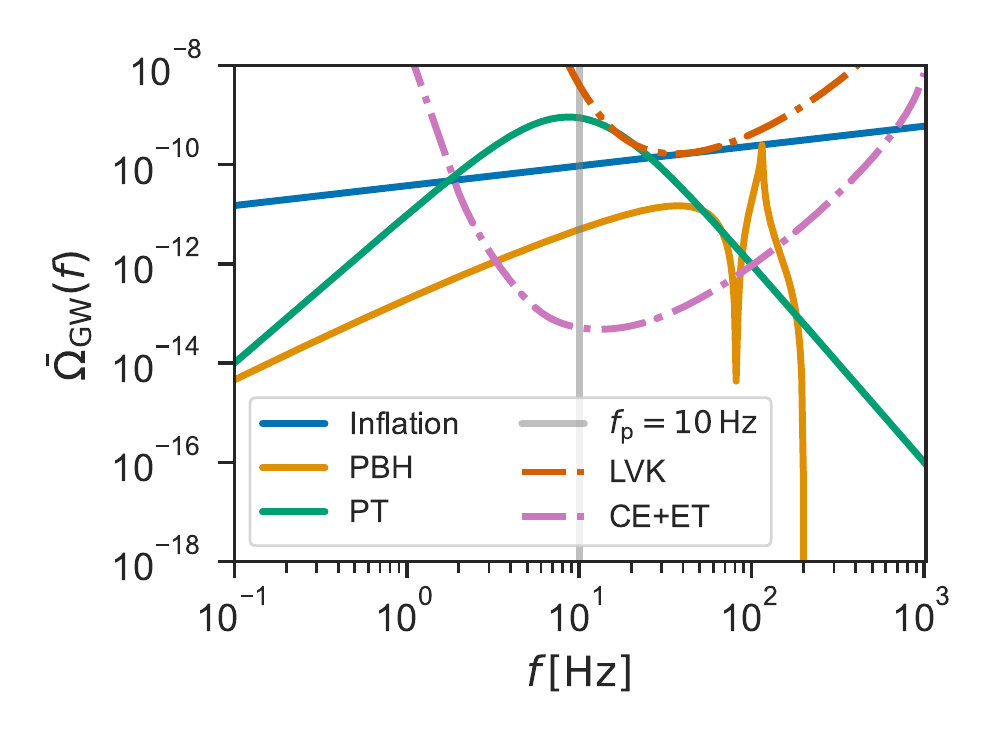}
    \end{subfigure}
    \begin{subfigure}{0.49\textwidth}
        \includegraphics[width=\textwidth]{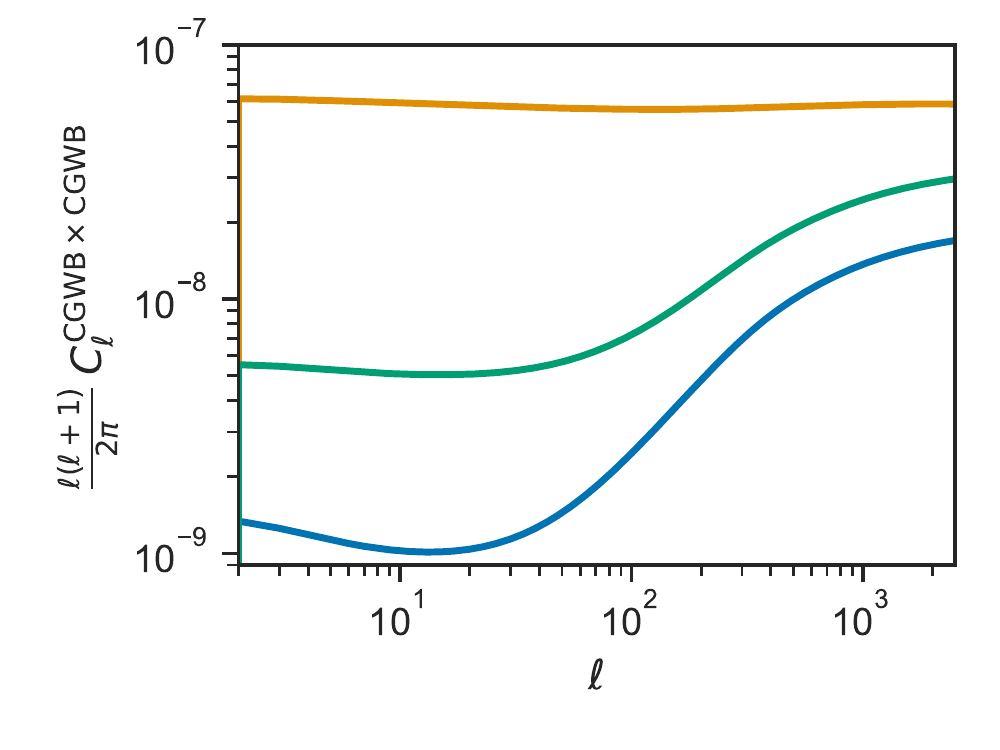}
    \end{subfigure}
    \caption{{\it Left:} Example of frequency spectra for the monopole of the CGWB generated either by: cosmological inflation (with a blue tilt $n_\mathrm{t}=0.4$), PBHs (with a peak adjusted to the frequency $f_* = 100 \, {\rm Hz}$), or a phase transitions (with a peak adjusted to $f_*=10 ~{\rm Hz}$). {\it Right:} angular power spectra for these three sources evaluated at $f_\mathrm{p} = 10 \,\rm Hz$, assuming adiabatic initial conditions in the inflation case, and non-adiabatic modes in the other two cases.
    See the text for more details on assumptions and parameters.}
        \label{fig:cgwb_different_sources}
\end{figure}

In the left panel of Figure~\ref{fig:cgwb_different_sources}, we plot the monopole of the CGWB generated by three different mechanisms:
\begin{itemize}
    \item cosmological inflation assuming a power-law tensor spectrum with $r=0.025$ and $n_\mathrm{gwb}=n_\mathrm{t}=0.4$,
    \item PBHs with an amplitude $A_*=2\times 10^{-5}$ at the peak frequency $f_*=100\, \rm Hz$,
    \item phase transitions assuming that the GW background is dominated by the sound wave contribution, with $n_1=3$, $n_2=-4$, $\Delta = 2$, $f_*=10\, \rm Hz$ and $\bar{\Omega}_*=1\times 10^{-8}$.
\end{itemize}
We choose these parameters in an optimistic way, that is, compatible with current CMB and interferometric bounds, but leading to a GW background potentially detectable by future interferometers. Thus, in the inflation case, we assumed a blue inflationary spectrum, and in the other two cases, we matched the location of the peaks to the frequency range probed by the CE and ET, while choosing amplitude parameters close to current bounds. The left panel of Figure~\ref{fig:cgwb_different_sources} also shows the combined sensitivity of CE+ET and the one of LVK. In all three cases, the signal clearly dominates the noise for $f\sim {\cal O}(10~ {\rm Hz})$, showing that such backgrounds would be detectable.

In the right panel of the same figure, we show the GW anisotropy angular power-spectra associated to these mechanisms, evaluated at a pivot frequency $f_\mathrm{p}=10 \,\rm Hz$ -- that is, close to the maximum sensitivity of the network CE$+$ET. 
We assume additionally $f_\mathrm{dec}(\eta_\mathrm{ini})=0$ and the following initial conditions:
\begin{itemize}
    \item For cosmological inflation, we take adiabatic initial conditions and $n_\mathrm{gwb}=n_\mathrm{t}$;
    \item For PBHs, on top of adiabatic initial conditions, we consider the non-adiabatic contribution generated by a (local) non-Gaussianity parameter $f_{\rm NL}=1$, and we infer $n_\mathrm{gwb}$ at the pivot scale from the background spectrum shown on the left panel ($n_\mathrm{gwb}\simeq 1.2$);
    \item For the phase transition, the simplest scenarii are expected to lead to adiabatic initial conditions, but non-adiabatic modes could arise in more complicated cases (see e.g. \cite{Kumar:2021ffi}). For illustrative purposes, we arbitrarily assume here that, on top of adiabatic initial conditions, the CGWB anisotropies include a non-adiabatic mode with the parametrization of Eq.~\eqref{eq:ic_nad}, taking $A_\mathrm{gwi}=1\times 10^{-10}$, $n_\mathrm{gwi}=0.0$, and computing $n_\mathrm{gwb}$ from the background spectrum shown on the left panel ($n_\mathrm{gwb}\simeq -0.5$). 
\end{itemize}
It is interesting to notice that the three signals considered here produce average monopole terms of the same order of magnitude,  but very different anisotropy spectra.
The features in the angular power spectra of Figure~\ref{fig:cgwb_different_sources} depend on the chosen initial condition and on the tensor tilt of the monopole signal (which is responsible for an enhancement/suppression of the angular power spectrum). 

For instance, in our examples, the angular power spectrum of the CGWB for the PT is one order of magnitude larger than that from inflation, because:
\begin{itemize}
    \item At the chosen frequency $f=10$~Hz, the AD+SW+ISW contribution to the $C_\ell^{\rm CGWB \times \rm CGWB}$ spectrum is enhanced by a factor 2.5, due to an increase in the factor $(4-n_\mathrm{gwb})^2$. To understand this in more details, one can note that, according to Eqs.~(\ref{eq:tot_cl_bis}, \ref{eq:transf2}), the SW and ISW terms get multiplied by $(4-n_\mathrm{gwb})^2$, while the AD term is independent of $n_\mathrm{gwb}$. The sign of the SW term is opposite to that of the  AD and ISW terms, but in absolute value, the SW term is the largest of the three. Thus, an increase of $(4-n_\mathrm{gwb})^2$ does lead to an increase of the total (AD+SW+ISW) contribution.
    \item Besides, our featured PT model includes a non-adiabatic contribution which is about four times larger than the adiabatic one.
\end{itemize}
Overall, this explains why the angular power-spectrum of the featured PT model is one order of magnitude larger than the inflation's one. For the PBH case, the enhancement with respect to the inflationary model is produced by similar reasons:
\begin{itemize}
    \item The decrease in the factor $(4-n_\mathrm{gwb})^2$ between the two cases reduces the (AD+SW+ISW) contribution by a factor $\sim 0.25$. %\nicola{not clear: decrease w.r.t. what?} 
    \item On the other hand, the non-adiabatic contribution is larger than the adiabatic one by two orders of magnitude, which is consistent with Eq.~\eqref{eq:C_ell_pbh_nad_enhancement}.
\end{itemize}
These two factors combine to enhance the PBH spectrum by one to two orders of magnitude  compared to the inflationary one.

The detectability of these spectra by future experiments will be discussed at length in section \ref{sec:forecast}. 

Another potential signal for GW interferometers is given by CSs, which are generated by phase transitions followed by a spontaneous breaking of symmetries, as relics of the previous more symmetric phase of the Universe. Such CSs can oscillate and give rise to a CGWB, that is typically characterized in terms of the string tension $G \mu$ (see e.g. \cite{Auclair:2019wcv,Maggiore:2019uih} and reference therein). Typically, the distribution of such CSs in the universe is not homogeneous which bring to the generation of anisotropies in the CGWB, which have been computed in \cite{Kuroyanagi:2016ugi,Jenkins:2018nty}.
Finally, also preheating models are typically characterized by anisotropies in the SGWB \cite{Bethke:2013aba, Bethke:2013vca}. In this paper we did not considered these last two generation mechanisms leaving a more dedicated analysis to a future study. 

\section{Cross-correlation spectra}
\label{sec:Cross-correlation spectra}

\subsection{Cross-correlation of CGWB at different frequencies}

\begin{figure}[t!]
    \centering
    \begin{subfigure}{0.49\textwidth}
        \includegraphics[width=\textwidth]{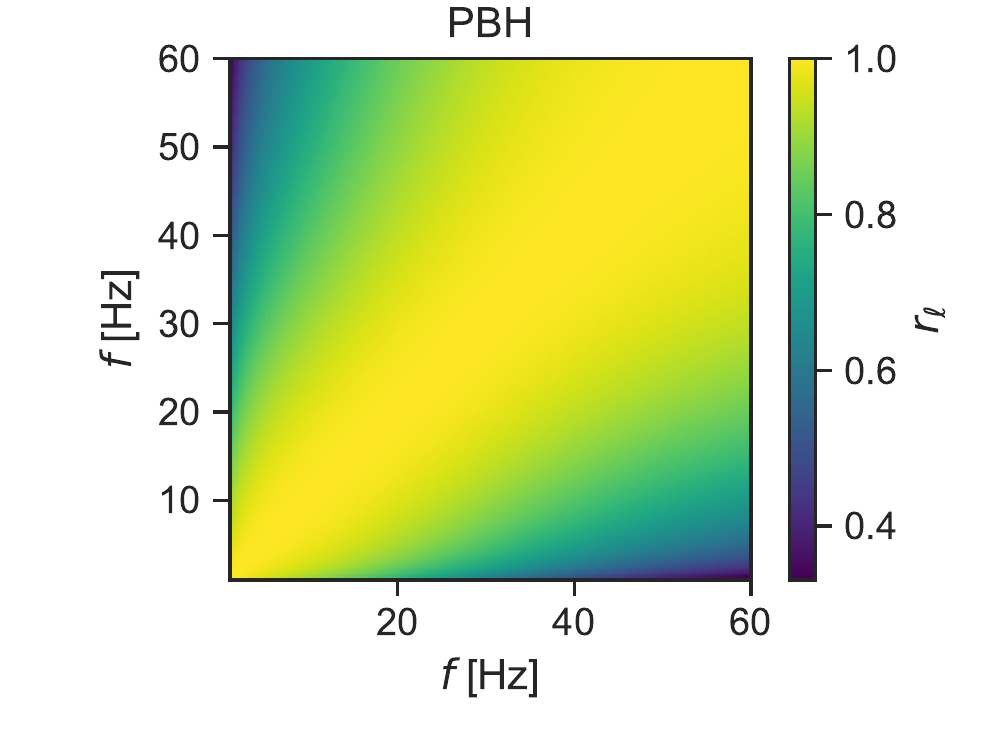}
    \end{subfigure}
    \begin{subfigure}{0.49\textwidth}
        \includegraphics[width=\textwidth]{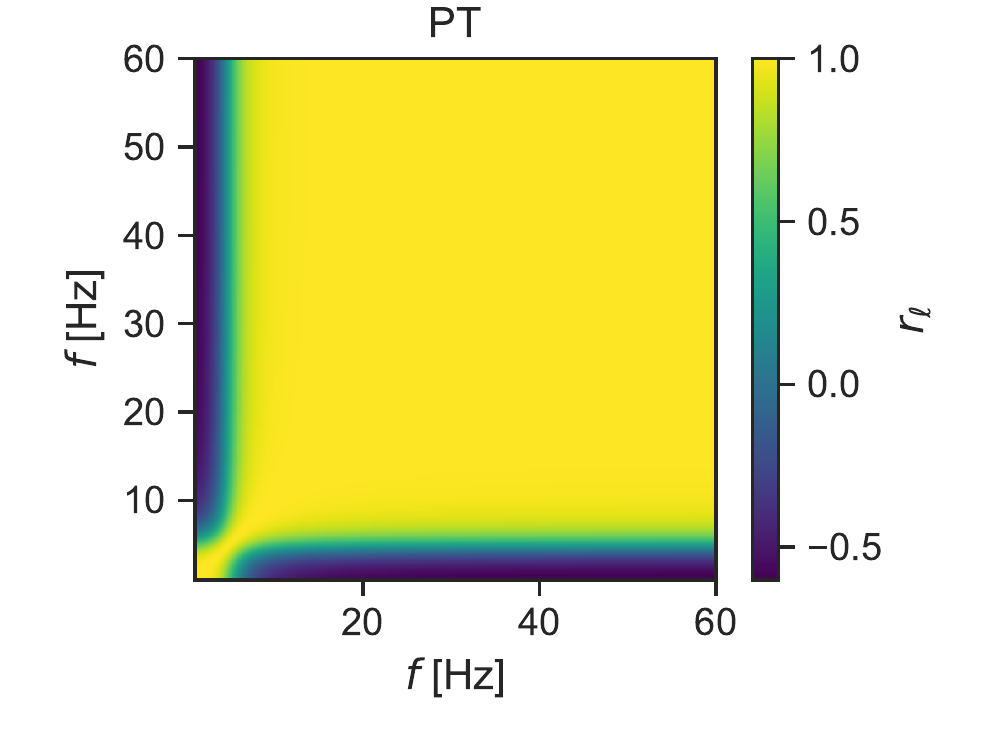}
    \end{subfigure}
    \caption{Correlation of the angular power spectrum $r_\ell(q_1,q_2)$ as a function of frequencies for $\ell = 2$, assuming two GW generation mechanisms: PBH with $f_\mathrm{NL}=10^{-2}$ (left) and PT (right).}
    \label{fig:C_ell_corr}
\end{figure}

In Eq.~(\ref{eq:tot_cl_bis}) we presented the general expression for the CGWB angular-power spectrum that would be inferred from a CGWB map at a given frequency. This expression depends on frequency for two reasons:
\begin{itemize}
    \item In general, the tensor tilt $n_\mathrm{gwb}$ depends on frequency $f$, that is, on momentum $q$. This function appears in the overall prefactor as well as in the expression of $\Delta_\ell^\mathrm{AD}$. Note that, in absence of non-adiabatic initial conditions, the factor $(4-n_\mathrm{gwb}(q))^2$ cancels in the contribution to $C_\ell^{\rm CGWB \times \rm CGWB}$ proportional to $[\Delta_\ell^\mathrm{AD}]^2$, but remains present in the terms containing SW and ISW contributions.
    \item The non-adiabatic primordial spectrum $P_\Gamma^\mathrm{NAD}(k,q)$ could depend on $q$ - and thus, so is the cross-correlation spectrum $P^{\cross}(k,q)$.
\end{itemize}
However, in future experiments, the CGWB angular spectrum is likely to be measured from the cross-correlation between pairs of CGWB maps obtained at two different frequencies, in order to reduce stochastic noise. In that case, the CGWB spectrum needs to be generalized to
\begin{equation} 
\begin{split}
    \delta_{\ell \ell'} \,\, \delta_{mm'}\, \, C_\ell^{\rm CGWB \times \rm CGWB}(q_1,q_2) \equiv \frac{1}{2}\Bigl\langle & \delta_{{\rm GW},\ell m}(q_1)\delta_{{\rm GW},\ell' m'}^*(q_2)\\
    &+\delta_{{\rm GW},\ell m}(q_2)\delta_{{\rm GW},\ell' m'}^*(q_1) \Bigl\rangle \; . 
\end{split} 
\end{equation}
The computation of the angular power spectrum of the cross-correlation of the CGWB at the frequencies $q_1$ and $q_2$ is done by using
\begin{align}
    C_\ell^{\rm CGWB \times \rm CGWB}(q_1,q_2) =&  4 \pi \left( 4 -  n_\mathrm{gwb}(q_1)\right)\left( 4 -  n_\mathrm{gwb}(q_2)\right) & \\
    &\int \frac{dk}{k} \Biggl\{P_\mathcal{R}(k)\prod_{i=1,2}\left[
    \Delta_\ell^\mathrm{AD}(k, \eta_0, \eta_\mathrm{in}, q_i) +\Delta_\ell^\mathrm{SW}(k, \eta_0, \eta_\mathrm{in}) +\Delta_\ell^\mathrm{ISW}(k, \eta_0,\eta_\mathrm{in})\right]
      \nonumber \\
    & \hspace{3em} +  
    \left[ j_\ell \left( k \left( \eta_0 - \eta_\mathrm{in} \right) \right) \right]^2 \left[P_\Gamma^\mathrm{NAD}(k,q_1,q_2)\right]
    \nonumber \\
    & \hspace{3em}+\sum_{i=1,2,j\neq i} 
    j_\ell \left( k \left( \eta_0 - \eta_\mathrm{in} \right) \right) 
    \left[
    \Delta_\ell^\mathrm{AD}(q_i) +\Delta_\ell^\mathrm{SW} +\Delta_\ell^\mathrm{ISW}\right]
    P^{\times}(k,q_j)
    \nonumber \\
    & \hspace{3em}+  \left[\Delta_\ell^{T}(k, \eta_0,\eta_\mathrm{in})\right]^2 \sum_{\lambda = \pm 2} P_{h_\lambda}(k) \Biggl\} \, ,
    \label{eq:tot_cl_bis_different_q}
\end{align}
where the dependence of $\Delta_\ell^\mathrm{AD}$ on each $q_i$ is still specified by Eq.~(\ref{eq:transf2}). In this generalization, the non-adiabatic spectrum becomes a function of the two frequencies involved in the cross-correlation,
\begin{align}
P_\Gamma^\mathrm{NAD}(k,q_1,q_2)
=
\frac{1}{2} & \left\langle \, \left[ \Gamma_0^\mathrm{NAD}(\eta_\mathrm{in}, \vec{k}, q_1)\Gamma_0^\mathrm{NAD\, *}(\eta_\mathrm{in}, \vec{k}, q_2)
\right. \right.
\nonumber \\
& ~~ \left. \left.
+\Gamma_0^\mathrm{NAD\, *}(\eta_\mathrm{in}, \vec{k}, q_1)\Gamma_0^\mathrm{NAD}(\eta_\mathrm{in}, \vec{k}, q_2) \right] \, \right\rangle~.
\end{align}

In the literature, when discussing signal-to-noise separation, it has always been assumed that CGWB anisotropy maps at different frequencies are fully correlated~\cite{Allen:1996gp,Mentasti:2020yyd,Alonso:2020rar,LISACosmologyWorkingGroup:2022kbp}, such that one could factorize the dependency on $q_1$ and $q_2$ as 
\begin{equation}
    C_\ell^{\rm CGWB\times CGWB}(q_1,q_2) = \frac{\mathcal{E}(q_1)\mathcal{E}(q_2)}{\mathcal{E}^2(q_\mathrm{p}) C_\ell^{\rm CGWB\times CGWB}(q_{\rm p})}\, .
    \label{eq:C_ell_corr_factorization}
\end{equation}
This condition is equivalent to stating that the correlation factor of the spectra at different frequencies, defined as
\begin{equation}
    r_\ell(q_1,q_2) \equiv \frac{C_\ell^{\rm CGWB \times CGWB}(q_1,q_2)}{\sqrt{C_\ell^{\rm CGWB \times CGWB}(q_1,q_1) C_\ell^{\rm CGWB \times CGWB}(q_2,q_2)}} \, ,
    \label{eq:cl_cross}
\end{equation}
is equal to one.

Here we stress that this assumption is only an approximation. For instance, when $n_\mathrm{gwb}$ is independent of $q$, and 
$P_\Gamma^\mathrm{NAD}(k,q)$ either vanishes or is independent of $q$, $r_\ell(q_1,q_2)=1$. This is typically the case when GWs are generated by inflation with purely adiabatic initial conditions and a negligible running of the tensor tilt. However, the condition is broken in other scenarii. Then, our formalism allows to compute explicitly the correlation factor according to Eqs.~(\ref{eq:tot_cl_bis_different_q}, \ref{eq:cl_cross}).

\begin{figure}
    \centering
    \includegraphics[scale=0.7]{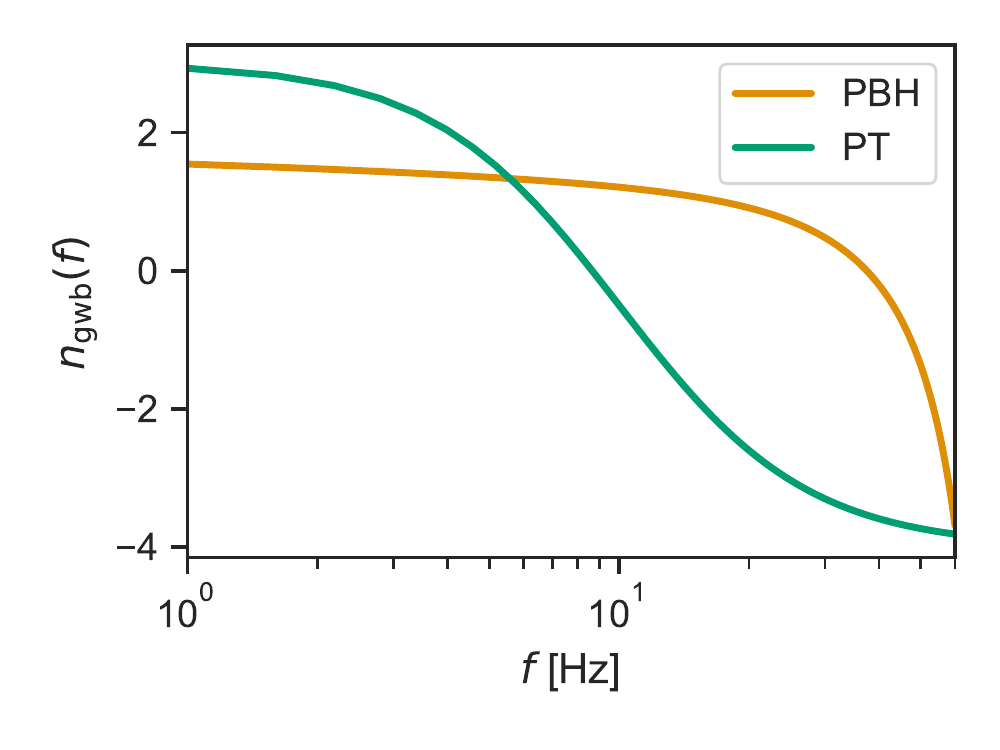}
    \includegraphics[scale=0.7]{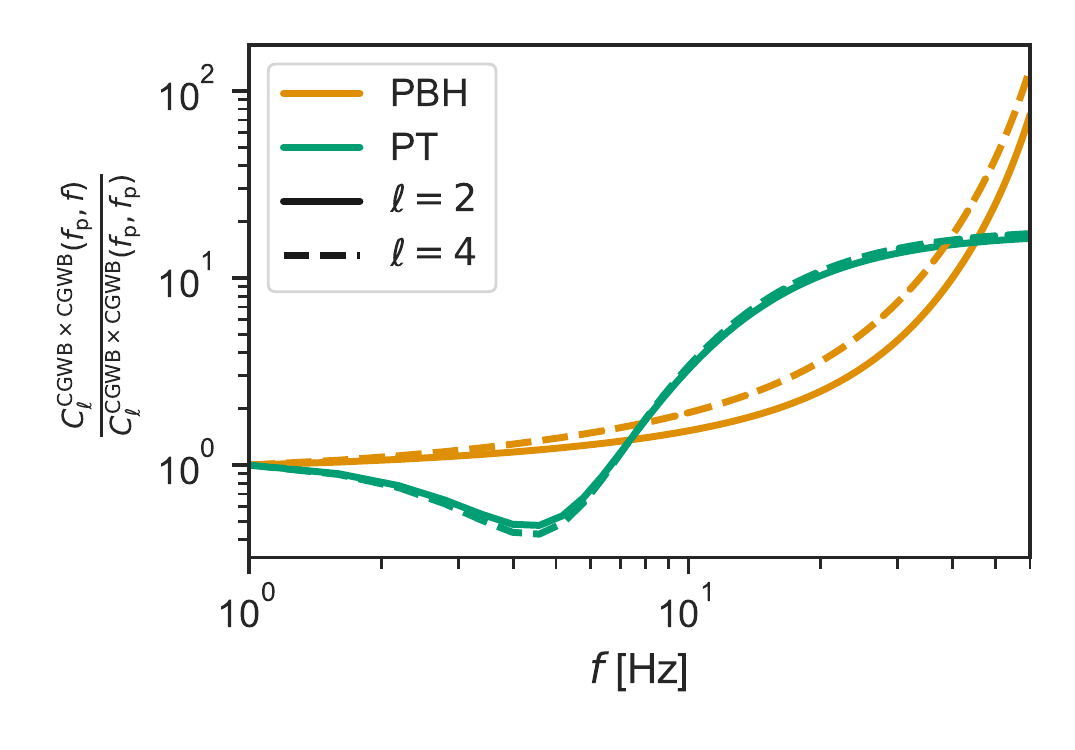}
    \caption{Left: dependence of the tensor tilt $n_\mathrm{gwb}$ over frequency. Right: dependence of the quadrupole and hexadecapole of the CGWB spectrum $C_\ell^{\rm CGWB\times CGWB}(q,q)$ over frequency (using the one-to-one correspondence between momentum $q$ and frequency $f$), normalized at the pivot frequency $f_p = 1\, \rm Hz$.}
    \label{fig:C_ell_freq}
\end{figure}

We illustrate this for two different cases in Figure~\ref{fig:C_ell_corr}.  These cases are similar to the examples picked up in Section \ref{sec:example} and shown in Fig.~\ref{fig:cgwb_different_sources}. However, in the PBH case, we consider $f_{\rm NL}=10^{-2}$ instead of $f_{\rm NL}=1$. This choice is motivated by the fact that at low multipoles, when $f_{\rm NL}$ is very large, the angular power spectrum is dominated by the NAD initial condition. Then, in very good approximation, the frequency dependence can be factorized and $r_\ell(q_1,q_2) \simeq 1$, like for the standard inflationary case. 

In Figure~\ref{fig:C_ell_freq} we show the angular power spectrum evaluated at $q_1=q_2$ for different frequencies, normalized at the pivot frequency $f_p = 1\, \rm Hz$. This shows how the anisotropies coming from different sources -- or from a given source but from different multipoles -- depend differently on frequency. This behaviour is potentially useful for an efficient component separation, as discussed in~\cite{ValbusaDallArmi:2022htu}.

\subsection{Cross-correlation of SGWB with CMB}
\label{sec:cross_correlations}

Since CMB photons and gravitons share the same geodesics, along which they get red-shifted or blue-shifted by the same metric fluctuations, we expect a significant correlation between CMB temperature anisotropies and GW energy density anisotropies. This cross-correlation has been studied in detail in~\cite{Ricciardone:2021kel} (sticking to the perturbations induced by scalar fluctuations on cosmological scales, which provide the dominant contribution; see also \cite{Galloni:2022rgg} where the anisotropies induced by tensor perturbations have been included too). An additional amount of correlation between the CMB and the CGWB, which is not considered in this work, could be caused by the existence of a non-trivial primordial scalar-tensor-tensor non-Gaussianity~\cite{Adshead:2020bji}.

The multipoles of GW anisotropies can be inferred from a line-of-sight integral according to Eqs.~(\ref{eq:solutionlm} -- \ref{eq:transf}). 
For the multipoles $a_{\ell m}$ of CMB temperature fluctuations induced by adiabatic scalar perturbations, the equivalent integral reads \cite{Seljak:1996is}
\begin{align}
    a_{\ell m} = 4\pi \, (-i)^\ell \int \frac{d^3k}{(2\pi)^3} & \, e^{i \vec{k} \cdot \vec{x}_0} \,  Y_{\ell m}^*(\hat{k}) \, {\cal R}(\vec{k}) \, \Theta^S_\ell(k,\eta_0)\, , 
    & \label{eq:alm} \\
    \Theta^S_\ell(k,\eta_0) = \int_{\eta_\mathrm{min}}^{\eta_0}d\eta
    \Bigl[ & g(\eta) \,\, \Bigl( T_{\Theta_0}(\eta,k) + T_\psi(\eta,k) \Bigr) j_\ell[k(\eta_0-\eta)] &\mathrm{(SW)} \nonumber \\
    & + g(\eta) \,\, k^{-1} T_{\theta_b}(\eta,k) j_\ell^\prime [k(\eta_0-\eta)] 
    &\mathrm{(DOP)} \nonumber \\
    & + e^{-\kappa(\eta)} \,\, \frac{\partial[ T_\psi(\eta,k)+ T_\phi(\eta,k)]}{\partial \eta} j_\ell[k(\eta_0-\eta)]\Bigr] \, , 
    &\mathrm{(ISW)} \nonumber
\end{align}
where $\kappa(\eta)$ is the photon optical depth, $g(\eta)$ the visibility function, $T_{\Theta_0}(\eta,k)$ the transfer function of the photon temperature monopole, and $T_{\theta_b}(\eta,k)$ the transfer function of the divergence of the baryon bulk velocity. The line-of-sight integral features three terms standing for the Sachs-Wolfe (SW), Doppler (DOP) and Integrated Sachs-Wolfe (ISW) contributions. Assuming adiabatic scalar perturbations only, we can write the CMB$\times$CGWB cross-correlation angular power spectrum as
\begin{equation}
    \delta_{\ell \ell^\prime}\delta_{m m^\prime} C_\ell^{\rm{CMB}\times \rm{CGWB}}(q)\equiv\frac{1}{2}\langle \delta_{{\rm GW},\ell m} (\eta,q) \, a^*_{\ell^\prime m^\prime}(\eta)+\delta^*_{{\rm GW},\ell m} (\eta,q) \, a_{\ell^\prime m^\prime}(\eta) \rangle  \,,
    \label{eq:definition_cross_correlation}
\end{equation}
where the adiabatic scalar contribution to $\delta_{{\rm GW},\ell m}$ can be inferred from Eq.~\eqref{eq:transf2}:
\begin{align}
    \delta_{{\rm GW}, \ell m} = 4\pi \, (-i)^\ell (4 - n_\mathrm{gwb}) & \int \frac{d^3k}{(2\pi)^3}  \, e^{i \vec{k} \cdot \vec{x}_0} \,  Y_{\ell m}^*(\hat{k}) \, {\cal R}(\vec{k}) \, \nonumber \\
    & \times \left[ \Delta^\mathrm{AD}_\ell(k,\eta_0,\eta_\mathrm{in})
    + \Delta^\mathrm{SW}_\ell(k,\eta_0,\eta_\mathrm{in})
    + \Delta^\mathrm{ISW}_\ell(k,\eta_0,\eta_\mathrm{in})
    \right]
    \, . 
    \label{eq:alm2}
\end{align}
This cross-correlation spectrum can be expanded as the sum of six terms, 
\begin{equation}
    C_\ell^{\rm{CMB}\times \rm{CGWB}}=C^{\rm{SW}\times \rm{SW}}_\ell + C^{\rm{SW}\times \rm{ISW}}_\ell+C^{\rm{ISW}\times \rm{SW}}_\ell+C^{\rm{ISW}\times \rm{ISW}}_\ell+C_\ell^{\rm DOP \times SW}+C_\ell^{\rm DOP \times ISW}\, ,
    \label{eq:cross_expansion}
\end{equation}
each of them involving at last one line-of-sight integral for the CMB part.\footnote{
{Here, for simplicity of notations, when referring to the SW of the CGWB, we include also the monopole of the (adiabatic) initial anisotropies, called AD in previous equations. Such a combination of AD+SW was referred to as the Free Streaming Monopole (FSM) in the notations of~\cite{Ricciardone:2021kel}.}} Below, we give approximate expressions for these six terms, based on the instantaneous decoupling approximation $g(\eta)=\delta(\eta-\eta_*)$, where $\eta_*$ is the conformal age of the universe at the time of photon decoupling:\footnote{This assumption implies that the optical depth is given by the Heaviside function $\kappa(\eta)=H(\eta-\eta_0)$.}
\begin{equation}
    \begin{split}
        \frac{C^{\rm{SW}\times \rm{SW}}_\ell}{4-n_{\rm gwb}} =& 4\pi\int \frac{dk}{k}P_\mathcal{R}(k) j_\ell[k(\eta_0-\eta_*)]j_\ell[k(\eta_0-\eta_\mathrm{in})] \\
        & \hspace{4em} \times \Bigl[ T_{\Theta_0}(\eta_*,k) + T_{\psi}(\eta_*,k)\Bigr] \left[ 
        %\frac{2-n_{\mathrm{gwb}(q)}}{4-n_{\mathrm{gwb}(q)}} 
        T_\Gamma^\mathrm{AD}(\eta_\mathrm{in},k,q) + 
        T_\psi(\eta_\mathrm{in},k) \right] \, ,
        \\
        \frac{C^{\rm{SW}\times \rm{ISW}}_\ell}{4-n_{\rm gwb}} =& 4\pi\int \frac{dk}{k}P_\mathcal{R}(k) j_\ell[k(\eta_0-\eta_*)] \Bigl[ T_{\Theta_0}(\eta_*,k) + T_{\psi}(\eta_*,k) \Bigr] \\
        & \hspace{4em} \times \int_{\eta_\mathrm{in}}^{\eta_0}d\eta \Bigl[ T_{\psi}^\prime(\eta,k) + T_{\phi}^\prime(\eta,k) \Bigr] j_\ell[k(\eta_0-\eta)] \, ,
        \\
        \frac{C^{\rm{ISW}\times \rm{SW}}_\ell}{4-n_{\rm gwb}} = &4\pi\int \frac{dk}{k}P_\mathcal{R}(k) j_\ell[k(\eta_0-\eta_\mathrm{in})] 
        \left[ 
        %\frac{2-n_{\mathrm{gwb}(q)}}{4-n_{\mathrm{gwb}(q)}} 
        T_\Gamma^\mathrm{AD}(\eta_\mathrm{in},k,q) + 
        T_\psi(\eta_\mathrm{in},k) \right]
        \\
        & \hspace{4em} \times \int_{\eta_*}^{\eta_0}d\eta \Bigl[ T_{\psi}^\prime(\eta,k) + T_{\phi}^\prime(\eta,k) \Bigr] j_\ell[k(\eta_0-\eta)] \, ,
        \\
        \frac{C^{\rm{ISW}\times \rm{ISW}}_\ell}{4-n_{\rm gwb}} =& 4\pi\int \frac{dk}{k}P_\mathcal{R}(k) \int_{\eta_*}^{\eta_0}d\eta \Bigl[ T_{\psi}^\prime(\eta,k) + T_{\phi}^\prime(\eta,k) \Bigr] j_\ell[ k(\eta_0-\eta)] \\
        & \hspace{4em} \times \int_{\eta_\mathrm{in}}^{\eta_0}d\tilde{\eta} \Bigl[ T_{\psi}^\prime(\tilde{\eta},k) + T_{\phi}^\prime(\tilde{\eta},k) \Bigr] j_\ell[k(\eta_0-\tilde{\eta})] \, , \\
        \frac{C_\ell^{\rm DOP\times SW}}{4-n_{\rm gwb}} =& 4\pi \int \frac{dk}{k}P_\mathcal{R}(k) 
        \, j_\ell^\prime [k(\eta_0-\eta_*)] \,\,
        j_\ell[k(\eta_0-\eta_\mathrm{in})] \\
        & \hspace{4em}
        \times k^{-1} T_{\theta_b}(\eta_*,k) 
        \left[ T_\Gamma^\mathrm{AD}(\eta_\mathrm{in},k,q) + 
        T_\psi(\eta_\mathrm{in},k) \right]]
        \, , \\
        \frac{C_\ell^{\rm DOP\times ISW}}{4-n_{\rm gwb}} =& 4\pi \int \frac{dk}{k}P_\mathcal{R}(k) \, j_\ell^\prime [k(\eta_0-\eta_*)] \,\,k^{-1} T_{\theta_b}(\eta_*,k) 
         \\
        & 
        \hspace{4em} \times \int_{\eta_\mathrm{in}}^{\eta_0}d\eta  
        \Bigl[ T_{\psi}^\prime(\eta,k) + T_{\phi}^\prime(\eta,k) \Bigr] 
        j_\ell[k(\eta_0-\eta)] \, .        
        \label{eq:cross_correlation_contributions}
    \end{split}
\end{equation}
Note that \CLASSGW does not rely on the instantaneous decoupling approximation and always computes the full CMB line-of-sight integral.

\begin{figure}[t!]
	\centering
	\includegraphics{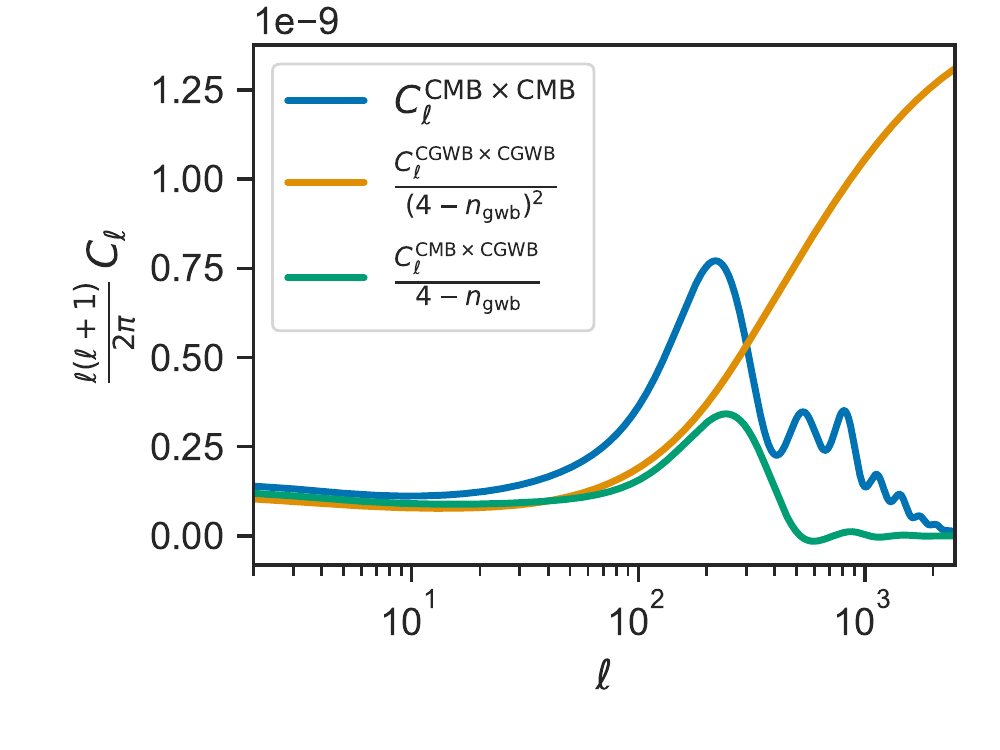}~~~~~~~
	\caption{Angular power spectrum of the CMB, of the CGWB with adiabatic initial condition, $n_{\rm gwb} = 0.4$ and $\fdec = 0$, and cross-correlation between the two.}
	\label{fig:comparison_cgwb_cmb_cross}
\end{figure}

In Fig. \ref{fig:comparison_cgwb_cmb_cross}, we compare the angular power spectra of CMB temperature auto-correlation, of the CGWB auto-correlation (sourced by inflation), and the cross-correlation spectrum. For a more straightforward comparison, in this figure, we divide the CGWB auto-correlation spectrum by $(4-n_\mathrm{gwb})^2$ and the cross-spectrum by $(4-n_\mathrm{gwb})$. In this case, the three spectra receive a contribution from nearly the same SW term, which explains their similar order of magnitude on large angular scales.

\begin{figure}[ht!]
    \centering
	~~~~~~~~~~~~~~~~~~~~~~~~~~~~~~~~\includegraphics[width=.85\textwidth]{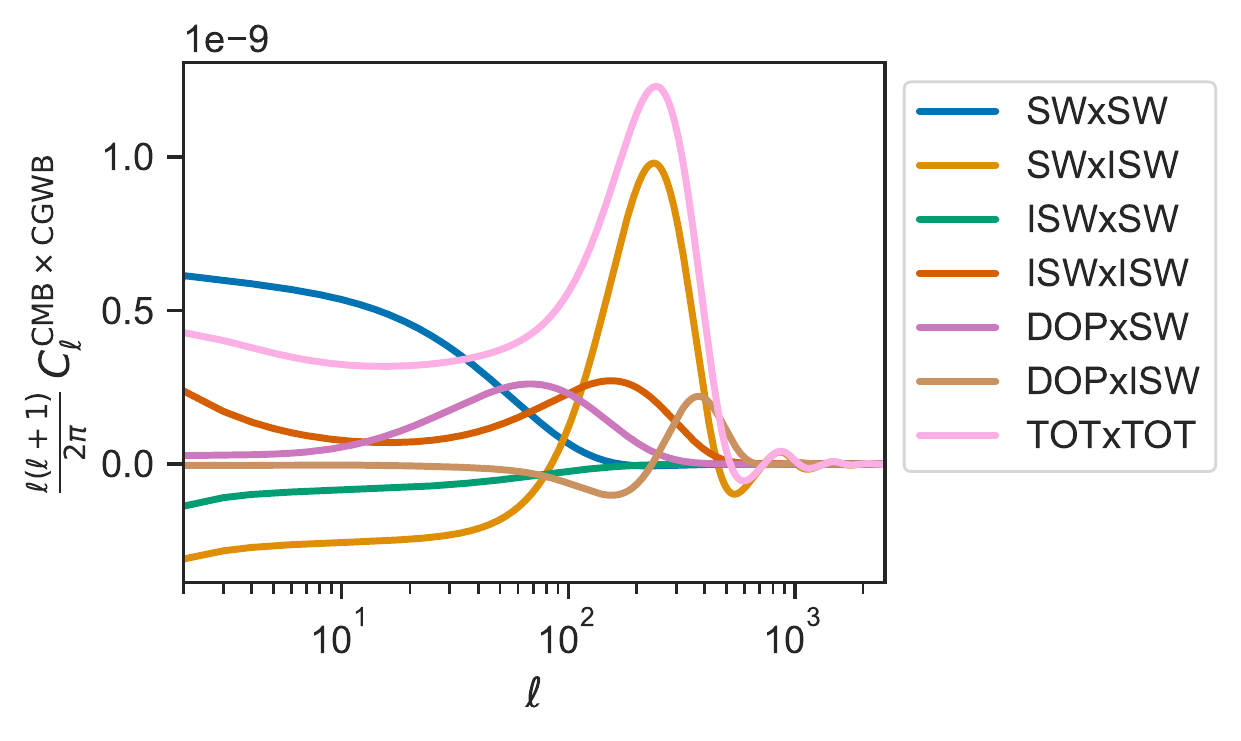}
	\caption{Contributions to the angular power spectrum of the cross-correlation between the CMB and the CGWB.}
	\label{fig:cross_contributions}
\end{figure}

The six contributions of Eq.~\eqref{eq:cross_correlation_contributions} to the cross-correlation spectrum are plotted in Fig.~\ref{fig:cross_contributions}, as well as the total cross-correlation spectrum. The overall behavior of each term can be understood qualitatively as follows.\\

\noindent {\bf SW $\times$ SW contribution.} This term gets contributions from Fourier modes propagating at precisely $\eta_*$ for the CMB perturbations and $\eta_\mathrm{in}$ for the GW perturbations. Thus, they originate from two different last scattering spheres, with a comoving radius given respectively by $r_* = \eta_0-\eta_*$ and $r_\mathrm{in}=\eta_0-\eta_\mathrm{in}$. Note that the decoupling times are extremely different, $\eta_* \gg \eta_\mathrm{in}$, but the radii of the two spheres are not, since $r_\mathrm{in}-r_* = \eta_*-\eta_\mathrm{in}$ is much smaller than $r_* \sim r_\mathrm{in} \sim \eta_0$. 

One may naively expect that the correlation between fluctuations observed on two different last scattering spheres are negligible. In reality, this is not the case. Indeed, primordial perturbations with a comoving wavelength $2\pi/k$ larger than the radii difference $(\eta_*-\eta_\mathrm{ini})$ imprint nearly the same patterns on the two spheres, and lead potentially to strong correlations. The correlation induced by each of these wavelengths is seen today mainly under an angle $\theta \simeq \frac{\pi}{k \eta_0}$, and contributes mainly to the multipole $\ell \simeq k \eta_0$. Thus, we expect the angular spectrum to be significant for  $\ell \ll \frac{2\pi\,  \eta_0}{\eta_*-\eta_\mathrm{in}} \sim 300$. Conversely, primordial perturbations with a comoving wavelength $2\pi/k$ smaller than $(\eta_*-\eta_\mathrm{ini})$ imprint different patterns on the two spheres and should leave negligible correlations. Thus, the angular spectrum should be suppressed for  $\ell \gg \frac{2\pi\,  \eta_0}{\eta_*-\eta_\mathrm{in}}$.

This expectation can be confirmed analytically. In the expression of ${C^{\rm{SW}\times \rm{SW}}_\ell}$ in Eq.~\eqref{eq:cross_correlation_contributions}, the transfer functions $T_{\Theta_0}$, $T_\psi$, $T_\Gamma^\mathrm{AD}$, which are independent of $k$ on super-Hubble scales, can be pulled out of the integral in first approximation.  We can do the same with the primordial curvature spectrum $P_\mathcal{R}(k)\simeq A_s$, assuming a tilt $n_\mathrm{s}$ close to one. Then, the shape of ${C^{\rm{SW}\times \rm{SW}}_\ell}$ as a function of $\ell$ depends on
\begin{equation}
    I_\ell=\int \frac{dk}{k} \, j_\ell[k(\eta_0-\eta_*)] \, j_\ell[k(\eta_0-\eta_\mathrm{in})]\, .
\end{equation}
A calculation summarized in Appendix~\ref{sec:Il} shows that this integral scales like
\begin{equation}
    I_\ell \propto \frac{1}{\sqrt{\ell}(\ell+\frac{1}{2})} \Bigl(\frac{\eta_0-\eta_*}{\eta_0-\eta_\mathrm{in}}\Bigl)^{\ell}
    \simeq \frac{1}{\sqrt{\ell}(\ell+\frac{1}{2})} \Bigl(1 - \frac{\eta_*-\eta_\mathrm{in}}{\eta_0}\Bigl)^{\ell} 
    \simeq \frac{1}{\sqrt{\ell}(\ell+\frac{1}{2})} e^{- \frac{\eta_*-\eta_\mathrm{in}}{\eta_0} \ell}, 
    \label{eq:sw_crosscorrelation_suppression}
\end{equation}
which, as expected, gets suppressed exponentially for $\ell \gg \frac{\eta_0}{\eta_*-\eta_\mathrm{in}}$.

We do observe this small-scale suppression in Fig. \ref{fig:cross_contributions}. On large angular scales, the amplitude of ${C^{\rm{SW}\times \rm{SW}}_\ell}$ is related to the positive value of $[T_{\Theta_0}+T_\psi][T_\Gamma^\mathrm{AD}+T_\psi]$ on super-Hubble scales. \\

\noindent {\bf SW $\times$ ISW contribution.} In this term, CMB perturbations contribute at the time $\eta_*$ and GW perturbations at all times in the range $\eta_\mathrm{in} \leq \eta^\prime \leq \eta_0$. The correlation between the CMB and GW perturbations peaks when the two types of perturbations are probed on the same sphere, that is, when $\eta^\prime \sim \eta_*$. Thus, this term depends on the product of the transfer functions $[T_{\Theta_0}(\eta_*,k) + T_{\psi}(\eta_*,k)][T_{\psi}^\prime(\eta_*,k) + T_{\phi}^\prime(\eta_*,k)]$, all evaluated around the time of photon decoupling. The behaviour of this product as a function of $k$ should determine the shape of ${C^{\rm{SW}\times \rm{ISW}}_\ell}$ as a function of $\ell$, using the angular projection relation $\theta \sim \frac{\pi}{k \eta_0}$ or $\ell \sim k \eta_0$. 

The factor $[T_{\Theta_0}(\eta_*,k) + T_{\psi}(\eta_*,k)]$ is constant on very large scales ($k \eta_* \ll 1$), and then has damped oscillations. The factor $[T_{\psi}^\prime(\eta_*,k) + T_{\phi}^\prime(\eta_*,k)]$, also responsible for the early ISW effect in the CMB auto-correlation spectrum, has a broad peak on scales similar to those of the first acoustic peak. As a result, ${C^{\rm{SW}\times \rm{ISW}}_\ell}$ has a plateau on large angular scales, a peak coinciding with the first acoustic oscillation, and then smaller damped oscillations. The plateau and the peak have different signs because $[T_{\Theta_0} +T_\psi]$ crosses zero near $k \sim 10^{-2}\,h/$Mpc at $\eta \sim \eta_*$ (due to the onset of acoustic oscillations).\\

\noindent {\bf ISW $\times$ SW contribution.} In this term, CMB perturbations contribute at all times in the range $\eta_*\leq \eta \leq \eta_0$ and GW perturbations at the precise time $\eta_\mathrm{in}$. Since these times never overlap, the correlation coming from this term is always very small. It is not exactly zero for the same reason as in the SW$\times$SW case: on the largest angular scales, the same primordial fluctuations contribute to the two last scattering spheres. Even though, the correlation is very small because the transfer function $[T_{\psi}^\prime(\eta_*,k) + T_{\phi}^\prime(\eta_*,k)])$ is negligible in the super-Hubble limit (in which $\phi$ and $\psi$ are nearly constant in time). Thus, the ${C^{\rm{ISW}\times \rm{SW}}_\ell}$ contribution is sub-dominant. \\

\noindent {\bf ISW $\times$ ISW contribution.} In this term, CMB perturbations contribute at times in the range $\eta_*\leq \eta \leq \eta_0$ and GW perturbations are all times in the range $\eta_\mathrm{in} \leq \eta^\prime \leq \eta_0$. The correlation comes mainly from all overlapping times $\eta \simeq \eta'$ with $\eta_* \leq \eta \leq \eta_0$. Actually, up to the prefactor $(4-n_\mathrm{gwb})$, this term is nearly identical to the ISW $\times$ ISW contribution to the CMB auto-correlation spectrum. Like the latter, it includes a tilted plateau at small $\ell$'s, corresponding to the late ISW effect, and a peak close to the scale of the first acoustic peak, corresponding to the early ISW effect.\\

\noindent {\bf DOP $\times$ SW contribution.} The discussion of this term is qualitatively similar to the SW $\times$ SW case. The main difference is that the transfer function associated to the Doppler term, $T_{\theta_b}$, vanishes on super-Hubble scales, unlike  $[T_{\Theta_0}+T_\psi]$. Thus, this term only has a broad peak for $\ell \sim 100$.\\

\noindent {\bf DOP $\times$ ISW contribution.} A reasoning similar to the SW $\times$ ISW case shows that the correlation now depends on the product of the transfer functions $T_{\theta_b}(\eta_*,k) \, [T_{\psi}^\prime(\eta_*,k) + T_{\phi}^\prime(\eta_*,k)])$, all evaluated around the time of photon decoupling. Compared to the SW transfer function $[T_{\Theta_0}(\eta_*,k) + T_{\psi}(\eta_*,k)]$, the Doppler transfer function $T_{\theta_b}(\eta_*,k)$ vanishes on scales larger than the sound horizon, and has oscillations on smaller scales, but with a phase shift compared to the SW case. Instead, the transfer functions $[T_{\psi}^\prime(\eta_*,k) + T_{\phi}^\prime(\eta_*,k)]$ are suppressed on scales smaller than the sound horizon. As a result, ${C^{\rm{DOP}\times \rm{ISW}}_\ell}$ has two small peaks on intermediate scales (comparable to the scales of the first two acoustic peaks in the CMB spectrum, but with a different phase).\\

\noindent {\bf Total contribution.} The total contribution ${C^{\rm{CMB}\times \rm{CGWB}}_\ell}$ is dominated by the SW$\times$SW and SW$\times$ISW contributions. It exhibits a tilted plateau on large angular scales, a peak on intermediate scales, and a few damped oscillations. Because of its origin in the SW $\times$ ISW contribution, the main peak does not originate simply for the first oscillation of the photon density transfer function, and reaches its maximum at a slightly larger $\ell$ than the first CMB peak.

\section{Sensitivity forecasts}
\label{sec:forecast}
In this section, we forecast the sensitivity of the ET+CE network to the information contained in the CGWB anisotropies. As a network configuration we assumed an ET triangular detector with 10-km arm-length placed in Sardinia, and 2 L-shaped CE detectors of 40 and 20 km arms placed in the actual two LIGO detectors.  Present and future GW interferometers, both on the ground and in space, are limited in angular sensitivity. Space-based interferometers that are expected to have high (angular) sensitivities, are the Big Bang Observer (BBO) \cite{Corbin:2005ny} and the DECI-hertz interferometer Gravitational wave Observatory (DECIGO) \cite{Kawamura:2020pcg} or network of detectors \cite{Orlando:2020oko}. So we use such reference sensitivities for our analysis, even if we are focusing on ground-based detectors. Having in mind that future noise upgrades of both ET and CE, could bring the sensitivity close to such values.  Of course, the \CLASSGW code can be easily used with other GW detectors like LISA and/or Taiji. 
Additionally, to showcase the theoretical limit on the constraining power of CGWB anisotropies, we consider a cosmic-variance-limited detection up to $\ell_\mathrm{max}=2500$, called CV.
Our forecasts rely on assuming a mock likelihood and fitting mock data with a Monte-Carlo-Markov-Chain (MCMC) method. Our pipeline is implemented as an extension of the Bayesian inference package \MontePy \cite{Audren:2012wb,Brinckmann:2018cvx}.

\subsection{Detector noise}
\label{sec:GW_Reconstructed_Map}

To perform a forecast on mock data, we need to make assumptions concerning instrumental noise. 

For this purpose, we rely on previous investigations of CGWB anisotropy map reconstruction techniques~\cite{Allen:1996gp,Mentasti:2020yyd,Alonso:2020rar,LISACosmologyWorkingGroup:2022kbp,Mentasti:2023icu,Mentasti:2023gmg}. The problem consists in finding optimal algorithms to go from raw data to CGWB maps at a given frequency. For GW interferometers, the raw data consists in a measurement of the time-delay required by light to complete a flight across the arms as a function of frequency and time - with each point in the timeline corresponding to a given orientation of the detector. The target is a map of CGWB density fluctuations at a given frequency, $\bar{\Omega}_{\rm CGWB}(q) \, \delta_\mathrm{CGWB} (q, \hat{n})$, see Eq.~(\ref{eq:gwdensitycontrast1}). The map can be expanded in harmonic space as  $\bar{\Omega}_{\rm CGWB}(q) \, a_{\ell m}^{\rm CGWB}$, where $a_{\ell m}^{\rm CGWB}$ is a Gaussian random field of zero mean and of covariance given -- in absence of detector noise -- by the angular power spectrum $C_\ell^{\rm CGWB\times CGWB}(q)$ of Eq.~(\ref{eq:tot_cl_bis}). 

Previous works approached this problem under the assumption that the dependence of the monopole $\bar{\Omega}_{\rm CGWB}(q)$ and of the power spectrum $C_\ell^{\rm CGWB\times CGWB}(q)$ over frequency are known, and also, that the correlation between anisotropies at different frequencies is exactly one or, equivalently, that the frequency dependence of the anisotropies can be factorized out of the stochastic part,
\begin{equation}
    a_{\ell m}^{\rm CGWB}(q) = \frac{\mathcal{E}(q)}{\mathcal{E}(q_{\rm p})}a_{\ell m}^{\rm CGWB}(q_{\rm p})\, ,
\end{equation}
where $q_{\rm p}$ is an arbitrary pivot scale. In section \ref{sec:cross_correlations}, we mentioned that this assumption is not exact, but in the forecasts presented in this work, we will stick to this approximation.

The measurement of the map in presence of detector noise and of other types of signals or foregrounds  is a typical component separation problem.
Refs.~\cite{Allen:1996gp,Mentasti:2020yyd,Alonso:2020rar,Contaldi:2020rht,LISACosmologyWorkingGroup:2022kbp} show how to build an unbiased estimator of the map at the pivot scale using a linear combination of raw data at different frequencies, with weights chosen to minimize the covariance of the map. This covariance can be split in two contributions:  the spectrum of the signal, $C_\ell^{\rm CGWB\times CGWB}(q_{\rm p})$, and the noise spectrum $N_\ell^\mathrm{GW}(q_{\rm p})$. The latter depends on:
\begin{itemize}
\item the assumed frequency dependence of the monopole $\bar{\Omega}_{\rm CGWB}(q)$ and of the angular power spectrum $\mathcal{E}(q)$,
\item the characteristics of the detector,
\item the number of years of observations.
\end{itemize}
Estimates of the angular noise spectrum can be performed using the code \texttt{schNell}~\cite{Alonso:2020rar}. The output of the code is the noise spectrum of the density fluctuation $\bar{\Omega}_{\rm CGWB}(q) \, a_{\ell m}^{\rm CGWB}$ rather than that of the density contrast $a_{\ell m}^{\rm CGWB}$. Thus, $N_\ell^\mathrm{GW}$ is given by the \texttt{schNell} output divided by $\bar{\Omega}_\mathrm{GW}^2$.
In the original version of the code, it was possible to compute the angular power spectrum of the noise only for monopoles that scale like a power law, $\bar{\Omega}_{\rm CGWB}(q)\propto q^{n_\mathrm{gwb}}$. We have generalized the algorithm for stochastic backgrounds with a non-trivial frequency dependence, like the CGWBs sourced by a PT or PBHs described in section~\ref{sec:CGWB_sources}. We have also implemented in the code the sensitivity of the network ET+CE. 

The results are displayed in Figures \ref{fig:Cosmology_spectrum}, \ref{fig:PBH_spectrum}, and \ref{fig:PT_spectrum} for various CGWB sourcing mechanisms. In each figure, the left plot compares each detector Power Law Sensitivity (PLS) to the monopole $\bar{\Omega}_{\rm CGWB}(q)$, while the right plot compares the noise spectrum $N_\ell^\mathrm{GW}(q_{\rm p})$ of each detector to the angular power spectrum $C_\ell^{\rm CGWB\times CGWB}(q_{\rm p},q_{\rm p})$. 
In the plots, the PLS curves and the noise spectra have been computed for ${\rm SNR}_{\rm thr}=1$ and five years of observation. Thus, on the left plots, when the monopole of a CGWB is tangent to the PLS curve of a given detector (network), this CGWB could be detected with a signal-to-noise ratio (SNR) of one after five years of observation. On the right plots, when the angular power spectrum stays above the noise spectrum at a given multipole $\ell$, the SNR for the measurement of this multipole would be larger than one after five years. For a longer time of observation, the noise spectrum can be simply rescaled by the square root of the observing time, $N_\ell^{\rm GW} \propto 1/\sqrt{T_{\rm obs}}$. We stress again that the PLS and noise curves depend on the frequency dependence of the signal, and thus, they slightly differ between each of Figures \ref{fig:Cosmology_spectrum}, \ref{fig:PBH_spectrum}, and \ref{fig:PT_spectrum}.

\subsection{Mock likelihood}
\label{sec:likelihood}

We model our data as a vector $\vec{a} = \left\{a_{\ell m}^{\rm CMB}, a_{\ell m}^{\rm GW}\right\}$ with a Gaussian likelihood $\mathcal{L}$ given by
\begin{equation}
    \mathcal{L}(\vec{a}|\Theta) = \frac{1}{\sqrt{(2 \pi)^2 \abs{\bar{\mathbf{C}}(\Theta)}}} \exp(-\frac{1}{2} \vec{a}^\dag \left[\bar{\mathbf{C}}(\Theta)\right]^{-1} \vec{a})~.
\end{equation}
The covariance matrix $\bar{\mathbf{C}}$ contains the theoretical prediction for the auto-correlation and cross-correlation spectrum of CMB and GW multipoles, respectively $\bar{C}_\ell^{\rm CMB \times \rm CMB}$, $\bar{C}_\ell^{\rm CGWB \times \rm CGWB}$, $\bar{C}_\ell^{\rm CMB \times \rm CGWB}$. According to the discussion of the previous section, the reconstructed map and the angular power spectrum of the cosmological background are evaluated at the pivot frequency $f_\mathrm{p}$. These spectra assume particular values of the model parameters $\Theta = (\theta_1, \theta_2, ...)$. The auto-correlation spectra also include the noise spectrum $N_\ell^\mathrm{CMB}$ (resp. $N_\ell^\mathrm{GW}$) of the assumed CMB (resp. GW) instrument. The determinant of the covariance matrix is denoted $\abs{\bar{\mathbf{C}}(\Theta)}$. 

It is well-known that such a likelihood can be written in a much more compact way in terms of the covariance matrix of the data, $\hat{\mathbf{C}} = \langle \vec{a}^\dag \vec{a} \rangle$. In the case of a forecast, this matrix contains the power spectra of the fiducial model $\hat{C}_\ell^{\rm CMB \times \rm CMB}$, $\hat{C}_\ell^{\rm CGWB \times \rm CGWB}$, $\hat{C}_\ell^{\rm CMB \times \rm CGWB}$, computed with fiducial parameter values, plus the noise spectra. After some calculations, one gets the following effective chi square:
\begin{equation}
    \chi_\mathrm{eff}^2 \equiv -2 \ln{\mathcal{L}} = \sum_\ell (2 \ell + 1) \left[ \frac{\abs{D_\ell}}{\abs{\bar{C}_\ell}} + \ln \frac{\abs{\bar{C}_\ell}}{\abs{\hat{C}_\ell}} - 2 \right]~,
\end{equation}
where, for each $\ell$, we defined the theory, data and mixed determinants:
\begin{align}
    \abs{\bar{C}_\ell} =& \left(\bar{C}_\ell^{\rm CMB \times CMB} + N_\ell^\mathrm{CMB} \right)\, \left(\bar{C}_\ell^{\rm CGWB \times CGWB} + N_\ell^\mathrm{GW} \right) - \left( \bar{C}_\ell^{\rm CMB \times CGWB} \right)^2, \\
    \abs{\hat{C}_\ell} =& \left(\hat{C}_\ell^{\rm CMB \times CMB} + N_\ell^\mathrm{CMB} \right) \, \left( \hat{C}_\ell^{\rm CGWB \times CGWB} + N_\ell^\mathrm{GW} \right) - \left( \hat{C}_\ell^{\rm CMB \times CGWB} \right)^2, \\
    \abs{D_\ell} =& \left( \bar{C}_\ell^{\rm CMB \times CMB} + N_\ell^\mathrm{CMB} \right) \, \left( \hat{C}_\ell^{\rm CGWB \times CGWB} + N_\ell^\mathrm{GW} \right) \nonumber\\& + \left( \hat{C}_\ell^{\rm CMB \times CMB} + N_\ell^\mathrm{CMB} \right) \, \left( \bar{C}_\ell^{\rm CGWB \times CGWB} + N_\ell^\mathrm{GW} \right) \nonumber\\& - 2 \bar{C}_\ell^{\rm CMB \times CGWB} \, \hat{C}_\ell^{\rm CMB \times CGWB}\, .
\end{align}
In the case of a forecast including only a GW detector, without CMB cross-correlation, the effective chi square simplifies to:
\begin{equation}
    \chi_\mathrm{eff}^2 = \sum_\ell (2 \ell + 1) \left[ \frac{\hat{C}_\ell^{\rm CGWB \times CGWB}+ N_\ell^\mathrm{GW}}{\bar{C}_\ell^{\rm CGWB \times CGWB}+ N_\ell^\mathrm{GW}} + \ln \frac{\bar{C}_\ell^{\rm CGWB \times CGWB}+ N_\ell^\mathrm{GW}}{\hat{C}_\ell^{\rm CGWB \times CGWB}+ N_\ell^\mathrm{GW}} - 1 \right] ~.
\end{equation}
For the CMB noise spectrum $N_\ell^\mathrm{CMB}$, we will always assume the Planck temperature sensitivity - thus our forecasts assume a cosmic-variance limited measurement of CMB temperature anisotropies up to approximately $\ell =1800$ \cite{Planck:2018nkj}.

\subsection{CGWB produced by inflation with a blue tilt}
\label{sec:forecast_Cosmology}

\begin{figure}[t!]
    \centering
    \includegraphics[width=0.49\textwidth]{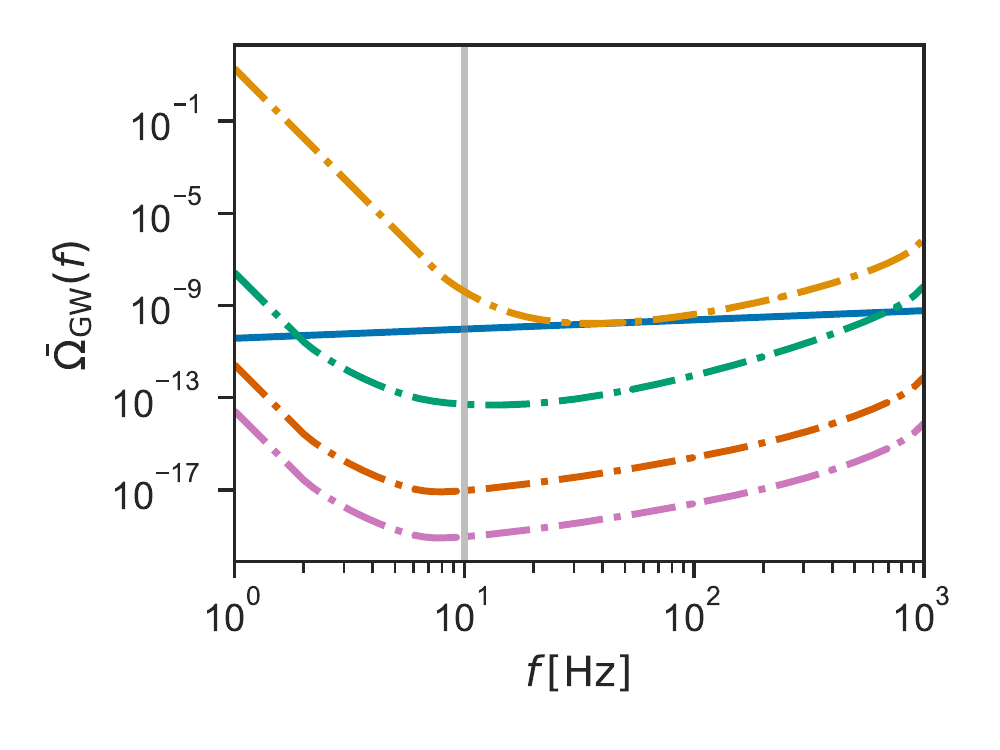}
    \includegraphics[width=0.49\textwidth]{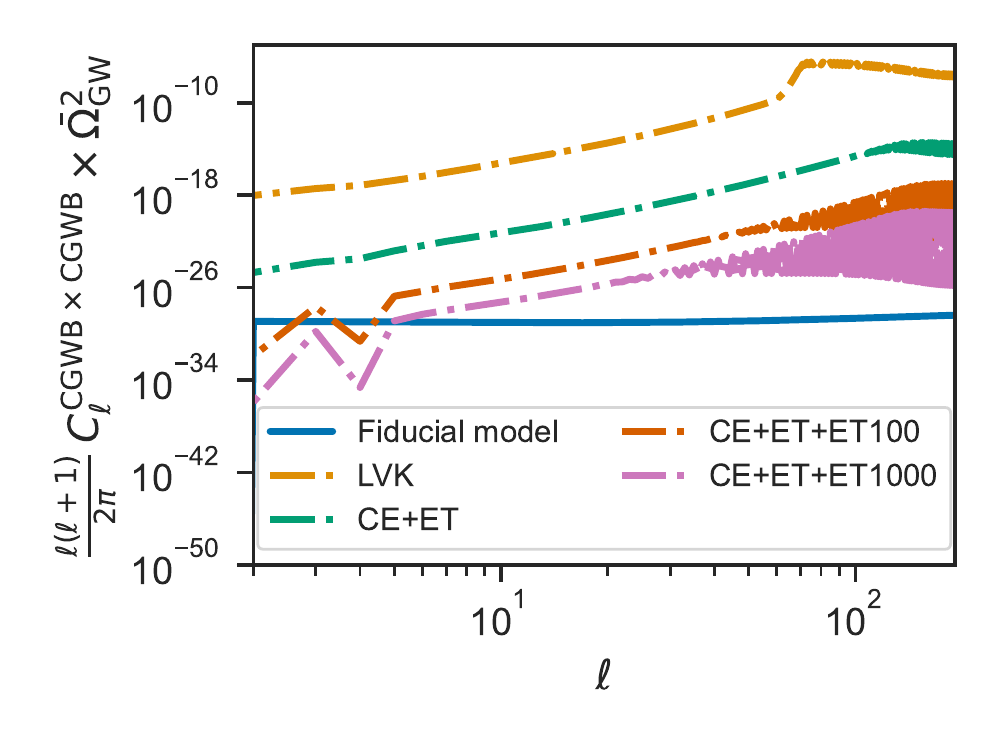}
    \vspace{-0.5cm}
    \caption{Fiducial model for the forecast assuming a
    CGWB produced by inflation with a blue tilt.
    We use a power-law CGWB with  $\OmGW = \num{9.5e-11}$ and $n_\mathrm{gwb} = 0.4$, corresponding to a CGWB generated by inflation with a blue tiled tensor spectrum:  $r = 0.025$ and $n_\mathrm{t} = 0.4$. 
    The fiducial values of all other parameters can be found in Table \ref{tab:forecast_Cosmology}.
    Left: Fiducial model (solid lines) and instrumental sensitivity (dot-dashed lines) for the CGWB background energy density (or monopole) $\OmGW(f)$. The vertical line shows the pivot frequency $f_{\rm p} = \SI{10}{Hz}$.
    Right: Power spectrum for the anisotropies in the CGWB, $\ClGW$ (solid lines) and detector noise $N_\ell^\mathrm{GW}$ (dot-dashed lines). Since the detector noise scales like $\OmGW^{-2}$, we multiplied all spectra by $\OmGW^2$.}
    \label{fig:Cosmology_spectrum}
\end{figure}

In this section, we assume that the universe can be described by the standard $\Lambda$CDM model with fiducial cosmological parameter values fixed to the Planck 2018 best fit, with two additional assumptions:
\begin{itemize}
    \item in order to get a sizeable CGWB, we assume that inflation takes place at an energy scale close to the current limit set by Planck+Bicep+Keck \cite{Galloni:2022mok}, with a tensor-to-scalar ratio $r=0.025$ and a power-law spectrum with a blue tilt $n_\mathrm{t}=0.4$. These optimistic assumptions allow us to consider a large GW background of the order of $\bar{\Omega}_\mathrm{GW}(f)\sim {\cal O}(10^{-10})$ at the frequency range best probed by LVK, CE and ET, see Fig.~\ref{fig:Cosmology_spectrum}, left plot. The assumption that the tensor spectrum follows a single power-law from CMB scales down to interferometer scales is not necessarily realistic, but it is not important either in the context of this forecast: any model with adiabatic initial conditions leading to the same $\bar{\Omega}_\mathrm{GW}(f)$ in the vicinity of $f_\mathrm{p}=10$~Hz would lead to similar results. 
    \item In standard analyses of cosmological data (not including CGWB anisotropies), the number $f_\mathrm{dec}(\eta_\mathrm{in})$ has no impact on observables. Instead, once we include CGWB anisotropy data, this parameter does play a role. In the case of GWs produced by inflation, $\eta_\mathrm{in}$ accounts for the Hubble-crossing time of the pivot frequency, see Eq.~\eqref{eq:def_eta_i}. Thus, we include $f_\mathrm{dec}(\eta_\mathrm{in})$ as an additional free cosmological parameter in our analysis. In our forecast, we choose a fiducial value $f_\mathrm{dec}(\eta_\mathrm{in})=0$, and we float this parameter during parameter inference in order to estimate the sensitivity of the mock data to this quantity.
\end{itemize}

When \MontePy is launched for the first time in the context of a forecast, it uses \texttt{GW\textunderscore CLASS} to compute the fiducial temperature, GW and cross-correlation spectra (taking fiducial parameter values from the input parameter file). When it is launched for the second time, it assumes that the fiducial spectra account for the spectra of the mock data, and it fits this data using the mock likelihood of Sec.~\ref{sec:likelihood}, while floating the free parameters of the model. In our case, the free parameters (with flat priors) are
$$
\{h, ~\omega_\mathrm{m},~ \omega_\mathrm{b}, ~\ln10^{10}A_s, ~n_\mathrm{s}, ~\tau_\mathrm{reio}, ~f_\mathrm{dec}(\eta_\mathrm{in})\},
$$
accounting for the Hubble rate, the density of non-relativistic matter and baryons, the primordial curvature spectrum amplitude and tilt, the optical depth to reionization and the number of decoupled relativistic degrees of freedom after inflation. Note that we are not including $\{r, n_\mathrm{t}\}$ to the list, because our mock data consists only in CMB temperature and CGWB maps. These observables have negligible sensitivity to the amplitude of tensor modes and to their spectral index. Of course, without a CGWB in the first place, there would be no CGWB anisotropies to measure. However, the anisotropies  themselves come dominantly from scalar fluctuations.\footnote{Equation~\eqref{eq:Cell-res} shows that the CGWB anisotropy spectrum gets a contribution from tensor modes. However, this effect is sub-dominant, thus we can safely neglect the tensor contribution in all our forecasts (in \CLASSGW, switching tensor modes on/off is an option).} In order to constrain $\{r, n_\mathrm{t}\}$, one would need to include CMB polarisation data (for large scales) and/or the measurements of the monopole $\bar{\Omega}_\mathrm{GW}(f)\sim {\cal O}(10^{-10})$ by GW interferometers (for small scales). Forecasts of the sensitivity of such data sets to $\{r, n_\mathrm{t}\}$ can easily be found in the literature. Instead, the goal of this paper is to focus on the information contained in the CGWB anisotropies and in their correlation with the CMB temperature. Thus, we may consider $\{r, n_\mathrm{t}\}$ as fixed parameters in the forecast, assuming that their value would be inferred directly from the monopole measured by GW interferometers. We also note that $\{\omega_\mathrm{b}, \tau_\mathrm{reio}\}$ are kept free in our analysis because they are determined from the CMB temperature spectrum - as however they have no direct sizeable effect on the GW spectrum.

In the left plot of Fig.~\ref{fig:Cosmology_spectrum}, our assumed monopole $\bar{\Omega}_\mathrm{GW}(f)$ is compared to the sensitivity of LVK, CE+ET, CE+ET+ET100 and CE+ET+ET1000. Intuitively, one expects that CGWB anisotropies could be observed by an instrument only if the background was detected with a very high signal-to-noise, typically of the order of $\sim 10^{5}$. According to the left plot of Fig.~\ref{fig:Cosmology_spectrum}, this would be the case only for CE+ET+ET100 and CE+ET+ET1000. This expectation is confirmed by the right plot, in which we compare the GW auto-correlation spectrum with the detector noise $N_\ell^\mathrm{GW}$ of each detector, computed with a modified version of the \texttt{schNell} code as described in Sec.~\ref{sec:GW_Reconstructed_Map}. We see that with the combination CE+ET, the signal is still a few orders of magnitude below the noise, while with the upgraded ET100 and ET1000 detectors the first few multipoles have a signal-to-noise larger than one (up to $\ell=4$ for CE+ET+ET100 and $\ell=5$ for CE+ET+ET1000). Thus, we will perform a sensitivity forecast only for CE+ET+ET1000 and for a cosmic-variance-limited experiment. All the analysis has been done assuming five years of continuous observation.

\begin{figure}[tp]
    \resizebox{\textwidth}{!}{%
	\begin{tabular}{|l|c l|c|c|c|}
		\firsthline
		Parameter & Fiducial & [Prior]& Planck + & Planck  + & Planck \\
        & & & CE+ET+ET1000 & CV($\ell_{\rm max} = 2500$) & alone \\
        \hline 
            $h$ & 0.6736 & [0.5 - 0.8] & $0.6741\pm 0.0096$ & $0.6756\pm 0.0068$ & $0.674\pm 0.010$ \\ \hline 
            $\omega_{m}$ & 0.143 & [0.1 - 0.2] & $0.1429\pm 0.0020$ & $0.1426\pm 0.0014$ & $0.1429\pm 0.0021$ \\ \hline 
            $\ln 10^{10}A_s$ & 3.044 & [1.7 - 5] & $3.044\pm 0.015$ & $3.0413\pm 0.0053$ & $3.044\pm 0.016$ \\ \hline 
            $n_s$ & 0.965 & [0.9 - 1] & $0.9654\pm 0.0051$ & $0.9662\pm 0.0028$ & $0.9653^{+0.0051}_{-0.0057}$ \\ \hline 
            $\omega_{b}$ & 0.02237 & [0.02 - 0.025] & $0.02238\pm 0.00020$ & $0.02240\pm 0.00016$ & $0.02238\pm 0.00022$ \\ \hline 
            $\tau_{\rm reio}$ & 0.0544 & [0.02 - 0.08] & $0.0547\pm 0.0071$ & $0.0536\pm 0.0012$ & $0.0545\pm 0.0073$ \\ \hline 
            $f_{\rm dec}(\eta_{\rm in})$ &    0 & [0 - 1] & $ < 0.597$ & $ < 0.159$ & - \\ \hline 
	\end{tabular}
    }
	\captionof{table}{Forecasted errors on parameters extracted from temperature anisotropy data from Planck, alone or in combination with mock GW anisotropy data from two futuristic GW detectors: CE+ET+ET1000, and an ideal cosmic-variance-limited instrument CV($\ell_{\rm max} = 2500$). We assume a CGWB produced by inflation with a blue tilt, like in Figure~\ref{fig:Cosmology_spectrum}.}
	\label{tab:forecast_Cosmology}
    \mbox{ }\\
	\includegraphics[width=1.\columnwidth]{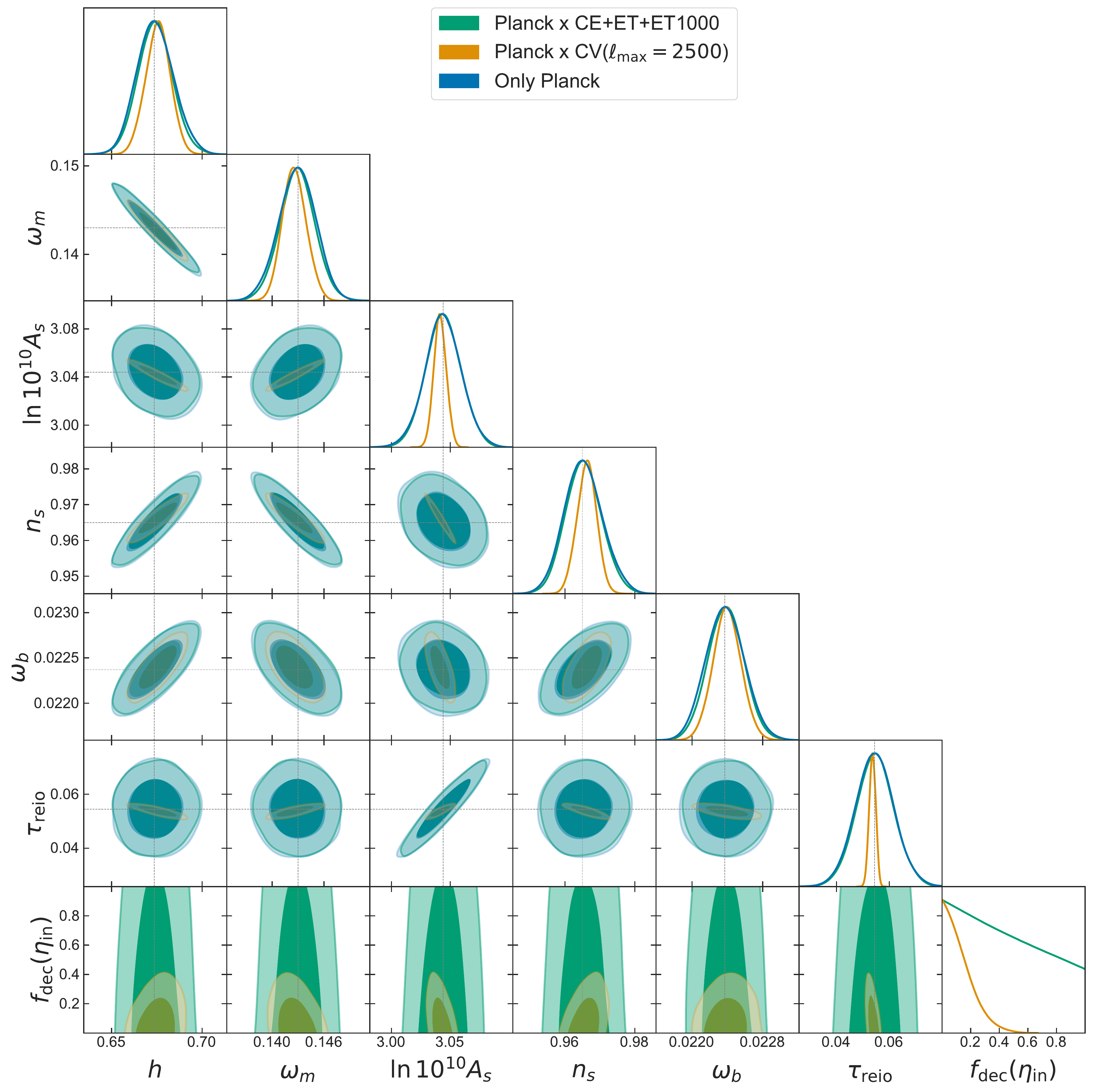}
	\captionof{figure}{For the same forecast as in Table~\ref{tab:forecast_Cosmology}, one-dimensional posteriors and two-dimensional 68\% / 95\% confidence limits on the reconstructed cosmological parameters.}
	\label{fig:forecast_Cosmology}
\end{figure}

In Table~\ref{tab:forecast_Cosmology} and Fig.~\ref{fig:forecast_Cosmology}, we show our forecasted errors and confidence contours for:
\begin{itemize}
    \item Planck (CMB temperature) alone, using a simplified version of the Planck likelihood dubbed \texttt{fake\textunderscore planck\textunderscore realistic} in \MontePy, to which any fiducial model can be passed instead of the real Planck data,
    \item Planck plus mock GW anisotropy data from CE+ET+ET1000, including the CMB$\times$GW cross-correlation,
    \item the same for Planck plus ideal cosmic-variance-limited GW anisotropy data up to $\ell=2500$, dubbed CV ($\ell_\mathrm{max}=2500$).
\end{itemize}

We first discuss the difference between the Planck alone and Planck+CV ($\ell_\mathrm{max}=2500$) forecasted errors:
\begin{itemize}
    \item The overall amplitude of the CMB (resp. GW spectrum) is fixed by $e^{-2 \tau_\mathrm{reio}}A_s$ (resp. $A_s$). Planck temperature data is sensitive to $\tau_\mathrm{reio}$ only through a small steplike feature at large angular scales, and is thus unable to measure each of $\tau_\mathrm{reio}$ or $A_\mathrm{s}$ accurately.  The combination of the two CMB temperature and GW spectra gives independent measurements of these two parameters, explaining why their errors shrink strongly in the combined case. We should remember however that this forecast does not include CMB polarisation data, which would also provide a good determination of $\tau_\mathrm{reio}$. Nevertheless, the sensitivity of the joint temperature+GW forecast, $\sigma(\tau_\mathrm{reio})=0.0014$, is about three times better than with Planck temperature+polarisation data, and twice better than in forecasts with future temperature+polarisation data from CMB-Stage-IV + LiteBIRD \cite{Brinckmann:2018owf}. The cosmic-variance-limited temperature+GW error is even competing with the sensitivity of $\tau_\mathrm{reio}$ measurements from future 21cm surveys like HERA or SKA \cite{Liu:2015txa}. This shows that an ideal CGWB detector would bring decisive information for the measurement of $\{A_\mathrm{s}, \tau_\mathrm{reio}\}$ -- and also potentially of the neutrino mass through the removal of parameter degeneracies \cite{Allison:2015qca,Brinckmann:2018owf}. 
    \item The Hubble rate and matter density parameters affect the shape of the CMB and GW spectra (tilt of the plateau due to the ISW effect, scale and shape of the acoustic oscillations in the CMB case, scale and shape of the raising of the GW spectrum for modes crossing the Hubble rate during radiation domination). Our forecast shows that the errors on $\{h,\omega_\mathrm{m}\}$ shrink roughly by a factor $\sqrt{2}$ in the combined fit, which suggests that the map of CMB temperature anisotropy and GW anisotropies contain roughly the same amount of information on $h$ and $\omega_\mathrm{m}$.
    \item The overall tilt of all spectra depends on $n_\mathrm{s}$. Thus, one may expect that the error on the tilt also shrinks by $\sqrt{2}$ in the combined case. The gain is actually a bit larger. This is due to the fact that the GW spectrum is not affected by acoustic oscillations and Silk damping, and thus, is a more direct probe of the primordial spectrum shape over a larger multipole range.
    \item The baryon density parameter $\omega_\mathrm{b}$ only affects the CMB spectrum. Still, its forecasted error decreases a tiny bit in the joint forecast. This comes for degeneracies between $\omega_\mathrm{b}$ and other parameters that are better determined by the GW data.
    \item The parameter $\fdec$ only affects the GW spectrum. With ideal GW anisotropy data, we get an error bar of about $\sigma(\fdec)=0.17$, showing that this parameter can in principle be measured.
\end{itemize}

These results quantify the amount of information that can be extracted from the data 
%in theory, \nicola{I would remove ``in theory"} 
with an extremely futuristic instrument. We now restore detector noise and compare the results obtained with Planck alone and with the combination Planck+CE+ET+ET1000. Table~\ref{tab:forecast_Cosmology} and Fig.~\ref{fig:forecast_Cosmology} show that the error bars are similar in these two cases, even for $\fdec$, which is unconstrained by Planck+CE+ET+ET1000 (see the flat posterior in Fig.~\ref{fig:forecast_Cosmology}). We have seen in Table.~\ref{tab:forecast_Cosmology} that a hypothetical ET1000 detector would measure the first few multipoles of the GW anisotropy spectrum with a signal-to-noise larger than one, and thus, would contain a non-zero amount of information on cosmological parameters like e.g. $A_\mathrm{s}$ or $n_\mathrm{s}$. However, the forecast shows that this amount is too small and that error bars are always dominated by the CMB temperature data. 

We can conclude that for a CGWB produced by inflation with a blue tilt, even a detector with a noise level $10^3$ better than the Einstein telescope would be unable to extract relevant information from the GW anisotropy map. In the next section, we will check whether similar conclusions apply to other mechanisms for the generation of the CGWB.

\subsection{CGWB produced by PBHs}
\label{sec:forecast_PBH}

\begin{figure}[tp]
    \centering
    \includegraphics[width=0.49\textwidth]{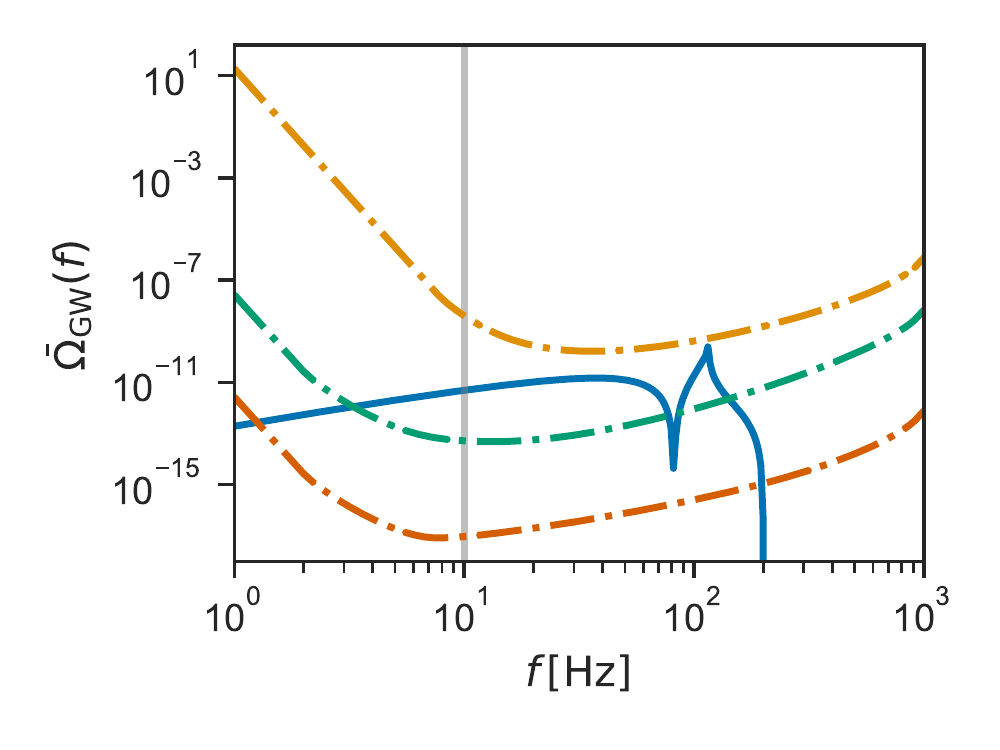}
    \includegraphics[width=0.49\textwidth]{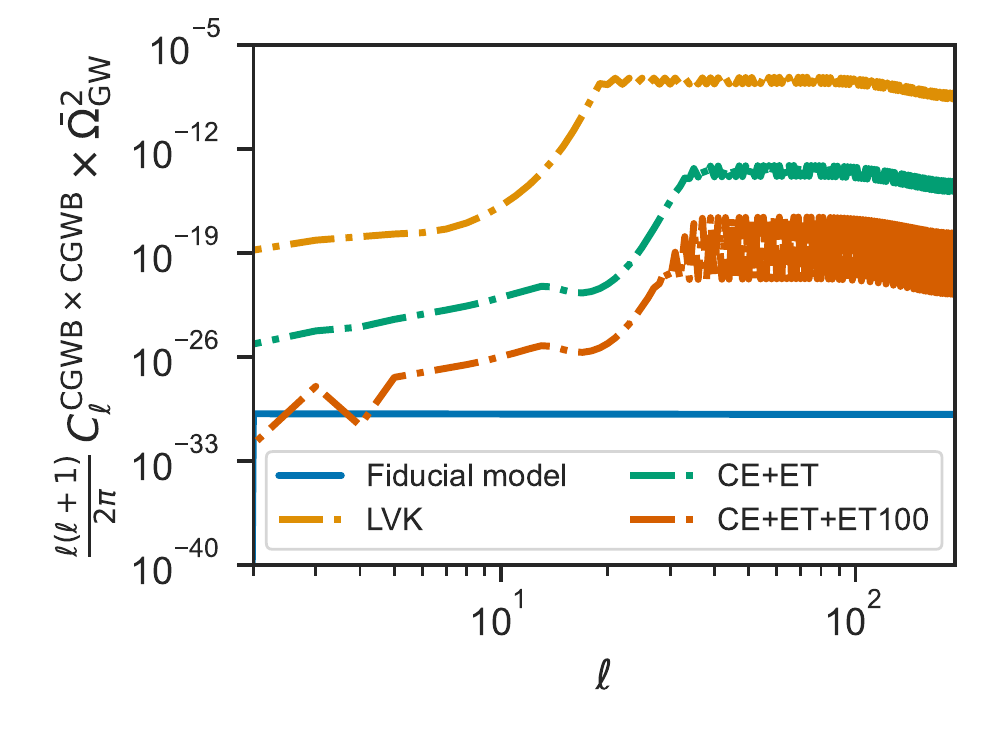}
    \vspace{-0.5cm}
    \captionof{figure}{Fiducial model for the forecast assuming a CGWB produced by PBHs, with a fiducial local non-Gaussianity parameter $f_{\rm NL}=1$.
    Left: Fiducial model (solid line) and detector sensitivity (dot-dashed lines) for the CGWB background energy density (i.e., monopole) $\OmGW(f)$. The vertical line shows the pivot frequency $f_{\rm p} = \SI{10}{Hz}$.
    Right: Angular power spectrum for the CGWB anisotropies $\ClGW$ and detector noise $N_\ell^\mathrm{GW}$. Since the detector noise scales like $\OmGW^{-2}$, we multiplied all spectra by $\OmGW$.}
    \label{fig:PBH_spectrum}
    \mbox{}\\\mbox{}\\
	\begin{tabular}{|l|c c|c|}
		\firsthline
		Parameter & Fiducial & [Prior]& CE+ET+ET100 \\ \hline 
            $f_{\rm NL}$ &    1 & [-11.1 - 9.3] & $1.17^{+0.23}_{-0.41}$, $-1.74^{+0.42}_{-0.23}$ \\ \hline 
            $f_{\rm dec}(\eta_{\rm in})$ &    0 & [0 - 1] & $---$ \\ \hline 
	\end{tabular}
	\captionof{table}{Forecasted errors on the cosmological parameters affecting  only the CGWB anisotropies, assumed to be measured by the GW detector combination CE+ET+ET100. We assume a CGWB produced by PBHs like in Figure~\ref{fig:PBH_spectrum}.}
	\label{tab:forecast_PBH}
 \mbox{}\\\mbox{}\\
	\includegraphics[width=0.45\textwidth]{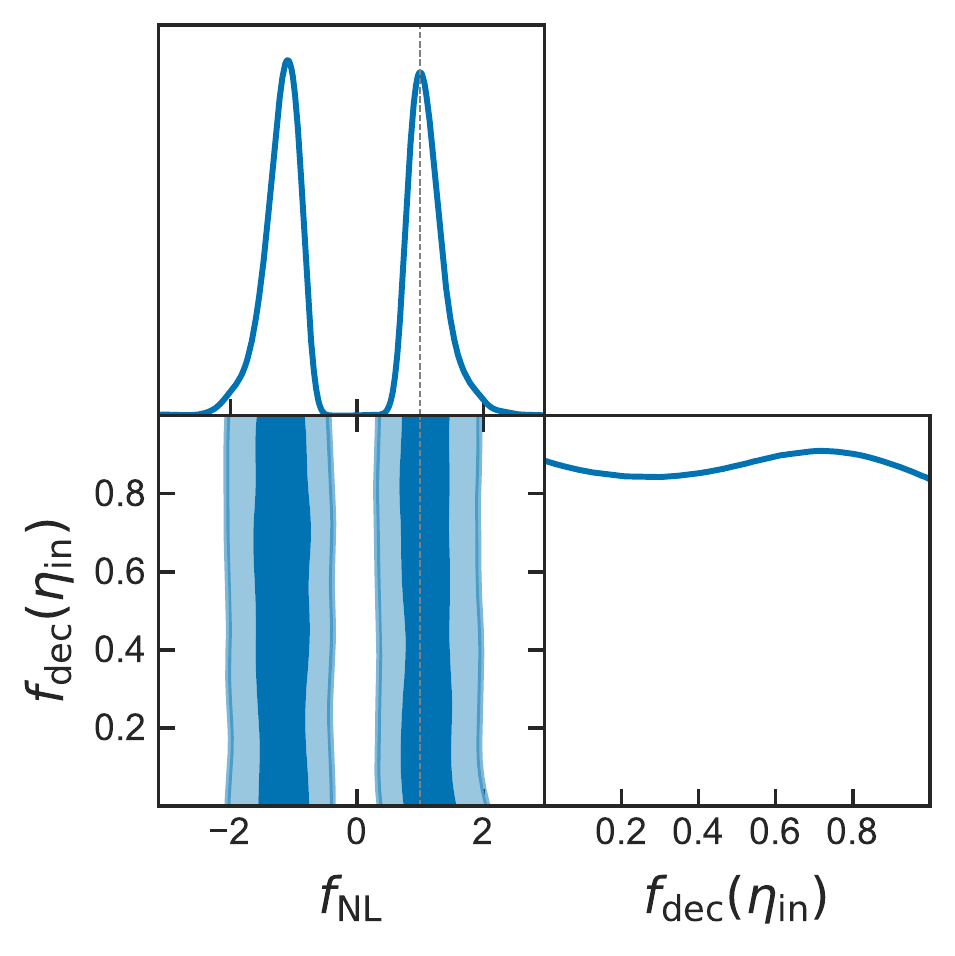}
 \vspace{-0.2cm}
	\captionof{figure}{For the same forecast as in Table~\ref{tab:forecast_PBH}, one-dimensional posteriors and two-dimensional 68\% / 95\% confidence limits on the reconstructed cosmological parameters.}
	\label{fig:forecast_PBH}
\end{figure}

We now consider the GWB generation mechanism discussed in Sec.~\ref{sec:PrimordialBlackHoles}, associated to primordial black hole formation caused by a sharp peak in the primordial spectrum with an underlying local non-Gaussian statistics. Our fiducial model has  a peak at a scale corresponding to the frequency $f_* = \SI{100}{Hz}$, with an amplitude $A_* = \num{2e-5}$ and a local non-Gaussianity parameter $f_{\rm NL} = 1$. We assume again a fiducial value $\fdec = 0$. We consider that the CGWB is measured at a frequency $f_{\rm p} = \SI{10}{Hz}$, and we estimate the tilt $n_{\rm gwb} = \pdv{\ln{\OmGW}}{\ln{f}}$ from the background spectrum shown on the left-hand side of Fig. \ref{fig:PBH_spectrum}.

Figure~\ref{fig:PBH_spectrum} shows that for such a fiducial model, CE+ET could detect the CGWB monopole, but the sensitivity of CE+ET+ET100 is needed to start probing the first multipoles of the anisotropy spectrum. For this reason, we perform our forecast for this combination only. Having seen in the last section that the measurement of the first few multipoles of the CGWB is not sufficient for bringing new information on the $\Lambda$CDM parameters, we fix them to the Planck best-fit values. We also fix $A_*$ and $k_*$, assuming that they would be inferred from data on the GW monopole. Thus, in our forecast, we only fluctuate the parameters that can be probed solely by the angular power spectra involving CGWB anisotropies, namely, $(f_{\rm NL}$ and $\fdec)$.

The forecast results are shown in Table~\ref{tab:forecast_PBH} and Fig.~\ref{fig:forecast_PBH}.
As can be seen in Fig. \ref{fig:forecast_PBH}, there is an ambiguity in the reconstructed value of $f_{\rm NL}$. This can be understood by noticing that the dominant non-adiabatic contribution to the $C_\ell$'s, $C_\ell^\mathrm{NAD}$, is sourced by $P_\Gamma^\mathrm{NAD}(k,q)$ and proportional to $f_\mathrm{NL}^2$ (see Eq.~\refeq{eq:PBH_NAD}). The degeneracy between models with opposite values of $f_\mathrm{NL}$ is lifted by the cross-correlation between the adiabatic and non-adiabatic modes, sourced by $P^\times(k,q)$ and proportional to $f_\mathrm{NL}$ (see Eq.~\refeq{eq:PBH_cross}). Since the non-adiabatic and cross-correlation terms depend differently on $\ell$, with perfect data, it would be possible to determine uniquely the value and the sign of $f_\mathrm{NL}$. However, the experiment of CE+ET+ET100 is only able to observe the first multipoles until $\ell \sim 4$ (see right plot of Fig. \ref{fig:PBH_spectrum}). Thus, the ambiguity cannot be removed, and the fit to the fiducial model (with $f_\mathrm{NL}=1$) is compatible with two values (close to $-2$ and $1$).

The results displayed in Fig.~\ref{fig:forecast_PBH} show that, in order to measure $\fdec$, it would also be necessary to measure more multipoles.

In conclusion, for the CGWB assumed here, the combination CE+ET+ET100 would be able to detect the anisotropies and use them to measure the order of magnitude of $|f_\mathrm{NL}|$, but a better sensitivity would be required to discriminate between the two reconstructed values of $f_\mathrm{NL}$ and to determine $\fdec$.

A complementary way to constrain non-Gaussianity using the SGWB$\times$CMB signal is through the astrophysical background anisotropies. A recent forecast for next generation detectors has been done in \cite{Perna:2023dgg}.

\subsection{CGBW produced by a PT with uncorrelated power-law non-adiabatic fluctuations}

\begin{figure}[tp]
    \centering
        \includegraphics[width=0.49\textwidth]{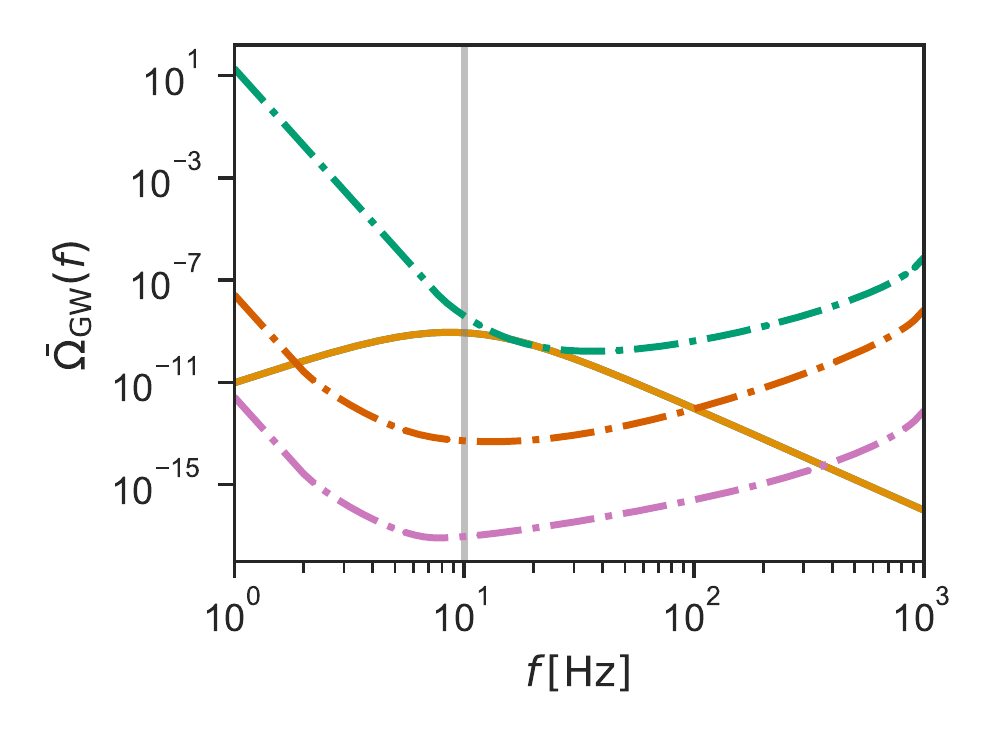}
        \includegraphics[width=0.49\textwidth]{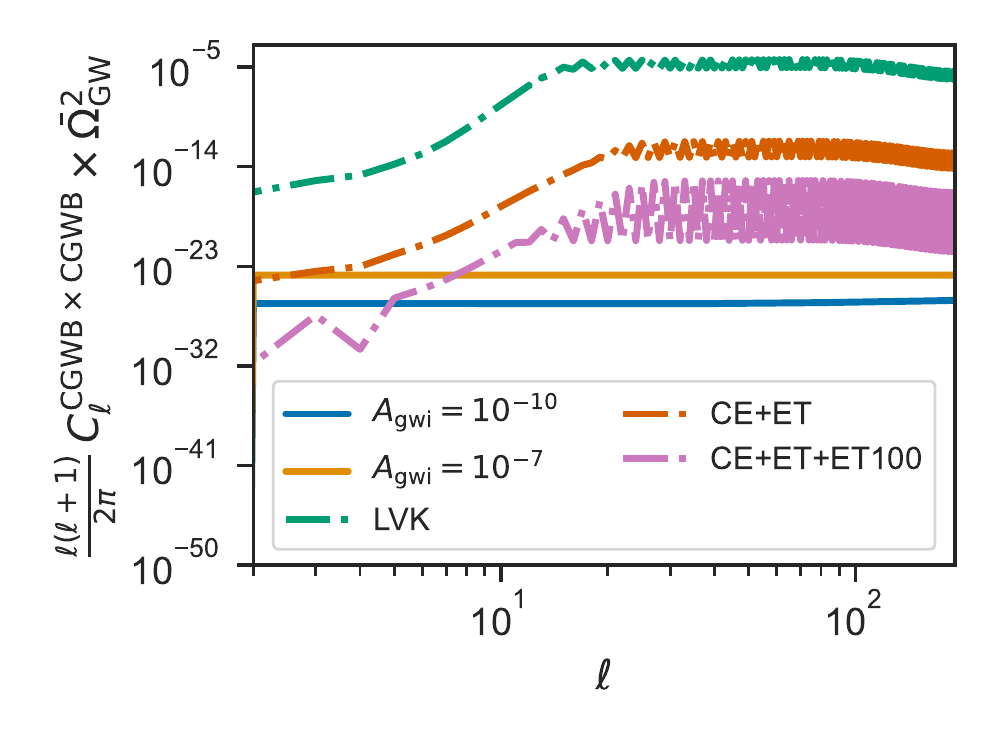}
        \vspace{-0.7cm}
    \captionof{figure}{Fiducial model for the forecast assuming a CGWB produced by sound waves during a PT. Left: fiducial model (solid lines) and experimental sensitivity (dot-dashed line) for the monopole $\OmGW(f)$. The vertical line shows the pivot frequency $f_{\rm p} = \SI{10}{Hz}$.
    Right: Power spectrum for the CGWB anisotropies $\ClGW$ and detector noise $N_\ell^\mathrm{GW}$, both multiplied by $\OmGW^{-2}$. For the anisotropy spectrum,  we assume an uncorrelated non-adiabatic contribution with amplitude $A_\mathrm{gwi}$.  The two curves correspond to the two fiducial models assumed here, featuring two different values of $A_\mathrm{gwi}$.}
    \label{fig:PT_spectrum}
    \mbox{}\\
	\begin{tabular}{|l|c l|c|}
		\firsthline
		Parameter & Fiducial & [Prior]& CE+ET+ET100 \\ \hline 
			$\ln 10^{10}A_{\rm gwi}$ &    0 & [-10 - 10] & $ < 0.927$ \\ \hline 
			$n_{\rm gwi}$ &    0 & [-2 - 2] & $ > 0.0344$ \\ \hline 
			$f_{\rm dec}(\eta_{\rm in})$ &    0 & [0 - 1] & $---$ \\ \hline 
            \hline
   			$\ln 10^{10}A_{\rm gwi}$ &  6.9 & [-10 - 30] & $6.6\pm 3.1$ \\ \hline 
			$n_{\rm gwi}$ &    0 & [-2 - 2] & $-0.06\pm 0.65$ \\ \hline 
			$f_{\rm dec}(\eta_{\rm in})$ &    0 & [0 - 1] & $---$ \\ \hline 
	\end{tabular}
	\captionof{table}{Forecasted errors on the cosmological parameters affecting only the CGWB anisotropy spectrum for the detector combination CE+ET+ET100. Like in Figure \ref{fig:PT_spectrum}, we assume a CGWB produced by sound waves during a PT, with an uncorrelated non-adiabatic contribution of amplitude $A_\mathrm{gwi}=\num{1.0e-10}$ (upper half) or $\num{1.0e-7}$ (lower half).}
	\label{tab:forecast_ISO}
\mbox{}\\
	\includegraphics[width=0.49\textwidth]{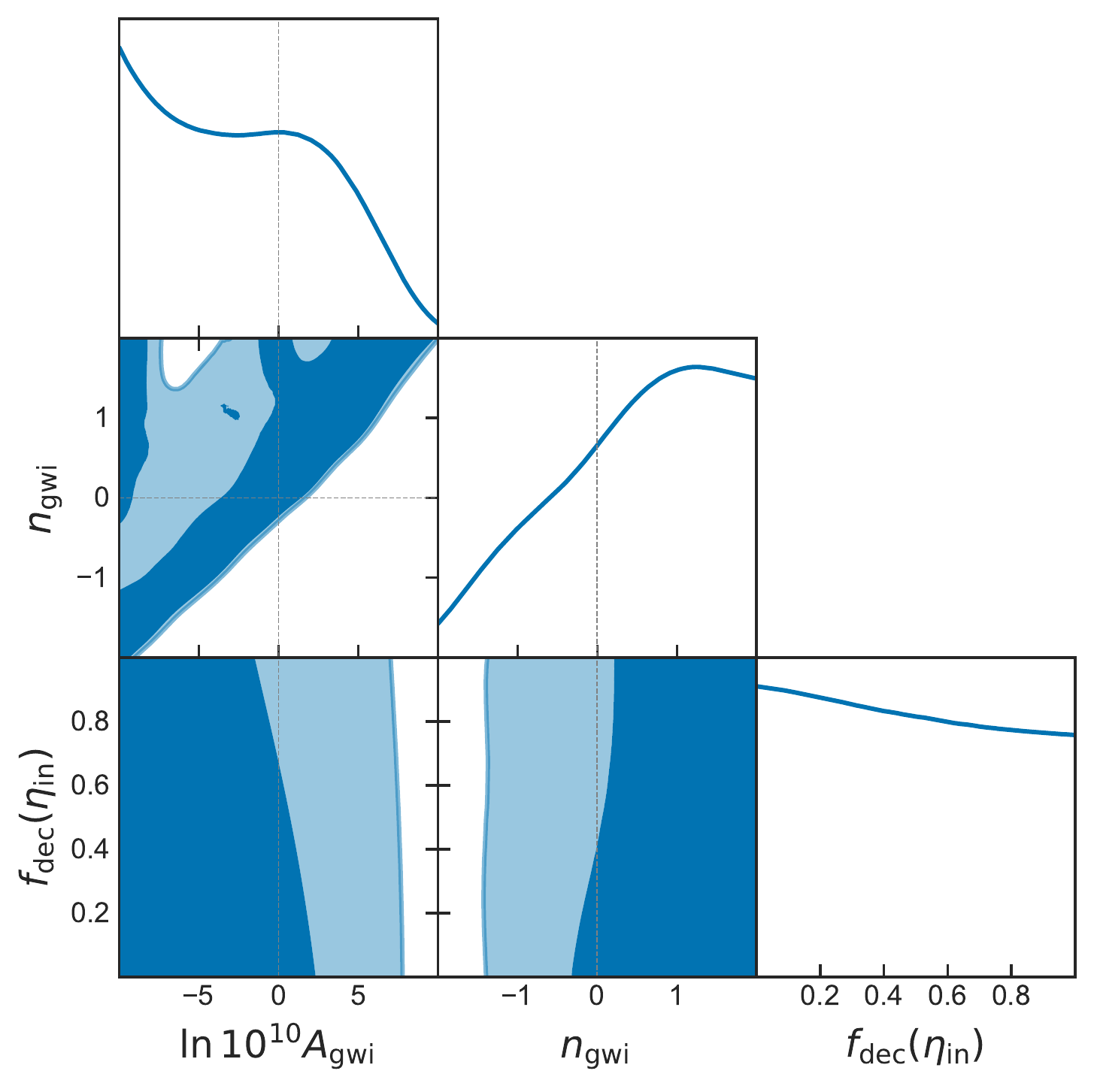}
    \includegraphics[width=0.49\textwidth]{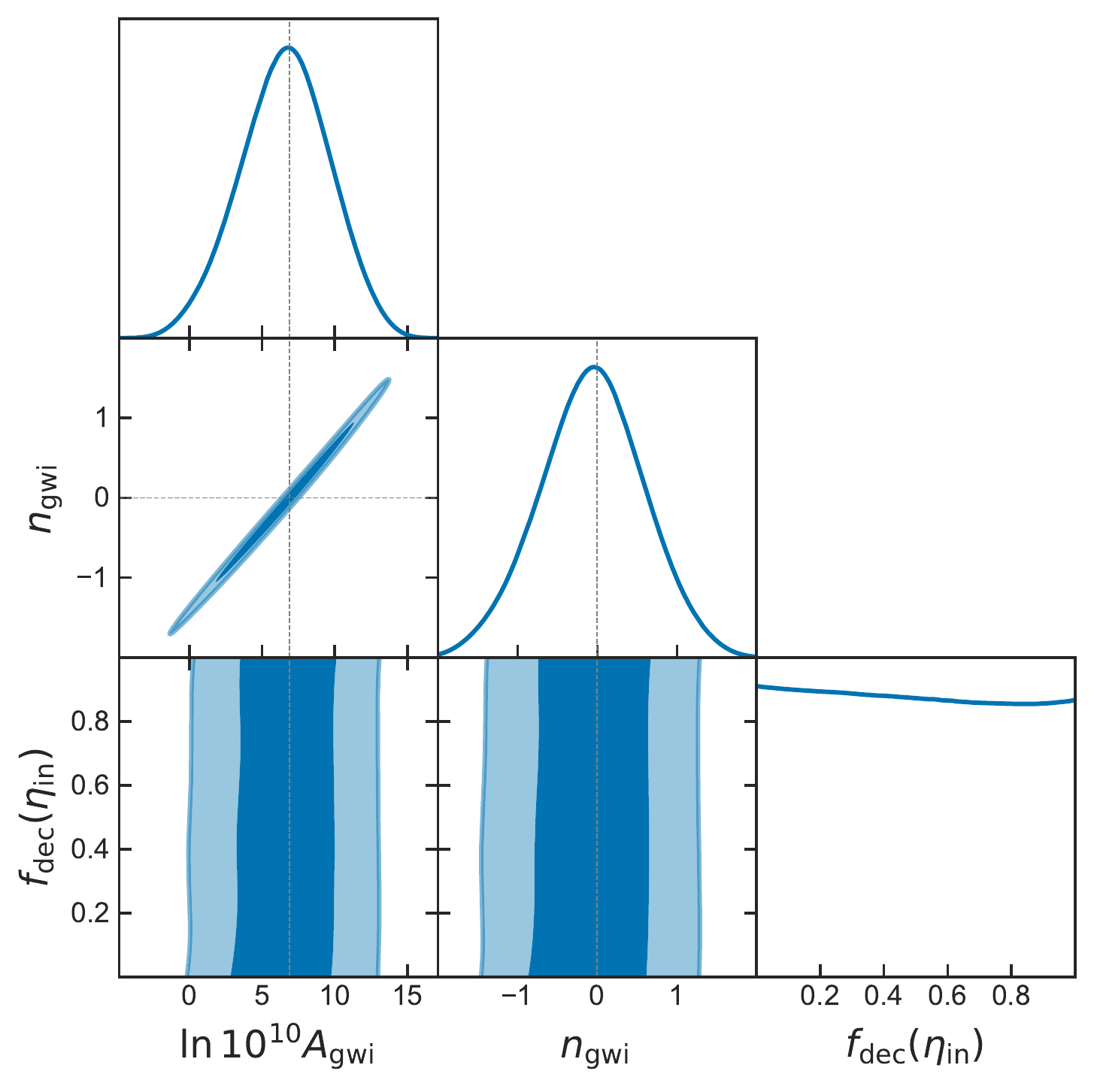}
    \vspace{-0.4cm}
	\captionof{figure}{For the same forecasts as in Table 
\ref{tab:forecast_ISO} (PT case with two fiducial values of $A_\mathrm{gwi}=\num{1.0e-10}$ (left) or $\num{1.0e-7}$ (right)), one-dimensional posteriors and two-dimensional 68\% / 95\% confidence limits on the reconstructed cosmological parameters.
}
	\label{fig:forecast_ISO}
\end{figure}

Like in section~\ref{sec:example}, we now assume a CGWB produced by sound waves during a phase transition with parameters $n_1=3$, $n_2=-4$, $\Delta=2$ and $\Omega_*=\num{1.0e-08}$. The left plot in Fig.~\ref{fig:PT_spectrum} shows that, for this model, the monopole would be marginally detectable by LVK and well detectable by CE+ET. However, at a pivot frequency of $f_p = \SI{10}{Hz}$, the CE+ET signal-to-noise ratio would be of the order of $10^4$, that is, not sufficient to detect CGWB anisotropies in presence of adiabatic initial conditions only. To get detectable CGWB anisotropies, we need to consider a more sensitive network like CE+ET+ET100 and/or assume that CGWB anisotropies are produced with intrinsic fluctuations beyond the adiabatic mode.

We thus assume like in section~\ref{sec:example} that the PT seeds non-adiabatic GW fluctuations, uncorrelated with the adiabatic one and parametrized with a power-law spectrum like in Eq.~\eqref{eq:ic_nad}. We assume that this spectrum is scale-invariant ($n_\mathrm{gwi}=0$) with a fiducial amplitude parameter given either by  $A_\mathrm{gwi} = \num{1.0e-10}$ or $A_\mathrm{gwi} = \num{1.0e-7}$. In the first case, the adiabatic and non-adiabatic modes have the same order of magnitude, while in the second case the non-adiabatic mode strongly dominates and enhances the total perturbation spectrum by three orders of magnitude. Next, we assume that the background is mapped at the pivot frequency $f_p = \SI{10}{Hz}$, and we infer $n_\mathrm{gwb}$ from the monopole frequency spectrum. Finally, we assume a fiducial value for the fraction of relativistic decoupled species $\fdec=0$. The right plot in Fig.~\ref{fig:PT_spectrum} shows that with the first fiducial model, CE+ET+ET100 could detect multipoles up to $\ell=5$, while for the second model it could probe anisotropies up to $\ell=7$.

The results of the two forecasts are presented in Table ~\ref{tab:forecast_ISO} and Figure~\ref{fig:forecast_ISO}. For the fiducial model with a small non-adiabatic contribution, the detector is only able to place an upper limit on the amplitude $A_\mathrm{gwi}$, which means that there would be no statistically significant detection of the anisotropies. For the model with an enhanced intrinsic GW fluctuations, there is a clear measurement of the amplitude of the non-adiabatic spectrum, and the detector could even put strong limits on its spectral index. We conclude that for a CGWB generated by a phase transition, the ability of future detectors to extract information from the spectrum of GW anisotropies depends mainly on the existence and amplitude of intrinsic GW fluctuations, beyond the unavoidable inhomogeneities corresponding to the adiabatic mode.

\section{Conclusions}
\label{sec:conclusions}

The detection of a CGWB is one of the most important objectives in cosmology, as it represents a unique probe of the physics operating at very high energy scales and of the early stages of evolution of the universe. The fact that GWs can be produced at very early times and that gravitons are collisionless below the Planck scale make it possible to inspect the structure of the universe since the time at which gravitons start propagating. 

Analogously to the CMB photons, primordial GWs are dominated by a homogeneous and isotropic contribution (the monopole) and exhibit tiny anisotropies of at least 1 part per $10^4$. In this paper we introduce a new version of \texttt{CLASS}, called \CLASSGW, which computes the angular power spectrum of the CGWB and its cross-correlation with the CMB. 

Although the features of the CGWB monopole depend heavily on the attributes of their generation mechanism, the CGWB anisotropy spectrum includes contributions that depend very weakly on the model. Indeed, some observable graviton fluctuations are always sourced by the modulation of the monopole by adiabatic curvature fluctuations, and then, additionally, by the propagation of the GWs through the large-scale perturbations of the universe. These contributions have essentially the same shape for all sourcing mechanisms, up to small differences triggered by the different times $\eta_\mathrm{in}$ at which the GW propagation starts, and by different values of the spectral index $n_\mathrm{gwb}$ of the monopole, which plays a role in the GW amplitude-to-energy-density conversion factor. These universal mechanisms tend to seed GW amplitude fluctuations of the same order of magnitude as curvature perturbations, that is, 1 part in $10^5$. Once converted into GW density fluctuations, the anisotropies are typically of the order of 1 part in $10^4$. 

The default contribution to the GW anisotropy spectrum includes some SW and ISW terms that are similar to those affecting the CMB temperature spectrum, but with slightly different features, since gravitons start propagating long before the last scattering of photons and the radiation-matter equality. In \CLASSGW, we carefully follow the evolution of the anisotropy source functions at early times in order to account for the primordial ISW effect that operates before recombination. This allows us to compute the observable effect of parameters that regulate the evolution of the background metric of the universe, such as the fractional energy density of relativistic and decoupled species $\fdec$ or the equation of state of the universe at $\eta_\mathrm{in}$. The measurement of CGWB anisotropies would provide a unique handle on complementary parameters which do not leave a direct signature on the monopole. We have included the effect of $\fdec$ in \CLASSGW, in order to study its detectability with future experiments.

Beyond these unavoidable contributions, there is a possibility that CGWB fluctuations get enhanced in a model-dependent way by some initial intrinsic inhomogeneities in the graviton phase-space distribution that could be imprinted by each particular generation mechanism. While the adiabatic mode is independent of frequency, the non-adiabatic mode requires a long and detailed model-dependent computation. In general, it should be parametrized as a non-factorizable function of the frequency of the GWs and of the wavelength of the large-scale perturbation considered. We include in \CLASSGW a few possible parametrizations of this model-dependent contribution, that we call the non-adiabatic mode. Our implementation takes into account the fact that the non-adiabatic mode can be arbitrarily correlated with the adiabatic mode.

In this work, we have considered and compared for the first time the angular power spectrum of the CGWB sourced by different mechanisms. In particular, we have described the anisotropies sourced by inflation, Primordial Black Holes (PBH) and Phase Transitions (PTs), with adiabatic and non-adiabatic initial conditions. These three mechanisms are considered to be the most promising candidates to produce GWs with amplitudes large enough to be detected by future ground- and space-based interferometers. In each of these three cases, we have reviewed the impact of initial conditions and of the spectral tilt $n_\mathrm{gwb}$ on the CGWB angular power spectrum, making clear how these two attributes can change by orders of magnitude the amplitude of the fluctuations in the energy density of the CGWB. This example illustrates how stochastic backgrounds are sensitive to the non-adiabatic initial conditions and to interesting cosmological parameters, such as $f_\mathrm{NL}$. In \CLASSGW, it is possible to select any of these backgrounds through various possible parametrization of the non-adiabatic initial conditions. 

In previous literature, the correlation of the CGWB at different frequencies has always been considered equal to one. In other words, the dependence of the angular power spectrum on frequency and multipole was assumed to be factorizable. This is however not the case when non-adiabatic initial conditions are taken into account, or when the tensor tilt changes with the frequency. A non-exact correlation of the spectrum has important consequences from a theoretical and experimental point of view, because it affects significantly the way in which we can perform the data analysis of future GW interferometers. Furthermore, a different frequency scaling of the GW anisotropies w.r.t. the monopole will be a crucial ingredient for performing component separation, reducing uncertainties in the reconstruction of GW maps. In this work, we show for the first time how to compute the correlation of the CGWB angular power spectra at different frequencies using \CLASSGW. \CLASSGW evaluates efficiently the angular power spectrum at multiple different frequencies for the auto- and cross-correlation.

We also studied in details the correlation between the CGWB and CMB anisotropy spectra. We provided a physical and mathematical interpretation of the angular power spectrum of the cross-correlation, in terms of projection effects from the last scattering surfaces of gravitons to that of photons. We found that the distance between these surfaces plays an important role at small scales. In particular, since photons and gravitons share their geodesics, there is a large correlation between their spectra on large angular scales (in absence of non-adiabatic perturbations). In the future, it would be possible to take advantage of this property to maximize the amount of information that can be extracted by a joint analysis of CMB and CGWB maps.

In order to establish the amount of information that can be extracted from CGWB observations by future experiments, we have presented some forecasts on the detectability of the CGWB angular power spectrum. Our forecasts show the sensitivity of future detectors to cosmological parameters through joint fits of mock CGWB and CMB data. Under the reasonable assumption that the correlation of the angular power spectrum of the CGWB at different frequencies is one, we have quantified the noise spectrum $N_\ell^\mathrm{GW}$ in CGWB maps reconstructed with various detectors. This noise depends on the amplitude of the monopole signal, on the frequency dependence of the monopole and of the anisotropy spectrum, and on the detector sensitivity and orientation. 

We have performed several forecasts, considering a CGWB produced either by inflation, PBHs or a PT, as well as various combination of detectors: First, the Einstein Telescope and Cosmic Explorer; Second, a more futuristic case in which the ET sensitivity is enhanced by a factor 100 or 1000, close to expectations for the future space-based interferometers BBO and DECIGO; and finally, a cosmic-variance-limited (CVL) detector.

Our Bayesian analysis shows that a CVL detector could put very tight constraints on the parameters of the minimal $\Lambda$CDM model, since the degeneracy between the primordial amplitude $A_s$ and the optical depth to reionization $\tau_\mathrm{reio}$ is broken by the CGWB. Besides, more realistic CGWB instruments offer a unique way to measure some parameters that are very difficult to probe otherwise, such as the relativistic decoupled density fraction $f_\mathrm{dec}(\eta_{\rm in})$, the non-Gaussianity parameter $f_\mathrm{NL}$  or the amplitude of non-adiabatic GW fluctuations generated during a PT. We leave for future works also the application of \texttt{GW\textunderscore CLASS} to other cosmological sources of interest for interferometers.

\section*{Acknowledgments}

We thank G. Galloni for useful discussions and comments on the draft. FS acknowledges support from both RWTH and MPIK computer clusters and from the IT staff.
FS acknowledges membership in the International Max Planck Research School for Astronomy and Cosmic Physics at the University of Heidelberg (IMPRS-HD).
NB, DB and SM acknowledge support from the COSMOS network (www.cosmosnet.it) through the ASI (Italian Space Agency) Grants 2016-24-H.0, 2016-24-H.1-2018 and 2020-9-HH.0. AR acknowledges financial support from the Supporting TAlent in ReSearch@University of Padova (STARS@UNIPD) for the project ``Constraining Cosmology and Astrophysics with Gravitational Waves, Cosmic Microwave Background and Large-Scale Structure cross-correlations''.

\newpage
\appendix

\section{Correspondence between notations}
\label{sec:notations}

We specify here the correspondence between the notations of previous papers on CGWB anisotropies \cite{Bartolo:2019oiq,Bartolo:2019yeu,ValbusaDallArmi:2020ifo,Ricciardone:2021kel} and those of this work, which are closer to the notations of the \CLASS code and papers (e.g. \cite{Blas:2011rf,Lesgourgues:2013bra}):\\

\begin{tabular}{|c|c|c|}
    \hline
    quantity & previous & current \\
    \hline
    scalar metric perturbations in Newtonian gauge & $\Phi, \Psi$ &
    $\psi, \phi$ \\
    comoving curvature perturbation &
    $\zeta$ &
    $\mathcal{R}$ \\
    tensor metric perturbations &
    $\chi_{ij}$ & 
    $h_{ij}$ \\
    tensor field (with polarisation $\lambda=\pm2$) &
    $\xi_\lambda$ &
    $h_\lambda$ \\ 
    tensor transfer function &
    $\chi$ &
    $T_h$ \\
    anisotropic stress of decoupled species &
    $2 {\cal N}_2$ &
    $\sigma_\mathrm{dec} $ \\
    adiabatic transfer function of the graviton p.s.d. $\Gamma$ &
    $\Xi$ &
    $T_\Gamma^\mathrm{AD}$\\
    \hline
\end{tabular}

\section{Adiabatic initial condition for the monopole}
\label{sec:adiabaticIC}

When we consider a system made of different particle species, it is useful to introduce the comoving density perturbation of each given species.
This quantity is defined in a gauge invariant way as~\cite{Kodama:1984ziu}
\begin{equation}
    \Delta_{c,j} = \delta^{\rm bol}_j -3(1+w_j)\mathcal{H}\left(v_{\parallel}+\omega_\parallel\right) \, ,
\end{equation}
where we have introduced the equation-of-state-parameter $w_j\equiv   p_j/\rho_j$ of the species $j$, the longitudinal part of the total velocity $v_\parallel$, related to individual velocities $v_j$ through
\begin{equation}
        v_\parallel \equiv  \frac{\sum_j (\rho_j+p_j)v_{\parallel \, j}}{\sum_j (\rho_j+p_j)}\, ,
\end{equation}
and the longitudinal component $\omega_\parallel$ of the metric perturbation $\delta g_{0i}$, which vanishes in the Newtonian gauge. The energy density perturbation\footnote{We write $\delta_j^{\rm bol}$ to distinguish between the ‘‘bolometric'' perturbation in the energy density, integrated over all the frequencies of the spectrum, and the perturbation in the energy density in a frequency bin, defined for the CGWB in Eq. \eqref{eq:prefactor_delta}.} $\delta_j^{\rm bol}$ in the Newtonian gauge is connected to the one evaluated in the comoving/synchronous gauge by exploiting the fact that, in the Newtonian gauge, $\omega_\parallel=0$. Thus, in this gauge,
\begin{equation}
    \delta_j^{\rm bol} =	\Delta_{c,j}+3(1+w_j)\mathcal{H}v_{\parallel}\, .
    \label{eq:delta_newtonian_to_comoving}
\end{equation}
The total comoving density is defined as
\begin{equation}
    \Delta \equiv \sum_j \frac{\rho_j}{\rho} \Delta_{c\,j} \, .
\end{equation}
It can be computed using the Poisson equation, 
\begin{equation}
    8\pi G \rho \, \Delta = -\frac{2k^2}{a^2}\phi\rightarrow  \Delta = -\frac{2 k^2}{3\mathcal{H}^2} \phi \, . 
    \label{eq:Poisson_two_gauges}
\end{equation}
For super-horizon modes, the ratio $k/\mathcal{H}$ goes to zero, therefore
\begin{equation}
    |\Delta| \ll |\phi|\, .
\end{equation}
The comoving density perturbation of a given species is related to the total comoving density perturbation through the relative entropy perturbation between pairs of species $S_{ij}$,
\begin{equation}
    \Delta_{c,i} = \frac{1+w_i}{1+w}\Delta + \sum_j \frac{\rho_j+p_j}{\rho+p}S_{ij}\, .
\end{equation}
Therefore, if we consider adiabatic perturbations such that $S_{ij}=0$, we get $\Delta_{c\,i}=\Delta$.
Then, we can infer $\delta^{\rm bol}_j$ from Eq. \eqref{eq:delta_newtonian_to_comoving} provided that we find an expression for $v_\parallel$.
The $(0,i)$ component of the Einstein equations gives
\begin{equation}
    \left(\phi^\prime+\mathcal{H}\psi\right)= -4\pi G a^2 v_{\parallel}\sum_j\left(\rho_j+p_j\right)\rightarrow v_\parallel =  -\frac{\phi^\prime+\mathcal{H}\psi}{2\mathcal{H}^2}\, ,
\end{equation}
which means that, for relativistic particles with $w_j=1/3$, the initial density perturbation is 
\begin{equation}
    \delta_j^{\rm bol} = 4\mathcal{H} v_\parallel = -\frac{2}{\mathcal{H}}\left(\phi^\prime+\mathcal{H}\psi\right) \, .
\end{equation}
In the case of the CMB, it is immediate to connect the perturbation of the energy density to the perturbation of the distribution function, because, under the assumption that the spectrum of photon is thermal, it is possible to show that the temperature fluctuations do not depend on $q$, finding $\delta^{\rm bol}_{\rm CMB}=4\Theta_0$, where $\Theta$ is the fractional temperature fluctuation of CMB photons. Since the energy spectrum of the CGWB is generally non-thermal, the relative perturbation of the energy density that appears in the Einstein equations is not the relative energy perturbation at the frequency $q$ defined in Eq. \eqref{eq:prefactor_delta}, but the bolometric density perturbation of the CGWB, which corresponds to the density perturbation at a frequency $q$ weighted over all the GW spectrum. These two density perturbations are connected by the following integral
\begin{equation}
   \delta^{\rm bol}_{\rm GW}(\eta,k) = \frac{\int \frac{d^3 q}{a^4(\eta)}q\bar{f}_{\rm GW}(q)\delta_{\rm GW}(\eta,k,q)}{\int \frac{d^3 q}{a^4(\eta)}q\bar{f}_{\rm GW}(q)} \, .
\end{equation}
The adiabatic initial condition for the bolometric energy overdensity is equivalent to an integral equation for the energy density perturbation at the frequency $q$,
\begin{equation}
    \int \frac{d^3 q}{a^4(\eta)}q\bar{f}_{\rm GW}(q)\left[\delta_{\rm GW}(\eta,k,q)+\frac{2}{\mathcal{H}}\left(\phi^\prime(\eta,k)+\mathcal{H}(\eta)\psi(\eta,k)\right)\right] = 0 \, .
\end{equation}
In order to have a reasonable solution for any distribution function $\bar{f}_{\rm GW}(q)$, the relative perturbation of the energy density at the frequency $q$ cannot depend on the frequency, therefore it is possible to use the relation $\delta_{\rm GW}^{\rm bol} = \delta_{\rm GW}$, finding
\begin{equation}
    \delta_{\rm GW}(\eta_{\rm in},k,q) = \delta_{\rm GW}(\eta_{\rm in},k)= -\frac{2}{\mathcal{H}}\left[\phi^\prime(\eta_{\rm in},k)+\mathcal{H}(\eta_{\rm in})\psi(\eta_{\rm in},k)\right] \, .
\end{equation}
A physical motivation for this result is that gravitons are decoupled and we can think of them as a collection of independent species, each corresponding to a given momentum bin, and not interacting with the other bins. The adiabatic initial condition in each redshift bin gives then
\begin{equation}
    \delta_{\rm GW}(\eta_{\rm in},k,q) = \delta_{\rm GW}(\eta_{\rm in},k,q^\prime) \rightarrow   \delta_{\rm GW}(\eta_{\rm in},k,q) = \delta_{\rm GW}(\eta_{\rm in},k) = \delta_{\rm GW}^{\rm bol}(\eta_{\rm in},k) \, .
\end{equation}
To compute the initial conditions for the Boltzmann equation, we use the relation  between $\delta_{\rm GW}$ and $\Gamma$ evaluated at initial time,
\begin{equation}
    \delta_{\rm GW}(\eta_\mathrm{in},k)  =\left(4 -  \frac{\partial \ln \, {\bar \Omega}_{\rm GW} \left( \eta_\mathrm{in} ,\, q \right)}{\partial \ln \, q}  \right)  \Gamma_0(\eta_\mathrm{in},k,q) \, , 
\end{equation}
which implies that
\begin{equation}
    \Gamma_0(\eta_\mathrm{in},k,q) = -\frac{2}{\mathcal{H}(\eta_\mathrm{in})}\frac{1}{\left( 4 -  \frac{\partial \ln \, {\bar \Omega}_{\rm GW} \left( \eta_\mathrm{in} ,\, q \right)}{\partial \ln \, q}  \right)}\left[\phi^\prime(\eta_\mathrm{in},k)+\mathcal{H}(\eta_\mathrm{in})\psi(\eta_\mathrm{in},k)\right]\, ,
\end{equation}
The source function for the initial condition contribution to the angular power spectrum is given by
\begin{equation}
    \delta_{\rm GW}^{\rm I}(\eta_0,k,q) = -\frac{2}{\mathcal{H}(\eta_\mathrm{in})}\left( 4 -  \frac{\partial \ln \, {\bar \Omega}_{\rm GW} \left( \eta_0 ,\, q \right)}{\partial \ln \, q}  \right) \frac{\phi^\prime(\eta_\mathrm{in},k)+\mathcal{H}(\eta_\mathrm{in})\psi(\eta_\mathrm{in},k)}{\left( 4 -  \frac{\partial \ln \, {\bar \Omega}_{\rm GW} \left( \eta_\mathrm{in} ,\, q \right)}{\partial \ln \, q}  \right)}\, .
\end{equation}
Similarly to the CMB case, if one neglects $\phi^\prime$ (and in this case consider a scale-invariant power tensor spectrum) the initial condition is $\delta_\gamma = 4\theta_\gamma = -2\psi$. If there are no variations in the number of relativistic and decoupled species at $\eta_\mathrm{in}$, $\phi^\prime(\eta_\mathrm{in},k)$ is negligible and we find the initial condition
\begin{equation}
    \delta_{\rm GW}^{\rm I}(\eta_\mathrm{in},k) = - 2 \psi(\eta_\mathrm{in},k)\, .  
\end{equation}
As shown in Appendix \ref{sec:Doppler}, there is no Doppler term in the case of the CGWB, because gravitons are decoupled. Therefore, the source function of the adiabatic initial condition defined in Eq. \eqref{eq:Gamma_ini_split} is 
\begin{equation}
    \label{eq:adiabtic_IC}
    T_\Gamma^\mathrm{AD}(\eta_\mathrm{in},k,q) = -\frac{2}{4 - n_{\rm gwb}(q)} T_\psi(\eta_\mathrm{in},k)= -\frac{2}{4 - n_{\rm gwb}(q)}\frac{2}{3}\left[1+\frac{4}{15}f_{\rm dec}(\eta_\mathrm{in})\right]^{-1} \, .
\end{equation}

\section{Adiabatic initial condition for the dipole}
\label{sec:Doppler}

The initial condition for the dipole/Doppler term has been derived for instance in~\cite{Dodelson:2003ft}. We compute it by combining the two Einstein equations
\begin{eqnarray}
    k^2\phi(\eta,k) +3\mathcal{H}\left[\phi^\prime(\eta,k)+\mathcal{H}(\eta)\psi(\eta,k)\right] &=& -16\pi G a^2(\eta)\sum_{i=\rm all} \rho_i(\eta)f_0^i(\eta,k) \, , \\
    k^2\phi(\eta,k) &=& -16\pi G a^2(\eta)\sum_{i = \rm all}\rho_i\left[f^i_0(\eta,k)+\frac{3\mathcal{H}(\eta)}{k}f^i_1(\eta,k)\right] \nonumber \, ,
\end{eqnarray}
where the $f^i_j$ represent the $j$-multipole of the distribution function of the species $i$.
Under adiabatic initial conditions, we have $f^i = f^{i^\prime}$. Therefore, it is easy to see that 
\begin{equation}
\begin{split}
    \Gamma_1(\eta_\mathrm{in},k,q) =& \frac{1}{4-n_{\rm gwb}}f_1(\eta_\mathrm{in},k,q) =  -\frac{1}{4 -n_{\rm gwb}}\frac{k}{6\mathcal{H}}\left(\frac{\phi^\prime}{\mathcal{H}}+\psi\right) \rightarrow |\Gamma_1(\eta_\mathrm{in},k,q)| \ll |\psi(\eta_\mathrm{in},k,q)| \, ,
\end{split}
\end{equation}
which means that the Doppler term in the initial conditions does not contribute to the angular power spectrum of the CMB anisotropies.

\section{Inflationary CGWB}
\label{sec:app_inflation}

Following \cite{Caprini:2018mtu} we introduce the characteristic GW strain $h_c(\eta, \vec{k})$ describing the variance of tensor perturbations in the metric,
\begin{equation}
    \langle h_r(\eta, \vec{k}) h^*_p(\eta, \vec{q}) \rangle = \frac{8 \pi^5}{k^3} h_c^2(\eta, k) \delta_{r p} \delta^{(3)}(\vec{k} - \vec{q})~.\label{eq:def_characteristic_strain}
\end{equation}
We compare this with the definition of the tensor power spectrum $P_T(k)$ for the inflationary GWs/tensor modes $h^{\rm inf}$ \cite[Eq. (188)]{Caprini:2018mtu}:
\begin{align}
    \langle h_{ij}^{\rm inf}(\vec{k}) \left( h^{\rm inf}_{ij}(\vec{q}) \right)^* \rangle =& \frac{2 \pi^2}{k^3} P_T(k) (2 \pi)^3 \delta^{(3)}(\vec{k} - \vec{q}) \label{eq:def_tensor_spectrum}\\
    =& \langle h_{ij}(\eta_\mathrm{in}, \vec{k}) h^*_{ij}(\eta_\mathrm{in}, \vec{q}) \rangle \\
    =& \sum_{r \, p} \langle h_r(\eta_\mathrm{in}, \vec{k}) h^*_p(\eta_\mathrm{in}, \vec{q}) \rangle \, 2 \delta_{r p} \\
    =& 4 \times \frac{8 \pi^5}{k^3} h_c^2(\eta_\mathrm{in}, k) \delta_{r p} \delta^{(3)}(\vec{k} - \vec{q})~.
\end{align}
Thus, we can identify
\begin{equation}
    \label{eq:rel_Pt_hc}
    P_T(k) = 2 h_c^2(\eta_\mathrm{in}, k) \equiv 2 \left( h^{\rm inf}(k) \right)^2~.
\end{equation}
For canonical models of inflation, one obtains \cite[Eq. (191)]{Caprini:2018mtu}
\begin{equation}
    P_T(k) \simeq \frac{2}{\pi^2} \frac{H_k^2}{m_{\rm Pl}^2} \, ,
    \quad {\rm for} \quad k = a_k H_k \, .
\end{equation}
Next, we want to connect the GW strain with the spectral GW energy density $\bar{\Omega}_{\rm GW}(k)$. The energy density $\rho_{\rm GW}$ is given by \cite[Eq. (83)]{Caprini:2018mtu}
\begin{equation}
    \rho_{\rm GW} = \frac{\langle \dot{h}_{ij} \dot{h}^{ij} \rangle}{32 \pi G} = \frac{\langle h'_{ij} h^{\prime \, ij} \rangle}{32 \pi G a^2}  =  \int\frac{d k}{k} \frac{d \rho_{\rm GW}}{d \log k} \, .
\end{equation}
We identify
\begin{equation}
    \bar{\Omega}_{\rm GW}(k) = \frac{1}{\rho_{c \, 0}} \frac{d \rho_{\rm GW}}{d \log k}
\end{equation}
with the critical energy density today, $\rho_{c \, 0} = \frac{3 H_0^2}{8 \pi G}$. Similar to the characteristic strain, we introduce the amplitude $h_c^{\prime \, 2}(\eta, \vec{k})$ \cite[Eq. (84)]{Caprini:2018mtu}:
\begin{equation}
    \langle h'_r(\eta, \vec{k}) h^{\prime \, *}_p(\eta, \vec{q}) \rangle = \frac{8 \pi^5}{k^3} \; \delta^{(3)}(\vec{k} - \vec{q}) \; \delta_{r p} \; h^{\prime \, 2}_c(\eta, k)~.
\end{equation}
Thus, we can identify
\begin{equation}
    \label{eq:rel_OmGW_hc}
    \bar{\Omega}_{\rm GW}(k) = \frac{1}{\rho_{c \, 0}} \frac{d \rho_{\rm GW}}{d \log k}
    = \frac{1}{\rho_{c \, 0}} \frac{h^{\prime \, 2}_c(\eta_0, k)}{16 \pi G \, a^2}
    = \frac{h^{\prime \, 2}_c(\eta_0, k)}{6 H_0^2 a^2} \, .
\end{equation}
Next, using the free evolution equation of GWs in an FLRW universe \cite[Eq. (64)]{Caprini:2018mtu},
\begin{equation}
    \label{eq:evolution_GWs}
    H''_r(\eta, \vec{k}) + \left( k^2 - \frac{a''}{a} \right) H_r(\eta, \vec{k}) = 0~,
\end{equation}
with $H_r(\eta, \vec{k}) = a(\eta) h_r(\eta, \vec{k})$, we find that the full solution for $a \propto \eta^n$ is given by \cite[Eq. (66)]{Caprini:2018mtu}
\begin{equation}
    \label{eq:general_solution_GW_propagation}
    h_r(\eta, \vec{k}) = \frac{A_r(\vec{k})}{a(\eta)} \, \eta \, j_{n-1}(k \eta) + \frac{B_r(\vec{k})}{a(\eta)} \, \eta \, y_{n-1}(k \eta) \, .
\end{equation}
The evolution on super-Hubble scales can be assumed to be constant, so $h^{\rm inf}(k)$ provides the correct initial condition for the evolution of tensor perturbations after Hubble crossing $k = a_k H_k$.
Here, we only consider modes that re-enter the horizon during radiation domination (RD), $k > k_{\rm eq} = \frac{1}{\eta_{\rm eq}}$, corresponding to $f > \SI{1.3e-17}{Hz}$.

During RD ($n=1$), the evolution is given by \cite[Eq. (201)]{Caprini:2018mtu}
\begin{equation}
    h_r^{\rm RD} (\eta, \vec{k}) = h^{\rm inf}(k) j_0(k \eta) \, .
\end{equation}
This equation can be matched to the one describing the propagation of GWs during matter domination (MD) ($n=2$), which we write as \cite[Eq. (204)]{Caprini:2018mtu}
\begin{equation}
    h_r^{\rm MD} (\eta, \vec{k}) = \tilde{A}(\vec{k}) \frac{j_1(k \eta)}{k \eta} + \tilde{B}(\vec{k}) \frac{y_1(k \eta)}{k \eta} \, ,
\end{equation}
using a more convenient parametrization of the prefactors compared to Eq. \eqref{eq:general_solution_GW_propagation}. Matching the solutions at matter-radiation equality, $\eta_{\rm eq}$, gives \cite[Eqs. (207), (208)]{Caprini:2018mtu}
\begin{align}
    \tilde{A}(k) &= h^{\rm inf}(k) \left[ \frac{3}{2} - \frac{\cos(2x)}{2} + \frac{\sin(2x)}{x} \right] \, , \\
    \tilde{B}(k) &= h^{\rm inf}(k) \left[ \frac{1}{x} - x - \frac{\sin(2x)}{2} + \frac{\cos(2x)}{x} \right] \, ,
\end{align}
with $x = k \, \eta_{\rm eq}$.
Now, the full solution today (assuming $\eta_0$ is still during MD) can simply be written as \cite[Eq. (209)]{Caprini:2018mtu}
\begin{equation}
    h_r(\eta_0, \vec{k}) = T(\eta_0, k) h^{\rm inf}(k) \, ,
\end{equation}
where $T(\eta_0, k)$ is the transfer function \cite[Eq. (210)]{Caprini:2018mtu}
\begin{equation}
    T(\eta_0, k) = \frac{\tilde{A}(k)}{h^{\rm inf}(k)} \frac{j_1(k \eta_0)}{k \eta_0} + \frac{\tilde{B}(k)}{h^{\rm inf}(k)} \frac{y_1(k \eta_0)}{k \eta_0} \, .
\end{equation}
Using this notation, the GW energy density Eq. \eqref{eq:rel_OmGW_hc} can be written as \cite[Eq. (212)]{Caprini:2018mtu}
\begin{equation}
    \bar{\Omega}_{\rm GW}(k) = \frac{1}{6 H_0^2 a_0^2} \left[ T'(\eta_0, k) \right]^2 \left( h^{\rm inf}(k) \right)^2 = \frac{1}{12 H_0^2 a_0^2} \left[ T'(\eta_0, k) \right]^2 P_T(k) \, .
\end{equation}
For sub-horizon scales $k \ll 1 / \eta_0 $ and averaging the oscillations $\sin \approx \cos \approx \frac{1}{\sqrt{2}}$, we only have to consider the leading terms in leading terms in $k \eta_0 \ll 1$ and $k \eta_{\rm eq} \ll 1$:
\begin{equation}
    T(\eta_0, k) \underset{k \eta_0 \ll 1}{\longrightarrow} \frac{k \eta_{\rm eq}}{(k \eta_0)^2} \sin(k \eta_0) \approx \frac{1}{\sqrt{2}} \frac{\eta_{\rm eq}}{k \eta_0^2} \, ,
\end{equation}
and \cite[Eq. (214)]{Caprini:2018mtu}
\begin{equation}
    T'(\eta_0, k) \underset{k \eta_0 \ll 1}{\longrightarrow} \frac{k \eta_{\rm eq}}{(k \eta_0)^2} k \cos(k \eta_0) \approx \frac{1}{\sqrt{2}} \frac{\eta_{\rm eq}}{\eta_0^2} = k \, T(\eta_0, k) \, .
\end{equation}
In summary, we have
\begin{equation}
    \bar{\Omega}_{\rm GW}(k) = \frac{1}{12 H_0^2 a_0^2} \frac{\eta_{\rm eq}^2}{2 \eta_0^4} P_T(k) \, .
\end{equation}

\section{Correlation between two last scattering spheres}
\label{sec:Il}

The cross-correlation angular power spectrum $C_\ell^{\mathrm{CMB}\times\mathrm{CGWB}}$ quantifies the correlation between two last scattering spheres with slightly different radii, which depends on the calculation of the integral
\begin{equation}
    I_\ell=\int \frac{dk}{k} \, j_\ell[k(\eta_0-\eta_1)] \, j_\ell[k(\eta_0-\eta_2)]\, .
    \label{eq:int_phase_shift}
\end{equation}
In our case, $\eta_0$ is the conformal age of the universe, while $\eta_1=n_*$ (resp. $\eta_2=\eta_\mathrm{in}$) represents the conformal time at CMB (resp. GW) decoupling. Thus, we are interested in the case $\eta_0 \gg \eta_1 \gg \eta_2$, with $(\eta_1-\eta_2) \ll \eta_0$. 

The general solution of this integral can be found by using
\begin{equation}
    \begin{split}
        I_\ell =& \int\frac{dk}{k}j_\ell[k(\eta_0-\eta_1)]j_\ell[k(\eta_0-\eta_2)]=\int \frac{dx}{x}j_\ell\Bigl(x \frac{\eta_0-\eta_1}{\eta_0-\eta_2}\Bigr)j_\ell(x) \\
        =& \frac{\pi}{2}\sqrt{\frac{\eta_0-\eta_2}{\eta_0-\eta_1}}\int \frac{dx}{x^2}J_{\ell+1/2}\Bigl(x \frac{\eta_0-\eta_1}{\eta_0-\eta_2}\Bigr)J_{\ell+1/2}(x)\, ,
    \end{split}
\end{equation}
valid for $\eta_2 < \eta_1 < \eta_0$ and defining $x=k(\eta_0-\eta_2)$. We have also used
\begin{equation}
    j_\ell(x)=\sqrt{\frac{\pi}{2x}}J_{\ell+1/2}(x)\, . 
\end{equation}
The solution is found in \citep{Gradshteyn:1702455}:
\begin{equation}
    \begin{split}
        \int \frac{dx}{x^2}J_{\ell+1/2}\Bigl(x \frac{\eta_0-\eta_1}{\eta_0-\eta_2}\Bigr)J_{\ell+1/2}(x) = 
        \frac{1}{4}&\Bigl(\frac{\eta_0-\eta_1}{\eta_0-\eta_2}\Bigr)^{\ell+1/2}\frac{\Gamma(\ell)}{\Gamma(3/2)\Gamma(\ell+3/2)}\times \\
        &\times F\Bigl[\ell,-\frac{1}{2};\ell+3/2;\Bigl(\frac{\eta_0-\eta_1}{\eta_0-\eta_2}\Bigr)^2\Bigr]\, ,
    \end{split}
\end{equation}
where the last term is given by the hypergeometric series (which converges in our case)
\begin{equation}
    F(\alpha,\beta;\gamma;z)=1+\frac{\alpha\cdot \beta}{\gamma\cdot 1}z+\frac{\alpha(\alpha+1)\beta(\beta+1)}{\gamma(\gamma+1)\cdot 1\cdot 2}z^2+\frac{\alpha(\alpha+1)(\alpha+2)\beta(\beta+1)(\beta+2)}{\gamma(\gamma+1)(\gamma+2)\cdot 1 \cdot 2 \cdot 3}z^3+\dots
\end{equation}
By using $\Gamma(\ell+3/2)=(\ell+1/2)\Gamma(\ell+1/2)$, we can write $I_\ell$ as
\begin{equation}
    I_\ell = \frac{\pi}{8}\Bigl(\frac{\eta_0-\eta_1}{\eta_0-\eta_2}\Bigr)^{\ell}\frac{1}{\ell+1/2} \frac{\Gamma(\ell)}{\Gamma(3/2)\Gamma(\ell+1/2)}F\Bigl[\ell,-\frac{1}{2};\ell+3/2;\Bigl(\frac{\eta_0-\eta_1}{\eta_0-\eta_2}\Bigr)^2\Bigr] \,.
    % \label{eq:phase_shift_solution}
\end{equation}
Thus, $I_\ell$ can be expressed as a convergent series in $y=\left( \frac{\eta_0-\eta_1}{\eta_0-\eta_2} \right)$, where $y$ is slightly smaller than one. The leading contribution is given by the lowest-order term in the series,
\begin{equation}
    I_\ell \simeq \frac{\pi}{8}\Bigl(\frac{\eta_0-\eta_1}{\eta_0-\eta_2}\Bigr)^{\ell}\frac{1}{\ell+1/2} \frac{\Gamma(\ell)}{\Gamma(3/2)\Gamma(\ell+1/2)}\,,
    \label{eq:phase_shift_solution}
\end{equation}
which, in the large-$\ell$ limit, scales like
\begin{equation}
    I_\ell \propto \Bigl(\frac{\eta_0-\eta_1}{\eta_0-\eta_2}\Bigl)^{\ell} \frac{1}{\sqrt{\ell}(\ell+1/2)} \, .
    \label{eq:phase_shift_dependence}
\end{equation}

\section{The \CLASSGW code}
\label{sec:CLASS_GW}

In this appendix, we present \CLASSGW in more detail. \CLASSGW will be made publicly available on \verb|GitHub| as a new branch.\footnote{The branch will be available under the name \texttt{GW\textunderscore CLASS} at \url{https://github.com/lesgourg/class_public/} once this paper has been accepted. The \CLASSGW branch will be up-to-date with the current base version of \CLASS, that is, with the {\tt master} branch.}
Like in the base version of \CLASS, \CLASSGW can be called using either the command line with a parameter file (\verb|.param|) or through the \verb|Python| interface \verb|classy|. 
The basic input and output specific to \CLASSGW are described in App.~\ref{sec:appendix_CLASSGW_output}, including the flags activating the CGWB computation and the format of the output files. App.~\ref{sec:appendix_CLASSGW_sources} gives a complete overview of all CGWB sources (that is, generation mechanisms) covered by \CLASSGW and of their parametrization. The most important modifications to the base \CLASS code are summarized in App.~\ref{sec:appendix_CLASSGW_details}, and a few precision tests establishing the validity and accuracy of our implementation are presented in App.~\ref{sec:appendix_CLASSGW_tests}.

\subsection{\CLASSGW input and output}
\label{sec:appendix_CLASSGW_output}

{\bf Input.} A comprehensive overview of the most important parameters specific to \CLASSGW is given in Table~\ref{tab:CLASSGW_general_parameter}.

To activate the computation of the CGWB at a given (list of) frequencies \verb|f_gwb|, one should include either \verb|gwCl| or \verb|OmGW| in the field \verb|output|, e.g. with \verb|output = gwCl, OmGW|. The key \verb|gwCl| activates the computation of the angular power spectrum $\ClGW$ using the line of sight integral, as described in Sec.~\ref{sec:BoltzmannApproach}, while \verb|OmGW| activates the computation of the spectral energy density $\OmGW(f)$ for a given source model (specified through the field \verb|gwb_source_type|). Note that \CLASSGW is designed primarily for the computation of $\ClGW$. Still, the computation of $\OmGW(f)$ is included, because the calculation of the $\ClGW$'s at a given frequency $f_i$ depend on it through the parameters $n_{\rm gwb}(f_i)$ and $\OmGW(f_i)$. As a consequence, the key \verb|gwCl| only works in combination with \verb|OmGW|.

\begin{table}[htp]
    \centering
    \begin{tabular}{|l|l|}
        \hline
        Parameter & \CLASSGW \\
        \hline \hline
        Compute $\ClGW$ and $\OmGW(f)$  & \verb|output = gwCl, OmGW| \\
        Compute cross correlations with CMB & \verb|output = tCl, gwCl, OmGW| \\
                \hline
        Pivot frequency $f_{\rm pivot}$ $[\si{Hz}]$ & \verb|f_pivot = 1| \\
        Minimum frequency $f_{\rm min}$ $[\si{Hz}]$ & \verb|f_min = 1e-3| \\
        Maximum frequency $f_{\rm max}$ $[\si{Hz}]$ & \verb|f_max = 1e2| \\
        \hline
        Observed CGWB frequencies $f$ $[\si{Hz}]$ & \verb|f_gwb = f_pivot| \\ 
            \quad can be a list of \verb|N| frequencies & \quad \verb|f_gwb = 1.0, 2.5, ..., 1e2|\\
        \hline
        Convert to GWB energy density & \verb|convert_gwb_to_energydensity = yes/no| \\
        \hline
        Contributions to $\ClGW$ & \verb|gravitational_wave_contributions =| \\
        \quad SW & \quad\verb|tsw,| \\
        \quad primordial ISW ($z \gg {\cal O}(10^5)$) & \quad\verb|pisw,|  \\
        \quad early ISW ($z \sim 10^3$) & \quad\verb|eisw,| \\
        \quad late ISW ($z \sim {\cal O}(1)$) & \quad\verb|lisw,| \\
        \quad adiabatic initial contribution & \quad\verb|ad,| \\
        \quad non-adiabatic initial contribution & \quad\verb|ini| \\
        Split redshift for early/late ISW & \verb|early_late_isw_redshift = 50| \\
        \hline
        $f_\mathrm{dec}(\eta_\mathrm{in})$ & \verb|f_dec_ini = 0| \\
        \hline
        CGWB source type & \verb|gwb_source_type =| \\
        \quad Power-law & \quad\verb|analytic_gwb| \\
        \quad Inflationary CGWB & \quad\verb|inflationary_gwb| \\
        \quad External CGWB & \quad\verb|external_gwb| \\
        \quad CGWB from PBHs & \quad\verb|PBH_gwb| \\
        \quad CGWB from PT & \quad\verb|PT_gwb| \\
        \hline
    \end{tabular}
    \caption{Most important parameters specific to \CLASSGW (and implemented default value when relevant).}
    \label{tab:CLASSGW_general_parameter}
\end{table}

\CLASSGW automatically computes the cross-correlations angular spectrum $C_\ell^{\rm CMB \times CGWB}(f_i)$ at all frequencies $f_i$ specified by \verb|f_gwb| if at least \verb|tCl| and \verb|gwCl| are simultaneously included in the \verb|output| field, e.g. with \verb|output = tCl, gwCl, OmGW|.

The minimum and maximum frequency $f_{\rm min}$ and $f_{\rm max}$ give the range for $\OmGW(f)$, while the pivot frequency $f_{\rm pivot}$ determines the pivot scale for $\OmGW(f)$. The CGWB anisotropies and their cross-correlations are evaluated at all frequencies $\vb{f_{\rm gwb}} = (f_1, ..., f_i, ..., f_j, ..., f_N)$. The parameter \verb|convert_gwb_to_energydensity| controls whether the power spectrum $\ClGW$ is the anisotropy spectrum of the density $\delta_{\rm GW}$ (when set to \verb|yes|, default) or of the graviton phase-space-distribution perturbation $\Gamma$ (when set to \verb|no|).\footnote{\CLASSGW computes the line-of-sight integral in terms of $\Gamma$, since this is more convenient (see Sec. \ref{sec:BoltzmannApproach}). The conversion to the energy density contrast $\delta_{\rm GW}$ is achieved by multiplying the transfer functions with $(4 - n_{\rm gwb})$.}

The different terms contributing to the power spectrum $\ClGW$ can be switched on/off thanks to the field \verb|gravitational_wave_contributions| in the same way as, in the base version of \CLASS, different contributions to $C_\ell^{\rm CMB}$ can be switched on/off with the field \verb|temperature_contributions|.

The user can adjust the GW initial conditions with several input fields. The initial conformal time defined in \eqref{eq:def_eta_i} is assumed to be $\eta_\mathrm{ini}=0$~Mpc. The number of relativistic decoupled degrees of freedom $\fdec$ at that time is specified through \verb|f_dec_ini|. In the text, we argued that a natural choice for this parameter is \verb|f_dec_ini=0| (default). However, to deactivate the effect of early decoupled species, one can set \verb|f_dec_ini = -1|. In this case, \CLASSGW will assume no decoupled species beyond those still present at recombination, that is, ordinary neutrinos plus possible extra relativistic relics passed in input for the CMB anisotropy calculations. Then, \CLASSGW will automatically infer $\fdec$ from the usual ``effective number of neutrinos'' $N_\mathrm{eff}$, base on input values for the density of ultra-relativistic relics \verb|ur| and light relics \verb|ncdm|. The type of CGWB source (or generation mechanism) can be set with the parameter \verb|gwb_source_type|. Each type leads to a different parametrization for $\OmGW(f)$. More details on the different sources are given in App.~\ref{sec:appendix_CLASSGW_sources} and in Chap.~\ref{sec:CGWB_sources}. On top of choosing a given source type, the user can activate intrinsic/non-adiabatic initial perturbations for the CGWB, labelled \verb|gwi| in \CLASSGW. The corresponding parameters are summarized in Tab.~\ref{tab:CLASSGW_gwi_mode}.
\begin{table}[t!]
    \centering
    \begin{tabular}{|l|l|}
        \hline
        Parameter & \CLASSGW \\
        \hline \hline
        initial condition: & \verb|ic =| \\
        \quad adiabatic mode & \quad\verb|ad,| \\
        \quad GW intrinsic/non-adiabatic initial mode & \quad\verb|gwi,|  \\
        \quad other isocurvature modes (baryon isocurvature, ...) & \quad\verb|bi, ...| \\
        \hline
        amplitude $A_{\rm gwi}$ & \verb|A_gwi = 0.| \\
        \quad or $\ln 10^{10} A_{\rm gwi}$ & \quad \verb|ln10^{10}A_gwi| \\
        spectral index $n_{\rm gwi}$ & \verb|n_gwi = 0.| \\
        running $\alpha_{\rm gwi}$ & \verb|alpha_gwi = 0.| \\
        \hline
        cross-correlations fraction with adiabatic mode & \verb|c_ad_gwi = 0.| \\
        its spectral index  & \verb|n_ad_gwi = 0.| \\
        its running & \verb|alpha_ad_gwi = 0.| \\
        \hline
    \end{tabular}
    \caption{Parameters describing a possible intrinsic/non-adiabatic initial mode for GWs in \CLASSGW, see Sec. \ref{sec:source_power_law} (the right column also shows the implemented default values). The syntax is very similar to that for ordinary isocurvature modes in the base version of \CLASS.}
    \label{tab:CLASSGW_gwi_mode}
\end{table}

\mbox{}\\

\noindent {\bf Output.} As usual in \CLASS, the output consists either in files (when running on a terminal) or python dictionaries (when calling the functions of the \verb|classy| python module).
\CLASSGW can output the angular power spectra $C_\ell$ (possibly including CMB, CGWB and cross-correlation spectra at all frequencies \verb|f_gwb|), the energy density spectra $\OmGW(f)$, and finally, if non-adiabatic modes are enabled, the primordial spectrum $P^\mathrm{NAD}(k)$. Table \ref{tab:CLASSGW_output} gives an overview of the labels corresponding to column names in the output files or to keys in output python dictionaries.
\begin{table}[htp]
    \centering
    \begin{tabular}{|l|l|l|}
        \hline
        Output & Terminal & \verb|classy| \\
        \hline \hline
        Power spectrum $C_\ell$ & \verb|cl.dat| & \verb|classy.raw_cl(l_max)| \\
        \quad multipole $\ell$ & \quad\verb|l| & \quad\verb|ell| \\
        \quad $C_\ell^{\rm CMB \times CMB}$ & \quad\verb|TT| & \quad\verb|tt| \\
        \quad $\ClGW(f_i, f_j)$ & \quad\verb|G[i]-G[j]| & \quad\verb|gg| (only $i=j=1$) \\
        \quad $C_\ell^{\rm CMB \times CGWB}(f_i)$ & \quad\verb|T-G[i]| & \quad\verb|tg| (only $i=1$)\\
        Alternative \verb|classy| function & -- & \verb|classy.cgwb_cl(l_max)| \\
        \quad multipole $\ell$ & & \quad\verb|ell| \\
        \quad frequencies $f_i$ [Hz] & & \quad\verb|f_gwb [Hz]| \\
        \quad $\ClGW(f_i, f_j)$ & & \quad\verb|gg| (3D-array \verb|[i][j][ell]|) \\
        \quad $C_\ell^{\rm CMB \times CGWB}(f_i)$ & & \quad\verb|tg| (2D-array \verb|[i][ell]|) \\
        \hline
        Spectral energy density $\OmGW(f)$ & \verb|OmegaGW.dat| & \verb|classy.get_omega_gw()| \\
        \quad frequency $f$ $[\si{Hz}]$ & \quad\verb|f [Hz]| & \quad\verb|f [Hz]| \\
        \quad $\OmGW(f)$ & \quad\verb|Omega_GW(f)| & \quad\verb|Omega_GW(f)| \\
        Alternative \verb|classy| function &  & \verb|classy.Omega_GW(f)| \\
        \hline
        Spectral tilt $n_{\rm gwb}(f)$ & -- & \verb|classy.n_gwb(f)| \\
        \hline
        Primordial spectrum $P(k)$ & \verb|primordial_Pk.dat| & \verb|classy.get_primordial()| \\
        \quad wave number $k$ $[\si{1/Mpc}]$ & \quad\verb|k [1/Mpc]| & \quad\verb|k [1/Mpc]| \\
        \quad scalar spectrum $P_\mathcal{R}(k)$ & \quad\verb|P_scalar(k)| & \quad\verb|P_scalar(k)| \\
        \quad CGWB isocurvature $P_{NAD}(k)$ & \quad\verb|P_gwi(k)| & \quad\verb|P_gwi(k)| \\
        \quad scalar $\times$ CGWB cross correlation & \quad\verb|ad x gwi| & \quad\verb|ad x gwi| \\
        \quad tensor spectrum $P_T(k) = 4 P_{h_\lambda}(k)$ & \quad\verb|P_tensor(k)| & \quad\verb|P_tensor(k)| \\
        \hline
    \end{tabular}
    \caption{Output files with their column names and output dictionaries with their associated keys.}
    \label{tab:CLASSGW_output}
\end{table}

\subsection{CGWB sources in \CLASSGW}
\label{sec:appendix_CLASSGW_sources}

Table~\ref{tab:CLASSGW_input_sources} summarizes the different schemes implemented in \CLASSGW to describe the various sources (or generation mechanisms) for the GW background. These schemes are labelled by the flag \verb|gwb_source_type|.

The simplest possible parametrization of the CGWB energy density $\OmGW(f)$ is a power-law. In \CLASSGW this case is activated with the flag \verb|gwb_source_type = analytic_gwb|. It assumes a parametrization of $\OmGW(f)$ with a syntax very similar to that of the scalar power spectrum in the case  \verb|analytic_Pk|:
\begin{equation}
    \OmGW(f) = \bar{\Omega}_* \left(\frac{f}{f_\mathrm{pivot}}\right)^{n_\mathrm{gwb} + \frac{1}{2} \alpha_{\rm gwb} \log\frac{f}{f_\mathrm{pivot}}} \, .
\end{equation}
\begin{table}[htp]
    \centering
    \begin{tabular}{|l|l|l|}
        \hline
        \texttt{gwb\textunderscore source\textunderscore type} & Parameter & \CLASSGW \\
        \hline \hline
        \texttt{analytic\textunderscore gwb} & Amplitude $\bar{\Omega}_*$ & \verb|Omega_gwb = 1e-10| \\
        & quad or $\ln(10^{10}\bar{\Omega}_*)$ & \quad\verb|ln10^{10}Omega_gwb| \\
        & tilt $n_{\rm gwb}$ & \verb|n_gwb = 0.| \\
        & running $\alpha_{\rm gwb}$ & \verb|alpha_gwb = 0.| \\
        \hline
        \texttt{inflationary\textunderscore gwb} & activate tensor modes & \verb|modes = t (, s)| \\
        & tensor to scalar ratio $r$ & \verb|r = 1.| \\
        & $n_\mathrm{t} ( = n_{\rm gwb} )$ & \verb|n_t = scc| \\
        & $\alpha_t$ & \verb|alpha_t = scc| \\
        \hline
        \texttt{external\textunderscore gwb} & command for $P(k)$ & \verb|command| \\
        & command for $\OmGW(f)$ & \verb|command_gwb| \\
        & argument1 & \verb|custom1| \\
        & ... & ... \\
        & argument10 & \verb|custom10| \\
        \hline
        \texttt{PBH\textunderscore gwb} & $A_*$ & \verb|A_star = 2e-5| \\
        & \quad or $\ln(10^{10}A_*)$ & \quad\verb|ln10^{10}A_star| \\
        & $f_*$ in $[\si{Hz}]$ & \verb|f_star = 10.| \\
        & $f_{\rm NL}$ & \verb|f_NL = 0.| \\
        \hline
        \texttt{PT\textunderscore gwb} & $\bar{\Omega}_*$ & \verb|OmegaPT_star = 1e-7| \\
        & $f_*$ $[\si{Hz}]$ & \verb|fPT_star = 1.| \\
        & $n_1$ & \verb|nPT_1 = 3.| \\
        & $n_2$ & \verb|nPT_2 = -4.| \\
        & $\Delta$ & \verb|deltaPT = 2.| \\
        \hline
    \end{tabular}
    \caption{Input parameters for the different CGWB sources.}
    \label{tab:CLASSGW_input_sources}
\end{table}
Details on this parametrization are provided in table~\ref{tab:CLASSGW_input_sources}. 

Second, one can assume that $\OmGW(f)$ is given by the primordial tensor spectrum generated by inflation, whose parametrization is already implemented in the base version of \CLASS. In that case, one should switch to \verb|gwb_source_type = inflationary_gwb|, check that tensor modes are activated with the usual \CLASS syntax (\verb|modes = t, ...|) and adjust the tensor-to-scalar ratio, tensor spectral index and tensor running, as described in Sec.~\ref{sec:sources_inflation} and table~\ref{tab:CLASSGW_input_sources}. As indicated, \CLASSGW uses $n_{\rm gwb} = n_{\rm t}$ in this parametrization.

Third, with the option \verb|gwb_source_type =  external_gw|, it is also possible to pass to \CLASSGW a spectrum $\OmGW(f)$ that has been tabulated in a file or that is being computed on-the-fly by an external code. This works in a similar way as for the primordial scalar/tensor spectra when the flag \verb|external_Pk| is activated in the base version of \CLASS (see Tab.~\ref{tab:CLASSGW_input_sources}).\footnote{\CLASSGW also allows to pass a tabulated non-adiabatic mode \texttt{gwi} together with the primordial power spectrum.}

Fourth, one can activate the sourcing of GW anisotropies by PBHs with non-Gaussian statistics described in Sec.~\ref{sec:PrimordialBlackHoles} with \verb|gwb_source_type  = PBH_gwb|. In this case, the parameters listed in Tab.~\ref{tab:CLASSGW_input_sources} match the definitions given in Sec.~\ref{sec:PrimordialBlackHoles}.

Fifth, the user may switch to GW anisotropies originating from a phase transition (see Sec.~\ref{sec:PhaseTransition}) with \verb|gwb_source_type  = PT_gwb|. The parameters shown in Tab.~\ref{tab:CLASSGW_input_sources} for this case are explained in Sec.~\ref{sec:PhaseTransition}.

\subsection{Numerical implementation}
\label{sec:appendix_CLASSGW_details}

The main modification in \CLASSGW consists in the implementation of new source functions for the scalar and tensor perturbations of the CGWB. For the ISW contribution to scalar and tensor perturbations, these source functions are defined in the \verb|perturbation| module:
\begin{lstlisting}[language=C,numbers=none,caption={Implementation of the source function \texttt{gwb0}, \texttt{gwb1} and \texttt{gwb2} in \texttt{perturbation.c:perturbations\_sources()}}]
    //Newtonian gauge
    // Integrand to be multiplied by j_ell for the ISW
    gwb0 = switch_gwb_isw * 2.*phi_prime;

    // Integrand to be multiplied by j^(1)_ell for the ISW
    gwb1 = switch_gwb_isw * k* (psi-phi);

    //Synchronous gauge
    // Integrand to be multiplied by j_ell for the ISW
    gwb0 = switch_gwb_isw * 2. * (eta_prime
            - a_prime_over_a_prime * alpha
            - a_prime_over_a * alpha_prime);

    // Integrand to be multiplied by j^(1)_ell for the ISW
    gwb1 = switch_gwb_isw * k * (alpha_prime
            + 2. * a_prime_over_a * alpha
            - eta);

    //Tensor modes
    gwb2 = -switch_gwb_isw * gwdot;
\end{lstlisting}
These source functions are integrated along the line of sight within the \verb|transfer| module. Since the other source functions are only meant to be evaluated at $\eta_\mathrm{in}$ or $\eta_\mathrm{min}$,\footnote{In this paper, we always denote conformal time as $\eta$, but inside the code it is denoted as \tt{tau}.} and do not need to be integrated over the line of sight, they are more conveniently defined directly within the \verb|transfer| module, in the lines:
\begin{lstlisting}[language=C,numbers=none,caption={Implementation of the source function \texttt{gwb\_sw0}, \texttt{gwb\_sw1}, \texttt{gwb\_ad} and \texttt{gwb\_ini} in \texttt{transfer.c:transfer\_sources()}}]
    // Combination of PISW and EISW: terms proportional to Phi
    gwb_sw0 = (-switch_gwb_pisw) 
            * phi[tau_ini_gwb] * ((1. + 2./5. * f_dec_ini) / (1. + 4./15. * f_dec_ini)) / ((1. + 2./5. * f_dec_late) / (1. + 4./15. * f_dec_late)) // this product gives phi(eta_ini) in the notations of the paper
            + (switch_gwb_pisw + switch_gwb_eisw)
            * phi[tau_ini_gwb]; // this is directly phi(eta_min) in the notations of the paper

    // Combination of SW, PISW, EISW: terms proportional to Psi
    gwb_sw1 = (switch_gwb_sw - switch_gwb_pisw)
            * psi[tau_ini_gwb] * (1. + 4./15. * f_dec_late) / (1. + 4./15. * f_dec_ini) // this product gives psi(eta_ini)
            + (switch_gwb_pisw - switch_gwb_eisw)
            * psi[tau_ini_gwb]; // this is directly psi(eta_min)

    // Adiabatic inital condition
    gwb_ad = switch_gwb_ad
            * psi[tau_ini_gwb] * (1. + 4./15. * f_dec_late) / (1. + 4./15. * f_dec_ini); // this product gives psi(eta_ini)

    gwb_ini = switch_gwb_ini * 1.;
\end{lstlisting}
The source functions \verb|gwb0, gwb_sw0, gwb_sw1, gwb_ad, gwb_ini| are then multiplied with the spherical Bessel function $j_\ell(k (\eta_\mathrm{in} - \eta_0))$ (associated to the flag \verb|SCALAR_TEMPERATURE_0| in the code), while \verb|gwb1| uses the derivative of this function $j_\ell^{(1)}(k (\eta_\mathrm{in} - \eta_0))$ (with flag \verb|SCALAR_TEMPERATURE_1|) and \verb|gwb2| uses the Bessel function usually associated to tensor modes (with flag \verb|TENSOR_TEMPERATURE_2|). Afterwards, in the \verb|harmonic| module, all terms are gathered to form the complete transfer function, possibly including non-adiabatic and/or tensor modes. 

In order to enable the computation of the source functions at early time $\eta \sim  \eta_\mathrm{min}$, the time sampling of the source functions needs to be extended to an earlier time in \CLASSGW than in the base code. This time is set by a precision parameter \verb|start_sources_at_tau_gwb| with default value 0.1~Mpc. Starting from this time, the source functions are sampled at every interval $\Delta \eta = \frac{a}{a'} \times \epsilon$, where $\epsilon$ is set like in the base version of \CLASS by the precision parameter \verb|perturbations_sampling_stepsize|.

The energy density $\OmGW$ is implemented in the \verb|primordial| module as a new variable \verb|lnOmGW|, together with the logarithmic frequency \verb|lnf|. The non-adiabatic initial spectrum $P^{\rm NAD}(k)$ is associated to a new type of initial condition with index \verb|index_ic_gwi|, defined on equal footing with usual isocurvature modes (baryon isocurvature, CDM isocurvature, etc.)

\subsection{Precision tests}
\label{sec:appendix_CLASSGW_tests}

We have performed an extended series of precision tests. First, we have compared in a systematic way the output of two versions of \CLASSGW developed independently by two of us. Second, we have checked the convergence of our results against precision parameter settings. We only highlight here the result of our most important  tests. 
 
\begin{figure}
    \centering
    \includegraphics[width=0.49\textwidth]{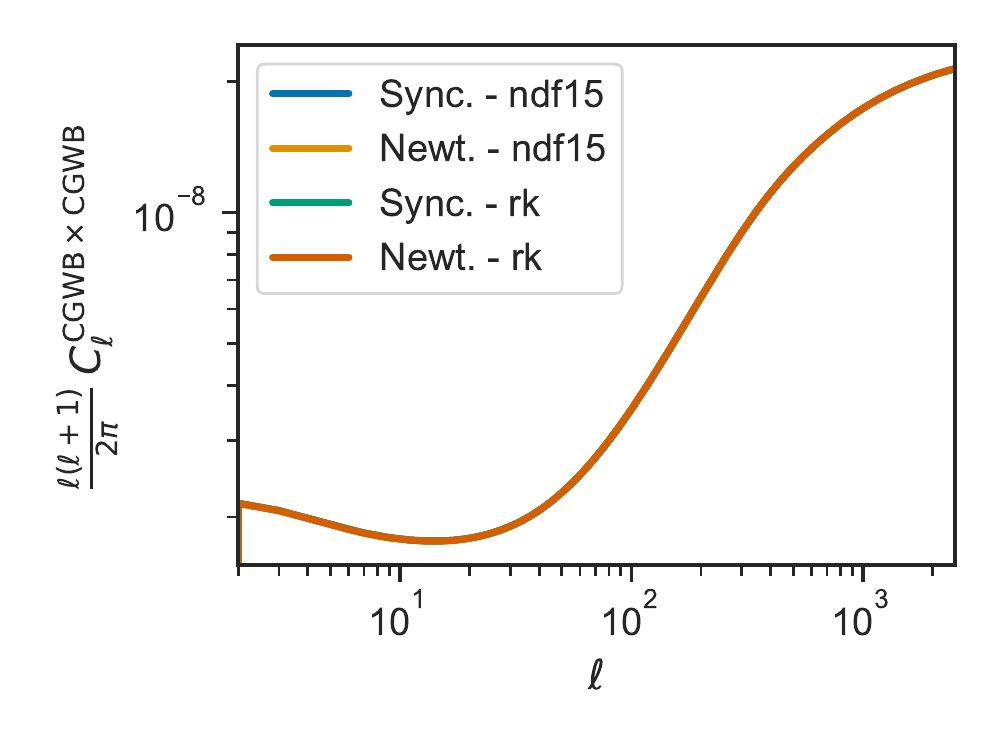}
    \includegraphics[width=0.49\textwidth]{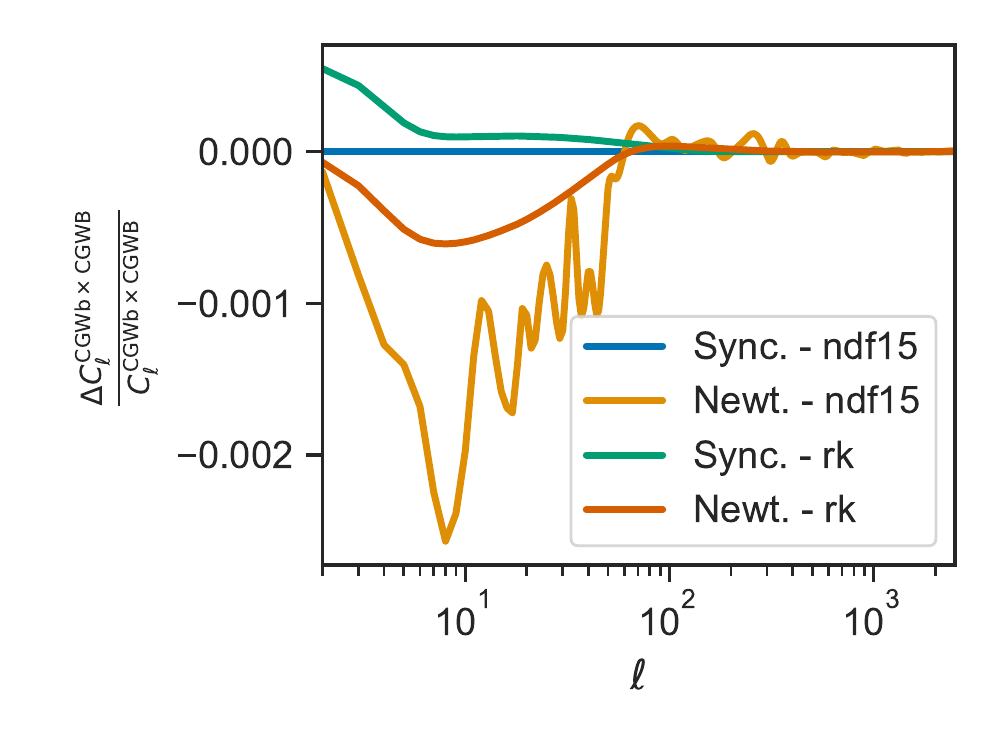}
    \caption{Comparison of \CLASSGW in two different gauges and using two different ODE evolvers. Left: power spectrum $\ClGW$, right: relative difference.}
    \label{fig:precision_gauges}
\end{figure}

Figure \ref{fig:precision_gauges} shows the comparison of the CGWB anisotropy spectrum in the Newtonian (Newt.) and synchronous (Sync.) gauge, using the two Ordinary Differential Equation (ODE) evolvers of \CLASS, \verb|ndf15| and \verb|rk|. We find that percent precision is achieved, although the Newtonian gauge is slightly less precise than the synchronous one.

It is worth highlighting the important role played by the precision parameter that controls the $k_{\rm max}$ value in \CLASSGW, called \verb|k_max_tau0_over_l_max_gwb|. This parameter fixes the ratio $R \equiv [k_{\rm max} \tau_0 / \ell_\mathrm{max}]$, such that $k_{\rm max}$ is inferred from the requested multipole $\ell_\mathrm{max}$ and from the conformal age of the universe $\tau_0$ computed for each cosmology (denoted as $\eta_0$ in this paper). This dimensionless parameter is fixed by default to $R=2.4$.

\begin{figure}
    \centering
    \includegraphics[width=0.49\textwidth]{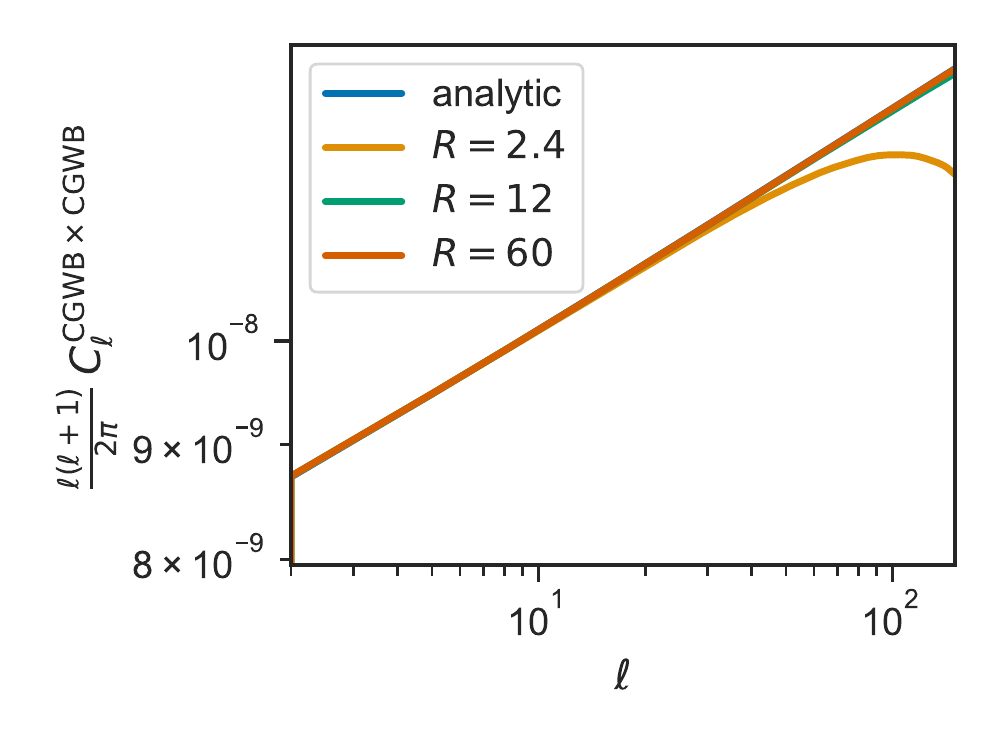}
    \includegraphics[width=0.49\textwidth]{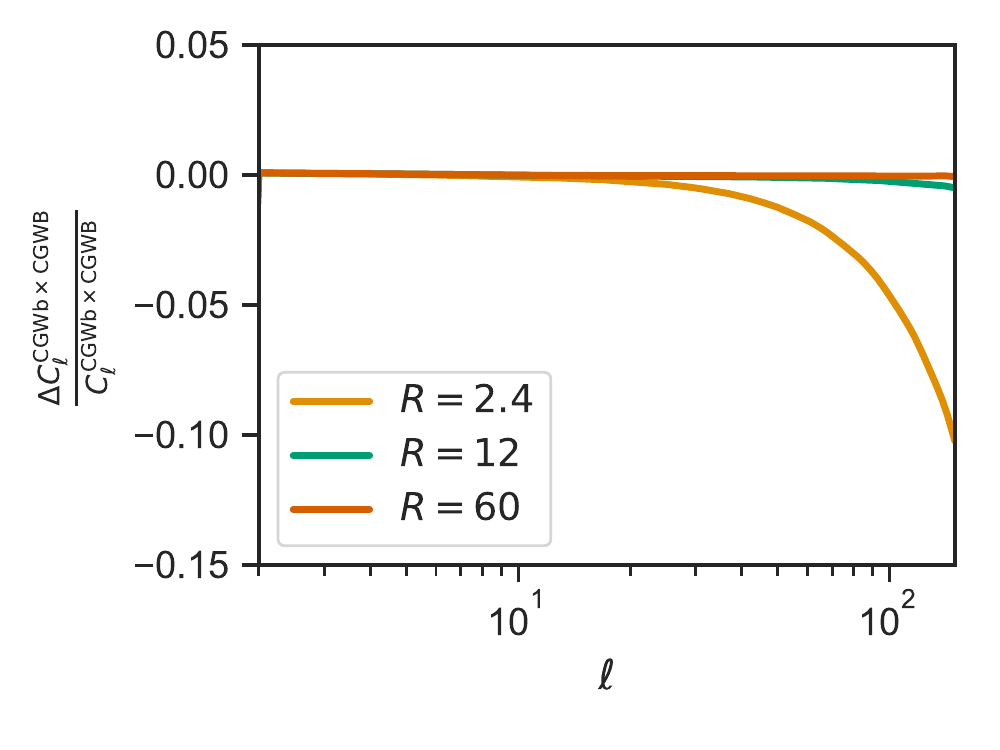}
    \caption{Comparison of the SW effect calculated analytically and with \CLASSGW for different choices of $k_{\rm max}$, controlled by the parameter $R \equiv [k_{\rm max} \tau_0 / \ell_\mathrm{max}]$. Left: power spectrum $\ClGW$, right: relative difference.}
    \label{fig:precision_kmax}
\end{figure}

Using the default value $R=2.4$ for the CGWB anisotropy spectrum leads to numerical artifacts for large $\ell$, see Fig. \ref{fig:precision_kmax}. The figure shows the SW contribution alone compared with its analytic solution (based on an analytic integral over the spherical Bessel function, valid for constant source functions on super-Hubble scales).
In \CLASSGW, the calculation of the $C_\ell^{\rm CGWB\times CGWB}$ spectrum relies on a new precision parameter \verb|k_max_tau0_over_l_max_gwb| set by default to $R=12$. With such a setting, the \CLASSGW spectrum is in excellent agreement with the analytic solution. A further increase of this parameter leads to even better accuracy at the expense of drastically increasing the computation time.

\begin{figure}
    \centering
    \includegraphics[width=0.49\textwidth]{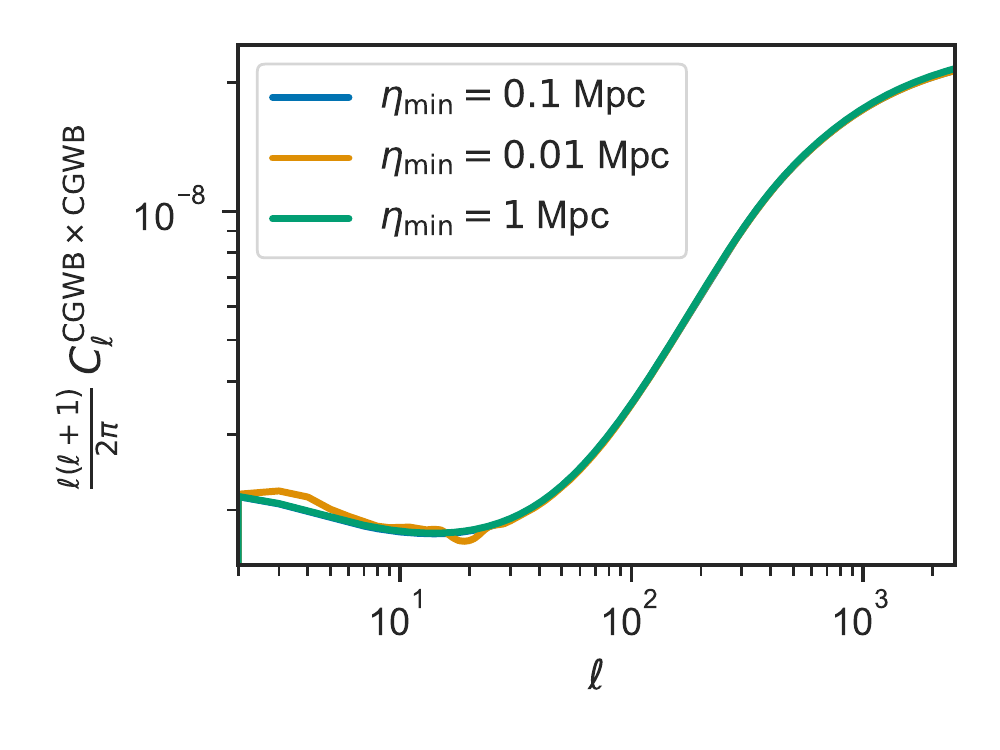}
    \includegraphics[width=0.49\textwidth]{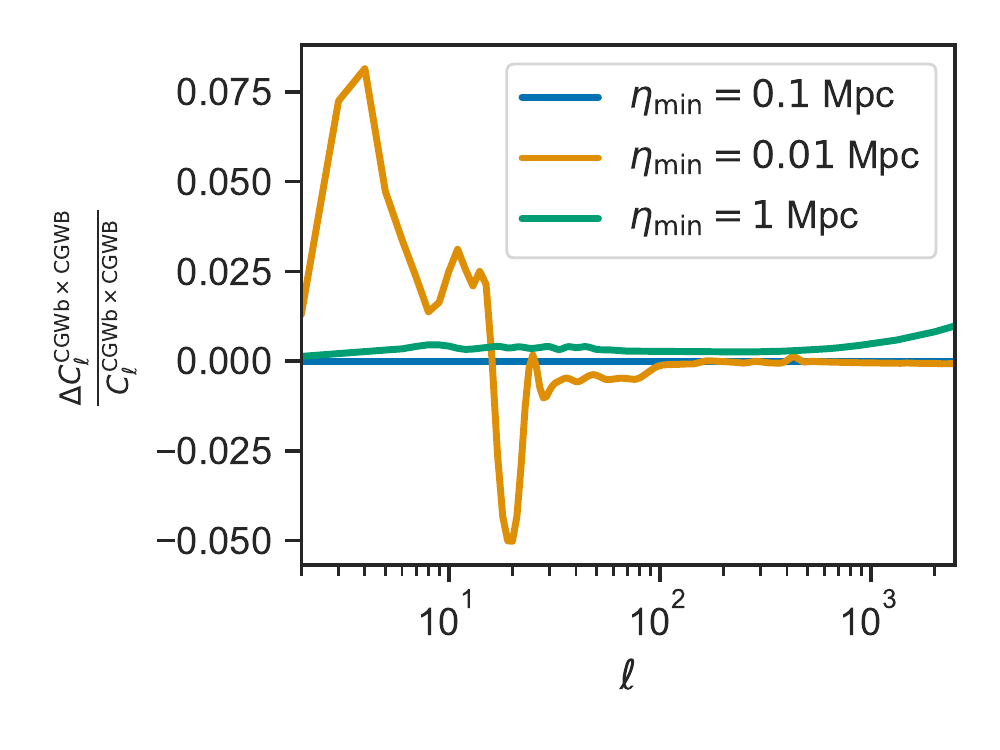}
    \caption{Impact of the parameter $\eta_{\rm min}$ that governs the starting time for of the line-of-sight integral of the ISW contribution, in the Newtonian gauge and using the \texttt{ndf15} evolver. Left: power spectrum $\ClGW$, right: relative difference. The results are more stable in the synchronous gauge.}
    \label{fig:precision_tau}
\end{figure}

Another important precision parameter is $\eta_{\rm min}$ (\verb|start_sources_at_tau_gwb|), which defines the conformal time at which the ISW contribution to GW perturbations starts to be stored in the \verb|perturbation| module -- and thus, starts to be integrated explicitly over the line of sight in the \verb|transfer| module. Starting at too late times $\eta > \SI{1}{Mpc}$ results in loosing contributions to the ISW term, and using imprecise initial conditions for the SW and initial terms. On the other hand, using too early times leads to numerical problems in \CLASS, which was originally developed for starting to store the CMB source functions slightly before photon decoupling. These problems are especially apparent in the Newtonian gauge, while the synchronous gauge is more stable. In \CLASSGW, this parameter is set to 0.1~Mpc. We find that this compromise gives stable and well-converged results in both gauges and using both evolvers. Fig.~\ref{fig:precision_tau} illustrates the impact of changing this parameter.

Finally, we gather in table~\ref{tab:CLASSGW_precision} the default settings for all precision parameters relevant specifically to \CLASSGW. These parameters are defined and further explained in \verb|input/precisions.h|.
\begin{table}[htp]
    \centering
    \begin{tabular}{|l|l|}
        \hline
        Parameter & \CLASSGW \\
        \hline \hline
        $R = [k_{\rm max} \tau_0 / \ell_\mathrm{max}]$ & \verb|k_max_tau0_over_l_max_gwb = 12.0| \\
        \hline
        Starting time of \CLASSGW $\eta_{\rm min}$ $[\si{Mpc}]$ & \verb|start_sources_at_tau_gwb = 0.1| \\
        \hline
        Logarithmic $f$ spacing & \verb|f_per_decade_primordial = 100.| \\
        \hline
        Trigger to neglect transfer function \verb|gwb0| & \verb|transfer_neglect_delta_k_S_gwb0 = 0.3| \\
        Trigger to neglect transfer function \verb|gwb1| & \verb|transfer_neglect_delta_k_S_gwb1 = 0.2| \\
        Trigger to neglect transfer function \verb|gwb2| & \verb|transfer_neglect_delta_k_T_gwb2 = 0.2| \\
        \hline
    \end{tabular}
    \caption{Precision parameter defined in \texttt{precisions.h}. These should not be changed, unless one is aware of the consequences.}
    \label{tab:CLASSGW_precision}
\end{table}

%\clearpage

\bibliographystyle{utphys}
\bibliography{Biblio.bib}

\end{document}